\documentclass{article}
\usepackage[utf8]{inputenc}
\usepackage{a4wide}
\RequirePackage{amsmath}
\usepackage{url}
\usepackage{bbding}
\usepackage{makeidx}  
\usepackage{mathpartir}
\usepackage{xspace}
\usepackage{framed}
\usepackage[dvipsnames]{xcolor}
\definecolor{shadecolor}{rgb}{0.95,0.95,0.95}
\usepackage{stmaryrd}
\usepackage{listings}

\usepackage{paralist}
\usepackage{placeins} 

\usepackage{amssymb}

\usepackage{pgfplots}
\pgfplotsset{width=7cm,compat=1.8}
\usepackage{pgfplotstable}

\usepackage{float}  
\floatstyle{plain}
\restylefloat{figure}
\usepackage{caption}
\DeclareCaptionFormat{myformat}{\hrulefill\\#1#2#3}
\newenvironment{myfigure}[1][]{%
    \begin{figure}[#1]\begin{center}\begin{shaded}\begin{minipage}{\columnwidth}%
}{%
    \end{minipage}\end{shaded}\end{center}\end{figure}%
}

\newcommand{\source}{\textsc{ImpEff}\xspace}
\newcommand{\target}{\textsc{ExEff}\xspace}
\newcommand{\erased}{\textsc{SkelEff}\xspace}
\newcommand{\noeff}{\textsc{NoEff}\xspace}

\newsavebox{\mybox}
\newcommand{\SystemFC}{System F$_\text{C}$\xspace}
\newcommand{\SystemF}{System F\xspace}

\definecolor{superlightgray}{gray}{0.80}
\newcommand{\highlight}[1]{\setlength{\fboxsep}{2pt}\colorbox{superlightgray}{\ensuremath{#1}}}

\newcommand{\vdashNamedD}[1]{\vdash_{{\scriptscriptstyle \hspace{-1mm}\mathtt{#1}}}}

\newcommand{\ruleform}[1]{\fbox{$#1$}}

\newcommand{\splashName}{\mathit{split}}
\newcommand{\splash}[3]{\splashName({#1},{#2},{#3})}

\newcommand{\ftv}[1]{\mathit{fv_\tyvar}({#1})}
\newcommand{\fdv}[1]{\mathit{fv_\dirtvar}({#1})}
\newcommand{\fsv}[1]{\mathit{fv_\evar}({#1})}
\newcommand{\fv}[1]{\mathit{fv}({#1})}

\newcommand{\constraint}{\pi}       

\newcommand{\keylet}{\mathtt{let}}

\newcommand{\keyin}{\mathtt{in}}
\newcommand{\keyreturn}{\mathtt{return}}
\newcommand{\keydo}{\mathtt{do}}
\newcommand{\keyfun}{\mathtt{fun}}

\newcommand{\bang}{\mathrel{!}}                   

\newcommand{\withdirt}[2]{{#1} \bang {#2}}                            
\newcommand{\letval}[3]{\keylet~{#1} = {#2}~\keyin~{#3}}              
\newcommand{\return}[1]{\keyreturn~{#1}}                              

\newcommand{\set}[1]{\{ #1 \}}

\newcommand{\kord}[1]{\mathtt{#1}}
\newcommand{\kop}[1]{\;\mathtt{#1}\;}
\newcommand{\kpre}[1]{\mathtt{#1}\;}
\newcommand{\kpost}[1]{\;\mathtt{#1}}

\newcommand{\tyInt} {\mathtt{Int}}
\newcommand{\hto}{\Rrightarrow}
\newcommand{\C}{\underline{C}}
\newcommand{\D}{\underline{D}}
\newcommand{\dirt}{\Delta}
\newcommand{\sig}{\Sigma}

\newcommand{\call}[3]{{{#1}\,{#2}\,{#3}}}
\newcommand{\case}{\mathop{\text{\texttt{|}}}}

\newcommand{\handler}[1]{\{ #1 \}}

\newcommand{\letin}[1]{\kpre{let} #1 \kop{in}}

\newcommand{\op}{\mathtt{Op}}
\newcommand{\ops}{\mathcal{O}}

\newcommand{\ret}{\kpre{return}}
\newcommand{\withhandle}[2]{\kpre{handle} #2 \kop{with} #1}
\newcommand{\longcases}{\call{\op_1}{x}{k} \mapsto c_{\op_1}, \ldots, \call{\op_n}{x}{k} \mapsto c_{\op_n}}
\newcommand{\shortcases}{[\call{\op}{x}{k} \mapsto c_\op]_{\op \in \ops}}
\newcommand{\shortcasesprimed}{[\call{\op}{x}{k} \mapsto c_\op']_{\op \in \ops}}

\newcommand{\shorthand}[1][\ret x \mapsto c_r]{\handler{#1, \shortcases}}

\newcommand{\ctx}{\Gamma}

\newcommand{\E}{\mathrel{!}}

\renewcommand{\le}{\leqslant}

\newcommand{\tyvar}{\alpha}
\newcommand{\covar}{\omega}
\newcommand{\dirtvar}{\delta}

\newcommand{\valRes}{v^R}
\newcommand{\compRes}{c^R}
\newcommand{\termVal}{v^T}
\newcommand{\termComp}{c^T}

\newcommand{\aVty}{\mathit{A}}
\newcommand{\bVty}{\mathit{B}}
\newcommand{\cCty}{\underline{\mathit{C}}}
\newcommand{\dCty}{\underline{\mathit{D}}}
\newcommand{\qualTy}{\mathit{K}}
\newcommand{\polyTy}{\mathit{S}}

\newcommand{\srcPartialCt}{\pi}
\newcommand{\srcFullCt}{\rho}

\newcommand{\trgPartialCt}{\pi}
\newcommand{\trgFullCt}   {\rho}

\newcommand{\fun} [2]{\keyfun~{#1} \mapsto {#2}}
\newcommand{\doin}[3]{\keydo~{#1} \leftarrow {#2} ;{#3}}
\newcommand{\cast}[2]{{#1} \vartriangleright {#2}}
\newcommand{\operation}[4][Op]{\mathtt{#1}~{#2}~({#3}. {#4})}

\newcommand{\vty}{\mathit{T}}             
\newcommand{\cty}{\underline{\mathit{C}}} 

\newcommand{\coercion}{\gamma} 

\newcommand{\skeletonOf}[1]{\mathit{skeleton}({#1})}

\newcommand{\tcSkeleton}[2]{{#1} \vdashNamedD{\ety} {#2}}

\newcommand{\tcTrgVal} [3]{{#1} \vdashNamedD{v} {#2} : {#3}}
\newcommand{\tcTrgComp}[3]{{#1} \vdashNamedD{c} {#2} : {#3}}
\newcommand{\tcTrgVty} [3]{{#1} \vdashNamedD{\vty}  {#2} : {#3}}
\newcommand{\tcTrgCty} [3]{{#1} \vdashNamedD{\cty}  {#2} : {#3}}

\newcommand{\tcTrgCo}[3]{{#1} \vdashNamedD{co} {#2} : {#3}}
\newcommand{\wfTrgCt}[2]{{#1} \vdashNamedD{\trgFullCt} {#2}}

\newcommand{\wfDirt}[2]{{#1} \vdashNamedD{\dirt} {#2}}

\newcommand{\tcNoEffTy}           [2]{{#1} \vdashNamedD{\mlTyA}       {#2}} 
\newcommand{\tcNoEffCoTy}         [2]{{#1} \vdashNamedD{\mlCoTy}     {#2}} 
\newcommand{\tcNoEffCoercion}     [3]{{#1} \vdashNamedD{co}       {#2} : {#3}} 
\newcommand{\tcNoEffTm}           [3]{{#1} \vdashNamedD{t}        {#2} : {#3}} 

\newcommand{\mlSig}{\Sigma}
\newcommand{\mlEnv}{\Gamma}

\newcommand{\mlTyA}{A}
\newcommand{\mlTyB}{B}
\newcommand{\mkMlCompTy}[1]{\mathtt{Comp}~{#1}} 
\newcommand{\mkMlCompCo}[1]{\mathtt{comp}~{#1}} 

\newcommand{\mlCoTy}{\pi}        
\newcommand{\mlCoercion}{\gamma} 

\newcommand{\mlTm}{\mathit{t}}
\newcommand{\mlHandler}{h}
\newcommand{\mlShortcases}{[\call{\op}{x}{k} \mapsto \mlTm_\op]_{\op \in \ops}}
\newcommand{\mlShorthand}[1][\return{(x : \mlTyA)} \mapsto \mlTm_r]{\handler{#1, \mlShortcases}}

\newcommand{\mlCoUnitRefl}{\langle \tyUnit \rangle}
\newcommand{\keyunsafe}{\mathtt{unsafe}}
\newcommand{\unsafe}[1]{\keyunsafe~{#1}}
\newcommand{\keyhandToFun}{\mathtt{handToFun}}
\newcommand{\handToFun}[2]{\keyhandToFun~{#1}~{#2}}
\newcommand{\keyfunToHand}{\mathtt{funToHand}}
\newcommand{\funToHand}[2]{\keyfunToHand~{#1}~{#2}}
\newcommand{\mlReturn}[1]{\return{#1}}

\newcommand{\mlSmallStep}[2]{{#1} \leadsto {#2}}

\newcommand{\mlValue}{\mlTm^R}

\newcommand{\fullDirt}[1]{\mathit{nonEmpty}({#1})}
\newcommand{\valTyToNoEff} [4]{{#1} \vdashNamedD{\vty} {#2} : {#3} \highlight{\rightsquigarrow {#4}}} 
\newcommand{\compTyToNoEff}[4]{{#1} \vdashNamedD{\cty} {#2} : {#3} \highlight{\rightsquigarrow {#4}}} 
\newcommand{\valToNoEff}   [4]{{#1} \vdashNamedD{v}    {#2} : {#3} \highlight{\rightsquigarrow {#4}}} 
\newcommand{\compToNoEff}  [4]{{#1} \vdashNamedD{c}    {#2} : {#3} \highlight{\rightsquigarrow {#4}}} 
\newcommand{\coToNoEff}    [4]{{#1} \vdashNamedD{co}  {#2} : {#3} \highlight{\rightsquigarrow {#4}}} 
\newcommand{\tyEnvToNoEff} [2]{\vdashNamedD{\tyEnv} {#1} \highlight{\rightsquigarrow {#2}}} 

\newcommand{\reflOf}[1]{\mathit{reflOf}({#1})}

\newcommand{\fromImpureVal} [4][\ctx]{\dirtvar \mapsto {#3}; {#1} \vdashNamedD{v}   {#2} \highlight{\rightsquigarrow {#4}}}
\newcommand{\fromImpureComp}[4][\ctx]{\dirtvar \mapsto {#3}; {#1} \vdashNamedD{c}   {#2} \highlight{\rightsquigarrow {#4}}}
\newcommand{\toImpureVal} [4][\ctx]{{#3} \mapsto \dirtvar; {#1} \vdashNamedD{v}   {#2} \highlight{\rightsquigarrow {#4}}}
\newcommand{\toImpureComp}[4][\ctx]{{#3} \mapsto \dirtvar; {#1} \vdashNamedD{c}   {#2} \highlight{\rightsquigarrow {#4}}}

\newcommand{\mlStuck}{\mlTm^S}

\newcommand{\bigStepVal} [2]{{#1} \leadsto_\mathrm{v}^{*} {#2}}
\newcommand{\bigStepComp}[2]{{#1} \leadsto_\mathrm{c}^{*} {#2}}

\newcommand{\smallStepVal} [3][]{{#2} \leadsto_\mathrm{v}^{#1} {#3}}
\newcommand{\smallStepComp}[3][]{{#2} \leadsto_\mathrm{c}^{#1} {#3}}
\newcommand{\congStepVal}[2]{{#1} \equiv^{\leadsto}_\mathrm{v} {#2}}
\newcommand{\congStepComp}[2]{{#1} \equiv^{\leadsto}_\mathrm{c} {#2}}
\newcommand{\trgShorthand}[1][\return{(x : \vty_x)} \mapsto c_r]{\handler{#1, \shortcases}}

\newcommand{\tyUnit}{\mathtt{Unit}}
\newcommand{\tmUnit}{\mathtt{unit}}

\newcommand{\tcElabVal} [4]{{#1} \vdash_{v} {#2} : {#3} \highlight{\rightsquigarrow {#4}}}
\newcommand{\tcElabComp}[4]{{#1} \vdash_{c} {#2} : {#3} \highlight{\rightsquigarrow {#4}}}
\newcommand{\tcElabPolyTy}[4]{{#1} \vdashNamedD{vty} {#2} : {#3} \highlight{\rightsquigarrow {#4}}}
\newcommand{\tcElabCompTy}[4]{{#1} \vdashNamedD{cty} {#2} : {#3}  \highlight{\rightsquigarrow {#4}}}
\newcommand{\tcElabConstraint}[3]{{#1} \vdashNamedD{ct} {#2} \highlight{\rightsquigarrow {#3}}}
\newcommand{\proveCt}[3]{{#1} \vdashNamedD{co} \highlight{{#2} :} {#3}}

\newcommand{\tyEnv}{\Gamma}

\newcommand{\tcVal} [3]{{#1} \vdash_{v} {#2} : {#3}}
\newcommand{\tcComp}[3]{{#1} \vdash_{c} {#2} : {#3}}
\newcommand{\tcProveCt}[2]{{#1} \vdashNamedD{co} {#2}}
\newcommand{\tcPolyTy}[3]{{#1} \vdashNamedD{vty} {#2} : {#3}}
\newcommand{\tcCompTy}[3]{{#1} \vdashNamedD{cty} {#2} : {#3}}
\newcommand{\tcConstraint}[2]{{#1} \vdashNamedD{ct} {#2}}

\newcommand{\inferStVal} [7]{{#1} ; {#2} \vdashNamedD{v} {#3} : {#4} \mid {#5} ; {#6} \highlight{\rightsquigarrow {#7}}}
\newcommand{\inferStComp}[7]{{#1} ; {#2} \vdashNamedD{c} {#3} : {#4} \mid {#5} ; {#6} \highlight{\rightsquigarrow {#7}}}

\newcommand{\constraints}{\mathcal{Q}}

\newcommand{\elabCty}[1]{\mathit{elab}_{\scriptscriptstyle\cCty}({#1})} 
\newcommand{\elabVty}[1]{\mathit{elab}_{\hspace{-0.2mm}\scriptscriptstyle\polyTy}({#1})} 

\newcommand{\elabConstraint}[1]{\mathit{elab}_{\hspace{-0.2mm}\scriptscriptstyle\srcFullCt}({#1})} 

\newcommand{\elabTyEnv}[1]{\mathit{elab}_{\hspace{-0.2mm}\scriptscriptstyle\tyEnv}({#1})}

\newcommand{\fcmp}{\mathbin{\raise 0.6ex\hbox{\oalign{\hfil$\scriptscriptstyle \mathrm{o}$\hfil\cr\hfil$\scriptscriptstyle\mathrm{9}$\hfil}}}}
\newcommand{\refl}[1]{\langle {#1} \rangle}
\newcommand{\trgUnitRefl}{\langle \tyUnit \rangle}
\newcommand{\dirtRefl}[1]{\langle {#1} \rangle}

\newcommand{\unify}{\mathtt{solve}}
\newcommand{\sol}{\sigma}
\newcommand{\cstr}{\mathcal{P}}
\newcommand{\queue}{\mathcal{Q}}

\newcommand{\erasedshortcases}{[\call{\op}{x}{k} \mapsto \eraseC{c_\op}]_{\op \in \ops}}

\newcommand{\ersShorthand}[1][\return{(x : \ety_x)} \mapsto c_r]{\handler{#1, \shortcases}}
\newcommand{\ety}{\tau}
\newcommand{\evar}{\varsigma}
\newcommand{\tcErsEty} [2]{{#1} \vdashNamedD{\ety} {#2}}
\newcommand{\tcErsVal} [3]{{#1} \vdashNamedD{ev} {#2} : {#3}}
\newcommand{\tcErsComp} [3]{{#1} \vdashNamedD{ec} {#2} : {#3}}
\newcommand{\eraseV}[2][\sigma]{\epsilon^{#1}_{\mathrm{v}}(#2)}
\newcommand{\eraseC}[2][\sigma]{\epsilon^{#1}_{\mathrm{c}}(#2)}
\newcommand{\eraseVT}[2][\sigma]{\epsilon^{#1}_{\mathrm{V}}(#2)}
\newcommand{\eraseCT}[2][\sigma]{\epsilon^{#1}_{\mathrm{C}}(#2)}
\newcommand{\eraseEnv}[2][\sigma]{\epsilon^{#1}_{\mathrm{E}}(#2)}

\newcommand{\keyleftinv}{\mathtt{left}}
\newcommand{\leftinv}[1]{\keyleftinv~({#1})}

\newcommand{\citep}[1]{\cite{#1}}
\newcommand{\shortcite}[1]{\cite{#1}}
\newcommand{\citet}[2]{{#1} \shortcite{#2}} 

\makeatletter 
\def\arcr{\@arraycr}
\makeatother

\lstdefinestyle{eff}{%
language=Caml,
moredelim=*[is][\itshape]{/@}{@/},
numbers=none,mathescape=true,showstringspaces=false,
morekeywords={handle,handler,with,return},
keywordstyle=\bfseries,
xleftmargin=1em,basicstyle=\ttfamily\small}

\lstdefinestyle{ocaml}{%
language=Caml,
moredelim=*[is][\itshape]{/@}{@/},
numbers=none,mathescape=true,showstringspaces=false,
keywordstyle=\bfseries,
xleftmargin=1em,basicstyle=\ttfamily\small}

\lstnewenvironment{efflisting}{\lstset{style=eff,basicstyle=\linespread{0.999}\ttfamily\small\color{magenta}}}{}
\lstnewenvironment{ocamllisting}{\lstset{style=ocaml,basicstyle=\linespread{0.999}\ttfamily\small\color{teal}}}{}

\begin{document}

\title
      {Explicit Effect Subtyping}

 \author{
        Georgios Karachalias\footnotemark[1]
        \and
        Matija Pretnar\footnotemark[2]
        \and
        Amr Hany Saleh\footnotemark[1]
        \and
        Stien Vanderhallen\footnotemark[1]
        \and
        Tom Schrijvers\footnotemark[1]
 }
\date{
\footnotemark[1]\ KU Leuven, Department of Computer Science, Belgium
\\
\footnotemark[2]\ University of Ljubljana, Faculty of Mathematics and Physics, Slovenia
}


\newtheorem{theorem}{Theorem}[section]
\newtheorem{corollary}[theorem]{Corollary}
\newtheorem{lemma}[theorem]{Lemma}
\newtheorem{conjecture}[theorem]{Conjecture}
\newtheorem{example}[theorem]{Example}


\maketitle

\begin{abstract}

As popularity of algebraic effects and handlers increases, so does a demand for
their efficient execution. Eff, an ML-like language with native support for
handlers, has a subtyping-based effect system on which an effect-aware
optimizing compiler could be built. Unfortunately, in our experience,
implementing optimizations for Eff is overly error-prone because its core
language is implicitly-typed, making code transformations very fragile.

To remedy this, we present an explicitly-typed polymorphic
core calculus for algebraic effect handlers with a subtyping-based
type-and-effect system. It reifies appeals to subtyping in explicit casts
with coercions that witness the subtyping proof, quickly exposing typing bugs
in program transformations.

Our typing-directed elaboration comes with a constraint-based inference algorithm
that turns an implicitly-typed Eff-like language into our calculus.
Moreover, all coercions and effect information can be erased in a straightforward way,
demonstrating that coercions have no computational content.
Additionally, we present a monadic translation from our calculus into a pure language without algebraic effects or handlers, using the effect information to introduce monadic constructs only where necessary.
\end{abstract}

\tableofcontents


\section{Introduction}\label{sec:intro}
Algebraic effect
handlers~\citep{DBLP:journals/acs/PlotkinP03,DBLP:journals/corr/PlotkinP13} are
quickly maturing from a theoretical model to a practical language feature for
user-defined computational effects. Yet, in practice they still incur a
significant performance overhead compared to native effects.

Our earlier efforts~\citep{optimization} to narrow this gap with an optimising
compiler from Eff~\citep{bauer15} to OCaml showed promising results, in some
cases reaching even the performance of hand-tuned code, but were very fragile and have been
postponed until a more robust solution is found.
We believe the main reason behind these and other\footnote{%
See issues \#11 and \#16 at \url{https://github.com/matijapretnar/eff/issues/}.
} problems lies in the complexity of
subtyping in combination with the implicit typing of Eff's core language,
further aggravated by the ``garbage collection''~\cite{pottier2001simplifying} of subtyping constraints (see Section~\ref{sec:conclusion}).

For efficient compilation, one must avoid the poisoning
problem~\citep{DBLP:conf/popl/WansbroughJ99},
where unification forces a pure computation to take the less precise
impure type of the context (e.g.\ a pure and an impure branch of a conditional
both receive the same impure type).
Since this rules out existing (and likely simpler) effect systems for handlers based on
row-polymorphism~\citep{DBLP:journals/corr/Leijen14,links_rows,DBLP:conf/popl/LindleyMM17}, we propose a polymorphic explicitly-typed calculus based on subtyping. More specifically, our contributions are as follows:

\begin{itemize}
\item First, in Section~\ref{sec:source} we present \source, a polymorphic
implicitly-typed calculus for algebraic effects and handlers with a subtyping-based
type-and-effect system. \source is essentially a (desugared) source language as
it appears in the compiler frontend of a language like Eff.

\item Next, Section~\ref{sec:target} presents \target, the core calculus, which
combines explicit System F-style polymorphism with explicit coercions for
subtyping in the style of \citet{Breazu-Tannen et al.}{coercions}. This calculus comes with a
type-and-effect system, a small-step operational semantics and a proof of type-safety.

\item Section~\ref{sec:inference} specifies the typing-directed elaboration of
\source into \target and presents a type inference algorithm for \source that
produces the elaborated \target term as a by-product. It also establishes that
the elaboration preserves typing, and that the algorithm is sound with respect
to the specification and yields principal types.

\item Finally, we present two different backends for \target:
\begin{itemize}
\item
Section~\ref{sec:erasure} defines \erased, which is a variant of
\target without effect information or coercions. \erased is also
representative of Multicore OCaml's support for algebraic effects and
handlers~\citep{ocaml}, which is a possible compilation target of Eff. By
showing that the erasure from \target to \erased preserves semantics, we
establish that \target's coercions are computationally irrelevant.
To enable erasure, \target annotates its types with \emph{(type)
skeletons}, which capture the erased counterpart and are, to our knowledge,
a novel contribution.
\item
Section~\ref{sec:elaboration-to-ocaml} defines \noeff, which is an alternative
backend of \target which tracks in its type system whether, but not which,
effects can happen. This backend is representative of pure OCaml or Haskell
code where effectful computations are represented with a free monad
implementation. Because \noeff lacks effect polymorphism, our type-preserving elaboration from
\target to \noeff needs to introduce unsafe coercions, though we claim that elaborated programs never get stuck.
\end{itemize}
\item
Our paper comes with two software artefacts:
an ongoing implementation\footnote{%
  \url{https://github.com/matijapretnar/eff/tree/explicit-effect-subtyping}
} of a compiler from Eff to OCaml with \target at its core,
and an Abella mechanisation\footnote{%
  \url{https://github.com/matijapretnar/proofs/tree/jfp-2019/explicit-effect-subtyping}
}
of Theorems~\ref{thm:safety},
\ref{thm:type_preservation},
\ref{thm:erasure_type_preservation},
\ref{thm:erasure_semantic_preservation},
\ref{thm:ml-preservation},
\ref{thm:ml-progress}
and \ref{thm:eff-to-ml-type-preservation}.
Remaining theorems all concern the inference algorithm,
and their proofs closely follow~\cite{DBLP:journals/corr/Pretnar13}.
\end{itemize}

\noindent
This article is an extended version of a paper that appeared at ESOP
2018~\cite{esop2018}.  There are two main novelties. Firstly, we have altered
the coercion forms available in \target.  Previously, it contained a range of
projection forms to support an operational semantics that never matches on the
coercions. Instead, we now do match on the coercions in the operational
semantics, and as a consequence no longer need the projections. This not only
reduces the size of the language but also has a considerable simplifying impact
on the metatheory proofs in Abella. Moreover, it reduces the gap between \target
and  \noeff. Secondly and most importantly, Section~\ref{sec:elaboration-to-ocaml}, on the elaboration of \target
to \noeff, is entirely new.

\section{Overview}\label{sec:overview}
This section presents an informal overview of the \target calculus, and the
main issues with elaborating to and erasing from it.

\subsection{Algebraic Effect Handlers}

The main premise of algebraic effects is that impure behaviour arises from a set
of \emph{operations} such as $\mathtt{Get}$ and $\mathtt{Set}$ for mutable
store, $\mathtt{Read}$ and $\mathtt{Print}$ for interactive input and output, or
$\mathtt{Raise}$ for exceptions~\citep{DBLP:journals/acs/PlotkinP03}. This
allows generalizing exception handlers to other effects,
to express backtracking, co-operative multithreading and other examples in a
natural way~\citep{DBLP:journals/corr/PlotkinP13,bauer15}.

Assume operations $\mathtt{Tick} : \tyUnit \rightarrow \tyUnit$ and $\mathtt{Tock} : \tyUnit \rightarrow \tyUnit$ that take a unit
value as a parameter and yield a unit value as a result. Unlike special built-in
operations, these operations have no intrinsic effectful behaviour, though we
can give one through handlers. For example, the handler
\begin{align*}
    \handler{
        &\call{\mathtt{Tick}}{x}{k} \mapsto (\mathtt{Print} \text{``tick''}; k~\tmUnit), \\
        &\call{\mathtt{Tock}}{x}{k} \mapsto \mathtt{Print} \text{``tock''}
    }
\end{align*}
replaces all calls of $\mathtt{Tick}$ by printing out ``tick'' and similarly for $\mathtt{Tock}$.
But there is one significant difference between the two cases. Unlike exceptions, which
always abort the evaluation, operations have a continuation waiting for their result.
It is this continuation that the handler captures in the variable~$k$ and potentially uses in
the handling clause. In the clause for $\mathtt{Tick}$, the continuation is resumed by passing it
the expected unit value, whereas in the clause for $\mathtt{Tock}$, the operation is discarded. Thus, if we handle a computation emitting the two operations, it will print out ``tick''
until a first ``tock'' is printed, after which the evaluation stops.
For a more thorough explanation of algebraic effect handlers, we refer the reader to Pretnar's tutorial~\citep{pretnar:tutorial},
which is conveniently based on a calculus with essentially the same term-level syntax and
operational semantics (but a far less involved type system).

\subsection{Elaborating Subtyping}

Consider the computation
$\doin{x}{\mathtt{Tick}~{\tmUnit}}{f~x}$
and assume that $f$ has
the function type $\tyUnit \to \tyUnit~!~\{\mathtt{Tock}\}$, taking unit values
to unit values and perhaps calling $\mathtt{Tock}$ operations in the process.
The whole computation then has the type
$\tyUnit~!~\{\mathtt{Tick},\mathtt{Tock}\}$ as it returns the unit value and may
call $\mathtt{Tick}$ and $\mathtt{Tock}$.

The above typing implicitly appeals to subtyping in several places. For instance,
$\mathtt{Tick}~{\tmUnit}$ has type $\tyUnit~!~\{\mathtt{Tick}\}$
and $f~x$ type $\tyUnit~!~\{\mathtt{Tock}\}$. Yet, because they are sequenced with
$\keydo$, the type system expects them to have the same set of effects. The discrepancies
are implicitly reconciled by the subtyping which admits both $\{\mathtt{Tick}\} \le \{\mathtt{Tick},\mathtt{Tock}\}$
and $\{\mathtt{Tock}\} \le \{\mathtt{Tick},\mathtt{Tock}\}$.

We elaborate the \source term into the explicitly-typed core language \target,
where such implicit appeals to subtyping turn into explicit casts using \emph{coercions}:
\[\doin{x}{(\cast{(\mathtt{Tick}~{\tmUnit})}{\coercion_1})}{\cast{(f~x)}{\coercion_2}}\]
A coercion $\coercion$ is a witness for a subtyping $A~!~\dirt \le A'~!~\dirt'$ and can be used
to cast a term $c$ of type $A~!~\dirt$ to a term $\cast{c}{\coercion}$ of type $A'~!~\dirt'$.
In the above term, $\coercion_1$ and $\coercion_2$ respectively witness
$\tyUnit~!~\{\mathtt{Tick}\} \le \tyUnit~!~\{\mathtt{Tick},\mathtt{Tock}\}$ and
$\tyUnit~!~\{\mathtt{Tock}\} \le
\tyUnit~!~\{\mathtt{Tick},\mathtt{Tock}\}$.

At this point, the reader
might wonder why coercions can influence value types, and not just effect sets.
This design allows us to flexibly cast types of higher-order functions and
handlers which would otherwise not be possible. For example, we can use a
coercion for $\dirtvar_3 \le \dirtvar_1$ to construct value type coercions that witnesses
\[
((\tyvar \to \tyvar'~!~\dirtvar_1) \to \tyvar''~!~\dirtvar_2)
\le
((\tyvar \to \tyvar'~!~\dirtvar_3) \to \tyvar''~!~\dirtvar_2)
\]
or
\[
(\tyvar'~!~\dirtvar_1 \hto \tyvar''~!~\dirtvar_2)
\le
(\tyvar'~!~\dirtvar_3 \hto \tyvar''~!~\dirtvar_2)
\]

\subsection{Polymorphic Subtyping for Types and Effects}
\label{sec:overview:polymorphism}

The above basic example only features monomorphic types and effects. Yet, our
calculus also supports polymorphism, which makes it considerably more expressive.
For instance the type of $\mathit{f}$ in $\letval{\mathit{f}}{(\fun{g}{g~\tmUnit})}{\ldots}$
is generalised to:
\[\forall \tyvar, \tyvar'.\forall \dirtvar,\dirtvar'.\tyvar \le \tyvar'\Rightarrow \dirtvar \le \dirtvar' \Rightarrow (\tyUnit \to \tyvar~!~\dirtvar) \to \tyvar' ~!~ \dirtvar'\]
This polymorphic type scheme follows the qualified types convention~\citep{markjones} where the
type $(\tyUnit \to \tyvar~!~\dirtvar) \to \tyvar' ~!~ \dirtvar'$ is subjected to
several qualifiers, in this case $\tyvar \le \tyvar'$ and $\dirtvar \le \dirtvar'$. The universal
quantifiers on the outside bind the type variables $\tyvar$ and $\tyvar'$, and the
effect set variables $\dirtvar$ and $\dirtvar'$.

The elaboration of $f$ into \target introduces explicit binders for both the
quantifiers and the qualifiers, as well as the explicit casts where subtyping
is used.
\[
\begin{array}{l}
  \Lambda \tyvar.\Lambda \tyvar'\!\!.\Lambda \dirtvar.\Lambda \dirtvar'\!\!.\Lambda (\covar\!:\!\tyvar \le \tyvar'). \Lambda (\covar'\!:\!\dirtvar \le \dirtvar'). \\
  \qquad\fun{(g\!:\!\tyUnit \to \tyvar \,!\, \dirtvar)\!}{\!(\cast{(g\,\tmUnit)\!}{\!(\covar\,!\,\covar')})} \\
\end{array}
\]
Here the binders for qualifiers introduce \emph{coercion variables}
$\covar$ between pure types and $\covar'$ between operation sets, which are then
combined into a \emph{computation coercion} $\covar~!~\covar'$ and used for casting the function application
$(g\,\tmUnit)$ to the expected type.

Suppose that $h$ has type $\tyUnit \to \tyUnit\,!\,\{\mathtt{Tick}\}$ and $f\,h$
type $\tyUnit\,!\,\{\mathtt{Tick},\mathtt{Tock}\}$. In the \target calculus the
corresponding instantiation of $f$ is made explicit through type and coercion
applications
\[ f\,\tyUnit\,\tyUnit\,\{\mathtt{Tick}\}\,\{\mathtt{Tick},\mathtt{Tock}\}\,\coercion_1\,\coercion_2\,h\]
where $\coercion_1$ needs to be a witness for $\tyUnit \le \tyUnit$ and
$\coercion_2$ for $\{\mathtt{Tick}\} \le \{\mathtt{Tick},\mathtt{Tock}\}$.

\subsection{Guaranteed Erasure with Skeletons}

One of our main requirements for \target is that its effect information and
subtyping can be easily erased. The reason is twofold. Firstly, we want to
show that neither plays a role in the runtime behaviour of \target programs.
Secondly and more importantly, we want to use a conventionally typed (System
F-like) functional language as a backend for the Eff compiler.

At first, erasure of both effect information and subtyping seems easy: simply
drop that information from types and terms. But by dropping the effect variables
and subtyping constraints from the type of $\mathit{f}$, we get
    $\forall \tyvar, \tyvar'. (\tyUnit \to \tyvar) \to \tyvar'$
instead of the expected type
    $\forall \tyvar. (\tyUnit \to \tyvar) \to \tyvar$.
In our naive erasure attempt we have carelessly discarded the connection between
$\tyvar$ and $\tyvar'$. A more appropriate approach to erasure would be to unify
the types in dropped subtyping constraints. However, unifying types may reduce
the number of type variables when they become instantiated, so corresponding
binders need to be dropped, greatly complicating the erasure procedure and its
meta-theory.

Fortunately, there is an easier way by tagging all bound type
variables with \emph{skeletons}, which are bare-bones types
without effect information. For example, the skeleton of a function type
$A \to B~!~\Delta$ is $\ety_1 \to \ety_2$, where $\ety_1$ is the
skeleton of $A$ and $\ety_2$ the skeleton of $B$. In \target every well-formed type
has an associated skeleton, and any two types
$A_1 \le A_2$ share the same skeleton. In particular, binders for type variables
are explicitly annotated with skeleton variables $\evar$. For instance, the actual
type of $\mathit{f}$ is:
\[\forall\evar.\forall (\tyvar:\evar), (\tyvar':\evar).\forall \dirtvar,\dirtvar'.\tyvar \le
\tyvar'\Rightarrow \dirtvar \le \dirtvar' \Rightarrow (\tyUnit \to
\tyvar~!~\dirtvar) \to \tyvar' ~!~ \dirtvar'\]
The skeleton quantifications
and annotations also appear at the term-level:
\begin{equation*}
\Lambda \evar. \Lambda (\tyvar:\evar).\Lambda (\tyvar':\evar).\Lambda \dirtvar.\Lambda \dirtvar'. \Lambda (\covar : \tyvar \le \tyvar'). \Lambda (\covar': \dirtvar \le \dirtvar').
\ldots
\end{equation*}
Now erasure is really easy: we drop not only effect and subtyping-related term
formers, but also type binders and application.
We do retain skeleton binders and applications, which take over the role of (plain) types in the backend
language. In terms, we replace types by their skeletons.
For instance, for $\mathit{f}$ we get:
\begin{equation*}
\Lambda \evar.  \fun{(g : \tyUnit \to \evar)}{g\,\tmUnit} ~~:~~ \forall\evar. (\tyUnit \to \evar) \to \evar
\end{equation*}

\subsection{Elaboration into a Pure Language}

We can drop effectful information only if the targeted language has a native implicit support for algebraic effects at any type. In a pure functional language, effectful computations that yield a result of type $A$ are represented with a user-defined type $\mkMlCompTy{A}$, which typically uses one of the known encodings, such as free monads~\cite{kammar,optimization}, delimited control~\cite{eff2ocaml}, or continuation-passing style~\cite{koka2017}.

Targeting such a language requires a more careful elaboration. For example, \target types $\tyInt~!~\{\mathtt{Tick}\}$ and $\tyInt~!~\{\mathtt{Tock}\}$ are both mapped to a type $\mkMlCompTy{\tyInt}$. The same could be done for the type $\tyInt~!~\emptyset$, but computations of that type are pure and do not require any encoding, so it is more efficient to avoid the library overhead and map the type to the pure type $\tyInt$ directly~\cite{koka2017, optimization}. This difference is the main complicating factor in the elaboration.

Since the computation $\return{5} : \tyInt~!~\emptyset$ is pure, it should be elaborated to $5$ of type $\tyInt$. But if we take a witness $\coercion$ for $\tyInt \le \tyInt$ and $\coercion_1$ for $\emptyset \le \{\mathtt{Tick}\}$, the coerced computation $\cast{(\return{5})}{(\coercion~!~\coercion_1)} : \tyInt~!~\{\mathtt{Tick}\}$ should be elaborated to the lifted value $\return{5} : \mkMlCompTy{\tyInt}$.

However, it is not simply a matter of replacing each cast with a $\keyreturn$. If we further take a witness $\coercion_2$ of $\{\mathtt{Tick}\} \le \{\mathtt{Tick},\mathtt{Tock}\}$, the computation
\[
    \cast{(\cast{(\return{5})}{(\coercion~!~\coercion_1)})}{(\coercion~!~\coercion_2)} : \tyInt~!~\{\mathtt{Tick}, \mathtt{Tock}\}
\]
also has to be elaborated to $\return{5} : \mkMlCompTy{\tyInt}$.
We will see that this is just one of the (smaller) issues that stem from the different treatment of pure and impure computation types, and show how to construct an appropriate elaboration (Section~\ref{sec:eff-to-ml}).

\subsection{Outline}

\begin{figure}[th!]
\pgfimage[width=.9\textwidth]{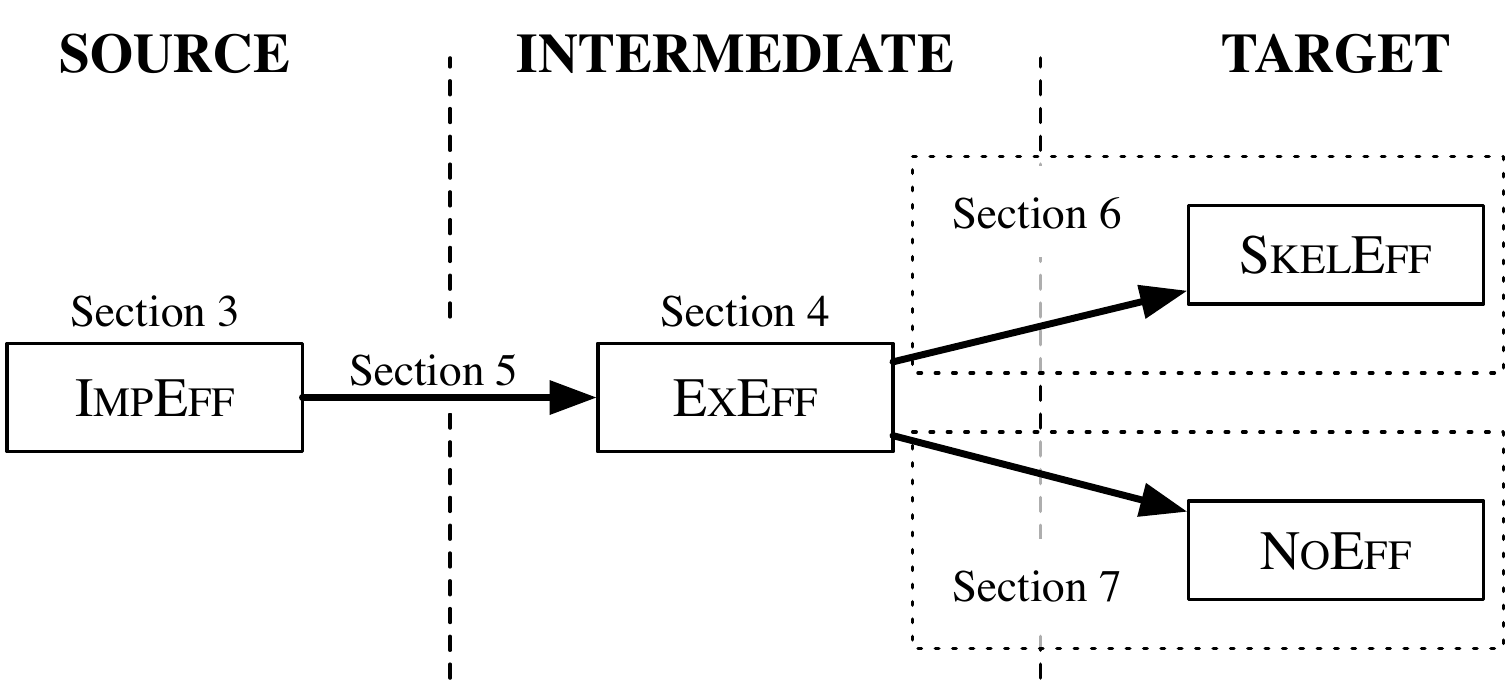}
\caption{Compiler and section structure}\label{fig:outline}
\end{figure}

The remainder of this article formalizes essentially a compiler pipeline for
Eff. Figure~\ref{fig:outline} depicts this pipeline and annotates the different
parts with the sections they are covered in.

\begin{description}
\item[Section~\ref{sec:source}:]
The starting point of the pipeline is \source, an implicitly-typed calculus for
algebraic effects and handlers with a subtyping-based type-and-effect system.
It is the core of the desugared source language as it appears in the compiler
frontend of Eff. We present its syntax and type system.
\item[Section~\ref{sec:target}:]
The heart of the compiler is \target, an intermediate language that 
is explicitly annotated with type and effect information. Its main novelty is that
it also makes appeals to subtyping explicit by means of coercions.
We present its syntax, type system and operational semantics.
\item[Section~\ref{sec:inference}:] 
We explain how to elaborate \source into \target, and provide a type inference algorithm
for \source that performs this elaboration. The algorithm is constraint-based, i.e., it
consists of two interleaved phases: constraint generation and constraint solving.
\item[Section~\ref{sec:erasure}:] Towards the end, the compiler forks to support two 
different compilation targets. The first compilation target is \erased. This
language is modelled after Multicore OCaml. In particular, it is a statically
typed language with built-in support for algebraic effects, but its type system
does not track effects. We provide its syntax and, in the appendix, also its type system
and operational semantics. Also, we explain how to elaborate the intermediate \target
into the \erased target language. Thanks to the skeleton-based setup of \target, this
elaboration is a fairly straightforward erasure procedure.
\item[Section~\ref{sec:elaboration-to-ocaml}:] The second compilation target is
\noeff, a statically typed calculus that distinguishes in its types between
pure and impure computations, but does not track which operations can happen in
impure computations. This models encodings of algebraic effects in languages
without native support. We present its syntax, type system and operational semantics.
Finally, we show how to elaborate \target into \noeff. This is much more involved than
the straightforward erasure procedure into \erased. Instead of
just throwing away all effect information and coercions, we have to
abstract it to the presence (pure) or absence (impure) of effects.
Unfortunately, polymorphism does not interact well with this abstraction
process. We show how to address this problem by conservatively assuming that polymorphic code
is impure and by adding unsafe coercions to obtain pure instantiations.

\end{description}

\section{The \source Language}\label{sec:source}
This section presents \source, a basic functional calculus
with support for algebraic effect handlers, which forms the core
language of our optimising compiler.

\begin{myfigure}[t]
\textbf{Terms}
\[
  \begin{array}{r@{~}c@{~}l}
    \text{value}~v                                                   & ::= &
      x                                                              \mid
      \tmUnit                                                        \mid
      \fun{x}{c}                                                     \mid
      h                                                              \\

    \text{handler}~h                                                 & ::= &
      \{ \return{x} \mapsto c_r, \longcases \}                       \\

    \text{computation}~c                                             & ::= &
      \return{v}                                                     \mid
      \operation{v}{y}{c}                                            \mid
      \doin{x}{c_1}{c_2}                                             \\
                                                                     & \mid &
      \withhandle{v}{c}                                              \mid
      v_1~v_2                                                        \mid
      \letval{x}{v}{c}                                               \\
  \end{array}
\]

\textbf{Types \& Constraints}
\[
  \begin{array}{r@{~}c@{~}l}
    \text{skeleton}~\ety                                             & ::= &
      \evar                                                          \mid
      \tyUnit                                                        \mid
      \ety_1 \to \ety_2                                              \mid
      \ety_1 \hto \ety_2                                             \\
    \\

    \text{value type}~\aVty, \bVty                                   & ::= &
      \tyvar                                                         \mid
      \tyUnit                                                        \mid
      \aVty \to \cCty                                                \mid
      \cCty \hto \dCty                                               \\

    \text{qualified type}~\qualTy                                    & ::= &
      \aVty                                                          \mid
      \srcPartialCt \Rightarrow \qualTy                              \\

    \text{polytype}~\polyTy                                          & ::= &
      \qualTy                                                        \mid
      \forall \evar. \polyTy                                         \mid
      \forall \tyvar\!:\!\ety. \polyTy                               \mid
      \forall \dirtvar. \polyTy                                      \\

    \text{computation type}~\cCty, \dCty                             & ::= &
      \aVty\,!\,\dirt                                                \\

    \text{dirt}~\dirt                                                & ::= &
      \dirtvar                                                       \mid
      \emptyset                                                      \mid
      \{ \op \} \cup \dirt                                           \\

    \\

    \text{simple constraint}~\srcPartialCt                           & ::= &
      \aVty_1 \le \aVty_2                                            \mid
      \dirt_1 \le \dirt_2                                            \\

    \text{constraint}~\srcFullCt                                     & ::= &
      \srcPartialCt                                                  \mid
      \cCty \le \dCty                                                
  \end{array}
\]

\vspace{-5mm}
\caption{\source Syntax}
\label{fig:source-syntax}
\end{myfigure}

\subsection{Syntax}

Figure~\ref{fig:source-syntax} presents the syntax of the source language.
There are two main kinds of terms: (pure) values $v$ and (dirty) computations
$c$, which may call effectful operations. Handlers $h$ are a subsidiary sort of
values. We assume a given set of \emph{operations}~$\op$, such as $\kord{Get}$
and $\kord{Put}$.
We abbreviate $\longcases$ as $\shortcases$, and write $\ops$ to denote the set
$\set{\op_1, \dots, \op_n}$.

Similarly, we distinguish between two basic sorts of types: the value types $A,
B$ and the computation types $\C, \D$. There are four forms of value types:
type variables $\tyvar$, function types $A \to \C$, handler types $\C \hto \D$
and the $\tyUnit$ type. Skeletons $\ety$ capture the shape of types, so, by
design, their forms are identical.
The computation type $A \E \dirt$ is assigned to a
computation returning values of type $A$ and potentially calling operations
from the \emph{dirt} set $\dirt$. A dirt set contains zero or more operations
$\op$ and is terminated either by an empty set or a dirt variable $\dirtvar$.
Though we use \texttt{cons}-list syntax, the intended semantics
of dirt sets $\dirt$ is that the order of operations $\op$ is irrelevant.
That is, $(\{ \op_1 \} \cup (\{ \op_2 \} \cup \dirt))$ denotes the same dirt as
$(\{ \op_2 \} \cup (\{ \op_1 \} \cup \dirt))$.
%
Similarly to all HM-based systems, we discriminate between value types (or
monotypes) $A$, qualified types $\qualTy$ and polytypes (or type schemes)
$\polyTy$.
(Simple) subtyping constraints $\srcPartialCt$ denote inequalities between
either value types or dirts. We also present the more general form of
constraints $\srcFullCt$ that includes inequalities between computation types (as we
illustrate in Section~\ref{subsec:source-typing} below, this allows for a single, uniform constraint entailment
relation).
Finally, polytypes consist of zero or more skeleton, type or dirt
abstractions followed by a qualified type.

\begin{myfigure}
\[
\begin{array}{r@{~}c@{~}l}
  \text{typing environment}~\ctx & ::= & \epsilon                                  \mid
                                         \tyEnv, \evar                             \mid
                                         \tyEnv, \tyvar : \ety                     \mid
                                         \tyEnv, \dirtvar                          \mid
                                         \tyEnv, x : \polyTy                       \mid
                                         \tyEnv, \highlight{\covar :} \srcPartialCt \\
\end{array}
\]
$\ruleform{\tcElabVal{\tyEnv}{v}{\aVty}{v'}}$ \textbf{Values}
\begin{mathpar}
\inferrule*[right=TmVar]
           { (x : \forall \bar{\evar}. \forall \overline{\tyvar:\ety}. \forall \bar{\dirtvar}. \bar{\srcPartialCt} \Rightarrow \aVty) \in \tyEnv \\
             \tcElabPolyTy{\tyEnv}{\bVty}{\ety}{\vty} \\
             \sol = [\overline{\ety'/\evar}, \overline{\bVty/\tyvar}, \overline{\dirt/\dirtvar}] \\
             \overline{\proveCt{\tyEnv}{\coercion}{\sol(\srcPartialCt)}}
           } 
           { \tcElabVal{\tyEnv}{x}{\sol(\aVty)}{x~~\bar{\ety}'~\bar{\vty}~\bar{\dirt}~\bar{\coercion}} }

\inferrule*[right=TmCastV]
           { \tcElabVal{\tyEnv}{v}{\aVty}{v'} \\
             \proveCt{\tyEnv}{\coercion}{\aVty \le \bVty}
           } 
           { \tcElabVal{\tyEnv}{v}{\bVty}{\cast{v'}{\coercion}} }

\inferrule*[right=TmUnit]
           { }
           { \tcElabVal{\tyEnv}{\tmUnit}{\tyUnit}{\tmUnit} }

\inferrule*[right=TmTmAbs]
           { \tcElabComp{\tyEnv, x : \aVty}{c}{\cCty}{c'} \\
             \tcElabPolyTy{\tyEnv}{\aVty}{\ety}{\vty}
           } 
           { \tcElabVal{\tyEnv}{(\fun{x}{c})}{\aVty \to \cCty}{\fun{(x : \vty)}{c'}} }

\inferrule*[right=TmHand]
           { \tcElabComp{\tyEnv, x : \aVty}{c_r}{\bVty~!~\dirt}{c_r'} \\
             \tcElabPolyTy{\tyEnv}{\aVty}{\ety}{\vty} \\
             \left[
               (\op : \aVty_\op \to \bVty_\op) \in \sig \quad
               \tcElabComp{\tyEnv, x : \aVty_\op, k : \bVty_\op \to \bVty~!~\dirt}{c_\op}{\bVty~!~\dirt}{c_\op'}
             \right]_{\op \in \ops} \\
             \highlight{c_\mathit{res} = \{ \return{(x : \vty)} \mapsto c_r', \shortcasesprimed \}}
           } 
           { \tcElabVal{\tyEnv}
                       {\{ \return{x} \mapsto c_r, \shortcases \}}
                       {\aVty~!~\dirt \cup \ops \hto \bVty~!~\dirt}
                       {c_\mathit{res}}
           } 
\end{mathpar}
$\ruleform{\tcElabComp{\tyEnv}{c}{\cCty}{c'}}$ \textbf{Computations}
\vspace{-4mm}
\begin{mathpar}
\inferrule*[right=TmCastC]
           { \tcElabComp{\tyEnv}{c}{\cCty_1}{c'} \\
             \proveCt{\tyEnv}{\coercion}{\cCty_1 \le \cCty_2}
           } 
           { \tcElabComp{\tyEnv}{c}{\cCty_2}{\cast{c'}{\coercion}} }

\inferrule*[right=TmTmApp]
           { \tcElabVal{\tyEnv}{v_1}{\aVty \to \cCty}{v_1'} \\\\
             \tcElabVal{\tyEnv}{v_2}{\aVty}{v_2'}
           } 
           { \tcElabComp{\tyEnv}{v_1~v_2}{\cCty}{v_1'~v_2'} }

\inferrule*[right=TmLet]
           { \polyTy = \forall \bar{\evar}. \overline{\tyvar:\ety}. \forall \bar{\dirtvar}. \bar{\srcPartialCt} \Rightarrow \aVty \\
             \tcElabVal{\tyEnv, \bar{\evar}, \overline{\tyvar:\ety}, \bar{\dirtvar}, \overline{\highlight{\covar :} \srcPartialCt}}{v}{\aVty}{v'} \\
             \tcElabComp{\tyEnv, x : \polyTy}{c}{\cCty}{c'}
           } 
           { \tcElabComp{\tyEnv}
                        {\letval{x}{v}{c}}{\cCty}
                        {\letval{x}{\Lambda \bar{\evar}.\Lambda \overline{\tyvar:\ety}. \Lambda \bar{\dirtvar}. \Lambda (\overline{\covar : \srcPartialCt}). v'}{c'}}
           } 

\inferrule*[right=TmReturn]
           { \tcElabVal{\tyEnv}{v}{\aVty}{v'} }
           { \tcElabComp{\tyEnv}{\return{v}}{\aVty~!~\emptyset}{\return{v'}} }

\inferrule*[right=TmOp]
           { (\op : \aVty_\op \to \bVty_\op) \in \sig \\
             \tcElabVal{\tyEnv}{v}{\aVty_\op}{v'} \\
             \tcElabComp{\tyEnv, y : \bVty_\op}{c}{\aVty~!~\dirt}{c'} \\
             \tcElabPolyTy{\tyEnv}{\bVty_\op}{\ety}{\vty_\op} \\
             \op \in \dirt
           } 
           { \tcElabComp{\tyEnv}{\operation{v}{y}{c}}{\aVty~!~\dirt}{\operation{v'}{y : \vty_\op}{c'}} }

\inferrule*[right=TmDo]
           { \tcElabComp{\tyEnv}{c_1}{\aVty~!~\dirt}{c_1'} \\
             \tcElabComp{\tyEnv, x : \aVty}{c_2}{\bVty~!~\dirt}{c_2'}
           } 
           { \tcElabComp{\tyEnv}{\doin{x}{c_1}{c_2}}{\bVty~!~\dirt}{\doin{x}{c_1'}{c_2'}} }

\inferrule*[right=TmHandle]
           { \tcElabVal{\tyEnv}{v}{\cCty \hto \dCty}{v'} \\
             \tcElabComp{\tyEnv}{c}{\cCty}{c'}
           } 
           { \tcElabComp{\tyEnv}{\withhandle{v}{c}}{\dCty}{\withhandle{v'}{c'}} }
\end{mathpar}

\vspace{-5mm}
\caption{\source Typing \& Elaboration}
\label{fig:source-typing}
\end{myfigure}

\subsection{Typing}\label{subsec:source-typing}

Figure~\ref{fig:source-typing} presents the typing rules for values and
computations, along with a typing-directed elaboration into our target
language~\target. In order to simplify the presentation, in this section we
focus exclusively on typing. The parts of the rules that concern elaboration
are highlighted in gray and are discussed in Section~\ref{sec:inference}.
In all the rules, we assume a global signature $\sig$ that captures
all defined operations along with their (well-formed) types.

\paragraph{\bf Values}

Typing for values takes the form $\tcVal{\tyEnv}{v}{\aVty}$, and, given
a typing environment $\tyEnv$, checks a value $v$ against a value type $A$.

Rule~\textsc{TmVar} handles term variables. Given that $x$ has type $(\forall
\overline{\evar}. \overline{\tyvar:\ety}. \forall \overline{\dirtvar}.
\overline{\srcPartialCt} \Rightarrow \aVty)$, we {\em appropriately}
instantiate the skeleton ($\overline{\evar}$), type ($\overline{\tyvar}$), and
dirt ($\overline{\dirtvar}$) variables, and ensure that the instantiated wanted
constraints $\overline{\sol(\srcPartialCt)}$ are satisfied, via side
condition $\overline{\tcProveCt{\tyEnv}{\sol(\srcPartialCt)}}$.
Rule~\textsc{TmCastV} allows casting the type of a value $v$ from $\aVty$ to
$\bVty$, if $\aVty$ is a subtype of $\bVty$
(upcasting).
As illustrated
by Rule~\textsc{TmTmAbs}, we omit freshness conditions by adopting the
Barendregt convention~\citep{barendregt}.
Finally, Rule~\textsc{TmHand} gives typing for handlers. It requires that
the right-hand sides of the return clause and all operation clauses have the same computation type
($\bVty\,!\,\dirt$), and that all operations mentioned are part of
the top-level signature $\sig$. The result type takes the form $\aVty~!~\dirt \cup \ops \hto
\bVty~!~\dirt$, capturing the intended handler semantics: given a computation
of type $\aVty~!~\dirt \cup \ops$, the handler
\begin{inparaenum}[(a)]
  \item produces a result of type $\bVty$,
  \item handles operations $\ops$, and
  \item propagates unhandled operations $\dirt$ to the output.
\end{inparaenum}

\paragraph{\bf Computations}

Typing for computations takes the form $\tcComp{\tyEnv}{c}{\cCty}$,
and, given a typing environment $\tyEnv$, checks a computation $c$ against a
type $\C$.

Rule~\textsc{TmCastC} behaves like Rule~\textsc{TmCastV}, but
for computation types.
Rule~\textsc{TmLet} handles polymorphic, non-recursive let-bindings.
%
Rule~\textsc{TmReturn} handles $\return{v}$ computations. Keyword
$\ret$ effectively lifts a value $v$ of type $\aVty$ into a computation of type
$\aVty~!~\emptyset$.
Rule~\textsc{TmOp} checks operation calls. First, we ensure that $v$ has the
appropriate type, as specified by the signature of $\op$. Then, the
continuation $(y. c)$ is checked. The side condition $\op \in \dirt$ ensures
that the called operation $\op$ is captured in the result type.
Rule~\textsc{TmDo} handles sequencing. Given that $c_1$ has type
$\aVty\,!\,\dirt$, the pure part of the result of type $\aVty$ is bound to term
variable $x$, which is brought in scope for checking $c_2$. As we mentioned
in Section~\ref{sec:overview}, all computations in a
$\texttt{do}$-construct should have the same effect set, $\dirt$.
Rule~\textsc{TmHandle} eliminates handler types, just as Rule~\textsc{TmTmApp}
eliminates arrow types.

\begin{myfigure}[t!]
$\ruleform{\proveCt{\tyEnv}{\coercion}{\srcFullCt}}$ \textbf{Constraint Entailment}
\begin{mathpar}
\inferrule*[right=CoVar]
           { (\highlight{\covar :} \srcPartialCt) \in \tyEnv }
           { \proveCt{\tyEnv}{\covar}{\srcPartialCt} }

\inferrule*[right=UCoRefl]
           { }
           { \proveCt{\tyEnv}{\trgUnitRefl}{\tyUnit \le \tyUnit} }

\inferrule*[right=ACoRefl]
           { (\tyvar : \ety) \in \tyEnv }
           { \proveCt{\tyEnv}{\refl{\tyvar}}{\tyvar \le \tyvar} }

\inferrule*[right=DCoRefl]
           { \wfDirt{\tyEnv}{\dirt} }
           { \proveCt{\tyEnv}{\dirtRefl{\dirt}}{\dirt \le \dirt} }

\inferrule*[right=VCoArr]
           { \proveCt{\tyEnv}{\coercion_1}{\bVty \le \aVty} \\
             \proveCt{\tyEnv}{\coercion_2}{\cCty \le \dCty}
           } 
           { \proveCt{\tyEnv}{\coercion_1 \to \coercion_2}{\aVty \to \cCty \le \bVty \to \dCty} }

\inferrule*[right=VCoHand]
           { \proveCt{\tyEnv}{\coercion_1}{\cCty_2 \le \cCty_1} \\
             \proveCt{\tyEnv}{\coercion_2}{\dCty_1 \le \dCty_2}
           } 
           { \proveCt{\tyEnv}{\coercion_1 \hto \coercion_2}{\cCty_1 \hto \dCty_1 \le \cCty_2 \hto \dCty_2} }

\inferrule*[right=CCoComp]
           { \proveCt{\tyEnv}{\coercion_1}{\aVty_1 \le \aVty_2} \\
             \proveCt{\tyEnv}{\coercion_2}{\dirt_1 \le \dirt_2}
           } 
           { \proveCt{\tyEnv}{\coercion_1~!~\coercion_2}{\aVty_1~!~\dirt_1 \le \aVty_2~!~\dirt_2} }

\inferrule*[right=DCoNil]
           { }
           { \proveCt{\tyEnv}{\emptyset_{\dirt}}{\emptyset \le \dirt} }

\inferrule*[right=DCoOp]
           { \proveCt{\tyEnv}{\coercion}{\dirt_1 \le \dirt_2} \\
             (\op : \aVty_\op \to \bVty_\op) \in \sig
           } 
           { \proveCt{\tyEnv}{\{\op\} \cup \coercion}{\{\op\} \cup \dirt_1 \le \{\op\} \cup \dirt_2} }
\end{mathpar}

\vspace{-5mm}
\caption{\source Constraint Entailment}
\label{fig:source-coercion-typing}
\end{myfigure}

\paragraph{\bf Constraint Entailment}

The specification of constraint entailment takes the form
$\tcProveCt{\tyEnv}{\srcFullCt}$ and is presented in
Figure~\ref{fig:source-coercion-typing}. Notice that we use $\srcFullCt$
instead of $\srcPartialCt$, which allows us to capture subtyping between two value
types, computation types or dirts, within the same relation. Subtyping can be
established in several ways:

Rule~\textsc{CoVar} handles assumptions.
Rules \textsc{UCoRefl}, \textsc{ACoRefl}, and \textsc{DCoRefl} express that
subtyping is reflexive, for the unit type, type variables, and dirts,
respectively. Notice that we do not have dedicated rules for reflexivity of
arbitrary computation or value types; as we illustrate below
(Section~\ref{subsec:target-syntax}), they can both be established using the
reflexivity of their subparts.
%
Rule~\textsc{VCoArr} establishes inequality of arrow types. As usual, the arrow
type constructor is contravariant in the argument type.
%
Rule~\textsc{VCoHand} is similar, but for handler types.
%
Rule~\textsc{CCoComp} captures the covariance of type constructor ($!$),
establishing subtyping between two computation types if subtyping is
established for their respective subparts.
%
Finally, Rules~\textsc{DCoNil} and~\textsc{DCoOp} establish subtyping between
dirts. Rule \textsc{DCoNil} captures that the empty dirty
set $\emptyset$ is a subdirt of any dirt $\dirt$ and Rule~\textsc{DCoOp} expresses that
dirt subtyping preserved under extension with the same operation
$\op$.

\paragraph{\bf Well-formedness of Types, Constraints, Dirts, and Skeletons}

The relations $\tcPolyTy{\tyEnv}{\aVty}{\ety}$ and
$\tcCompTy{\tyEnv}{\cCty}{\ety}$ check the well-formedness of value
and computation types respectively. Similarly, relations
$\tcConstraint{\tyEnv}{\srcFullCt}$ and
$\wfDirt{\tyEnv}{\dirt}$ check the well-formedness of constraints and dirts,
respectively. They are all
defined in
Appendix~\ref{appendix:source-additional}.

\begin{example}
\label{exa:running-source}
Recall the definition $\letval{\mathit{f}}{(\fun{g}{g~\tmUnit})}{\ldots}$
of a polymorphic $f$ from Section~\ref{sec:overview:polymorphism}. Under different rule applications,
$f$ can be given different typings, including simple $(\tyUnit \to \tyUnit~!~\emptyset) \to \tyUnit~!~\emptyset$
under the typing
\[
  \inferrule*[Right=TmTmAbs]{
    \inferrule*[Right=TmTmApp]{
      \inferrule*[right=TmVar]{ }{
        \tcVal{g : (\tyUnit \to \tyUnit~!~\emptyset)}{g}{\tyUnit \to \tyUnit~!~\emptyset}
      }
      \and
      \inferrule*[Right=TmUnit]{ }{
        \tcVal{\tyEnv}{\tmUnit}{\tyUnit}
      }
    }{
      \tcComp{g : (\tyUnit \to \tyUnit~!~\emptyset)}{g~\tmUnit}{\tyUnit~!~\emptyset}
    }
  }{
    \tcVal{\epsilon}{(\fun{g}{g~\tmUnit})}{(\tyUnit \to \tyUnit~!~\emptyset) \to \tyUnit~!~\emptyset}
  }
\]%
and the more involved polytype
\[
  \polyTy = \forall \evar. \forall \tyvar:\evar, \tyvar':\evar.\forall \dirtvar,\dirtvar'.\tyvar \le \tyvar'\Rightarrow \dirtvar \le \dirtvar' \Rightarrow (\tyUnit \to \tyvar~!~\dirtvar) \to \tyvar' ~!~ \dirtvar'
\]
obtained by generalizing
\[
  \inferrule*[Right=TmTmAbs]{
    \inferrule*[Right=TmCastC]{
      \inferrule*[Left=TmTmApp]{
        \cdots
      }{
        \tcComp{\tyEnv, g : (\tyUnit \to \tyvar~!~\dirtvar)}{g~\tmUnit}{\tyvar ~!~ \dirtvar}
      }
      \and
      \inferrule*[Right=CCoComp]{
        \inferrule*[Left=CoVar]{ }{
          \tcProveCt{\tyEnv}{\tyvar \le \tyvar'}
        }
        \and
        \inferrule*[Right=CoVar]{ }{
          \tcProveCt{\tyEnv}{\dirtvar \le \dirtvar'}
        }
      }{
        \tcProveCt{\tyEnv}{\tyvar~!~\dirtvar \le \tyvar'~!~\dirtvar'}
      }
    }{
      \tcComp{\tyEnv, g : (\tyUnit \to \tyvar~!~\dirtvar)}{g~\tmUnit}{\tyvar' ~!~ \dirtvar'}
    }
  }{
    \tcVal{\underbrace{\evar, \tyvar : \evar, \tyvar' : \evar, \dirtvar, \dirtvar', \tyvar \le \tyvar', \dirtvar \le \dirtvar'}_{\tyEnv}}{(\fun{g}{g~\tmUnit})}{(\tyUnit \to \tyvar~!~\dirtvar) \to \tyvar' ~!~ \dirtvar'}
  }
\]
Using the latter typing, $f$ may be applied to a pure $\mathit{id} = \fun{x}{\return{x}}$ as
\[
  \inferrule*[Right=TmTmApp]{
    \inferrule*[Left=TmVar]{
      \tcPolyTy{\tyEnv}{\tyUnit}{\tyUnit} \\\\
      \tcProveCt{\tyEnv}{\emptyset \le \emptyset} \\\\
      \sol = [\tyUnit / \evar, \tyUnit / \tyvar, \tyUnit / \tyvar', \emptyset / \dirtvar, \emptyset / \dirtvar']
    }{
      \tcVal{\tyEnv}{f}{(\tyUnit \to \tyUnit~!~\emptyset) \to \tyUnit~!~\emptyset}
    }
    \and
    \inferrule*[Right=TmTmAbs]{
      \cdots
    }{
      \tcVal{\tyEnv}{\mathit{id}}{\tyUnit \to \tyUnit~!~\emptyset}
    }
  }{
    \tcComp{\underbrace{f : \polyTy}_{\tyEnv}}{f~\mathit{id}}{\tyUnit ~!~ \emptyset}
  }
\]
We can also apply $f$ to an impure $\mathit{tick} = \fun{x}{\operation[\mathtt{Tick}]{x}{y}{\return{y}}}$,
and even enlarge the final dirt as
\[
  \inferrule*[Right=TmTmApp]{
    \inferrule*[Left=TmVar]{
      \tcPolyTy{\tyEnv}{\tyUnit}{\tyUnit} \\\\
      \tcProveCt{\tyEnv}{\{\mathtt{Tick}\} \le \{\mathtt{Tick}, \mathtt{Tock}\}} \\\\
      \sol = [\tyUnit / \evar, \tyUnit / \tyvar, \tyUnit / \tyvar', \{\mathtt{Tick}\} / \dirtvar, \{\mathtt{Tick}, \mathtt{Tock}\} / \dirtvar']
    }{
      \tcVal{\tyEnv}{f}{(\tyUnit \to \tyUnit~!~\{\mathtt{Tick}\}) \to \tyUnit~!~\{\mathtt{Tick}, \mathtt{Tock}\}}
    }
    \and
    \inferrule*[Right=TmTmAbs,vdots=4em,leftskip=6em]{
      \cdots
    }{
      \tcVal{\tyEnv}{\mathit{tick}}{\tyUnit \to \tyUnit~!~\{\mathtt{Tick}\}}
    }
  }{
    \tcComp{\tyEnv}{f~\mathit{tick}}{\tyUnit ~!~ \{ \mathtt{Tick}, \mathtt{Tock} \}}
  }
\]
\end{example}

\section{The \target Language}\label{sec:target}
\begin{myfigure}[t]
\textbf{Terms}
\[
  \begin{array}{r@{~}c@{~}l}
    \text{value}~v                                                   & ::= &
      x                                                              \mid
      \tmUnit                                                        \mid
      \fun{(x:\vty)}{c}                                              \mid
      h                                                              \\
    & \mid &
      \Lambda \evar. v                                               \mid
      v~\ety                                                         \mid
      \Lambda \tyvar : \ety. v                                       \mid
      v~\vty                                                         \mid
      \Lambda \dirtvar. v                                            \mid
      v~\dirt                                                        \mid
      \Lambda (\covar : \trgPartialCt). v                            \mid
      v~\coercion                                                    \mid
      \cast{v}{\coercion}                                            \\

    \text{handler}~h                                                 & ::= &
      \{ \return{(x : \vty)} \mapsto c_r, \longcases \}                \\

    \text{computation}~c                                             & ::= &
      \return{v}                                                     \mid
      \operation{v}{y : \vty}{c}                                     \mid
      \doin{x}{c_1}{c_2}                                             \\
    & \mid &
      \withhandle{v}{c}                                              \mid
      v_1~v_2                                                        \mid
      \letval{x}{v}{c}                                               \mid
      \cast{c}{\coercion}                                            \\
  \end{array}
\]

\textbf{Types}
\[
  \begin{array}{r@{~}c@{~}l}
    \text{skeleton}~\ety                                             & ::= &
      \evar                                                          \mid
      \tyUnit                                                        \mid
      \ety_1 \to \ety_2                                              \mid
      \ety_1 \hto \ety_2                                             \mid
      \forall \evar. \ety                                            \\

    \\

    \text{value type}~\vty                                           & ::= &
      \tyvar                                                         \mid
      \tyUnit                                                        \mid
      \vty \to \cty                                                  \mid
      \cty_1 \hto \cty_2                                             \mid
      \forall \evar. \vty                                            \mid
      \forall \tyvar\!:\!\ety. \vty                                  \mid
      \forall \dirtvar. \vty                                         \mid
      \trgPartialCt \Rightarrow \vty                                 \\

   \text{simple coercion type}~\trgPartialCt                         & ::= &
     \vty_1 \le \vty_2                                               \mid
     \dirt_1 \le \dirt_2                                             \\

   \text{coercion type}~\trgFullCt                                   & ::= &
     \trgPartialCt                                                   \mid
     \cty_1 \le \cty_2                                               \\

   \\

    \text{computation type}~\cty                                     & ::= &
      \vty~!~\dirt                                                   \\

    \text{dirt}~\dirt                                                & ::= &
      \dirtvar                                                       \mid
      \emptyset                                                      \mid
      \{ \op \} \cup \dirt                                           \\
  \end{array}
\]

\textbf{Coercions}
\[
\begin{array}{@{\hspace{0mm}}r@{~}c@{~}l}
    \coercion & ::=  &
      \covar                                                         \mid 
      \trgUnitRefl                                                   \mid 
      \refl{\tyvar}                                                  \mid 
      \dirtRefl{\dirt}                                               \mid 
      \coercion_1 \to \coercion_2                                    \mid 
      \coercion_1 \hto \coercion_2                                   \mid 
      \emptyset_\dirt                                                \mid 
      \{\op\} \cup \coercion                                         \mid 
      \forall \evar. \coercion                                       \mid 
      \forall (\tyvar : \ety). \coercion                             \mid 
      \forall \dirtvar. \coercion                                    \mid 
      \trgPartialCt \Rightarrow \coercion                            \mid 
      \coercion_1~!~\coercion_2                                      \\   
\end{array}
\]

\vspace{-5mm}
\caption{\target Syntax}
\label{fig:target-syntax}
\end{myfigure}

\subsection{Syntax}\label{subsec:target-syntax}

Figure~\ref{fig:target-syntax} presents \target's syntax. \target is a
type theory akin to \SystemF~\citep{systemf}, where every term
encodes its own typing derivation. In essence, all abstractions and
applications that are implicit in \source, are made explicit in \target via
new syntactic forms. Additionally, \target supports impredicative and higher-rank polymorphism, which is reflected
in the lack of discrimination between value types, qualified types and type
schemes; all non-computation types are denoted by $\vty$. While this design choice
is not strictly required for the purpose at hand, it makes for a cleaner system.

In short, \target relates to \source the same way that System
F~\cite{girardthesis,reynolds-systemf-1,reynolds-systemf-2} relates to the
Hindley-Damas-Milner system~\cite{hindley,milner,DamasMilner}.

\paragraph{\bf Coercions}
Of particular interest is the use of explicit {\em subtyping coercions},
denoted by $\coercion$. \target uses these to replace the
implicit casts of \source (Rules~\textsc{TmCastV} and~\textsc{TmCastC} in
Figure~\ref{fig:source-typing}) with explicit casts $(\cast{v}{\coercion})$ and
$(\cast{c}{\coercion})$. Essentially, coercions $\coercion$ are explicit
witnesses of subtyping derivations: each coercion form corresponds to a
subtyping rule.

The first coercion form, $\covar$, is a coercion variable, that is, a yet
unknown proof of subtyping. Forms $\trgUnitRefl$, $\refl{\tyvar}$, and
$\dirtRefl{\dirt}$ witness reflexivity for the $\tyUnit$ type, type variables,
and dirts $\dirt$, respectively.

Most of the remaining coercion forms are simple congruences;
subtyping for skeleton abstraction, type abstraction, dirt abstraction, and
qualification is witnessed by forms $\forall \evar. \coercion$, $\forall
\tyvar. \coercion$, $\forall \dirtvar. \coercion$, and $\trgPartialCt
\Rightarrow \coercion$, respectively;
similarly, syntactic forms $\coercion_1 \to \coercion_2$ and $\coercion_1 \hto
\coercion_2$ capture injection for the arrow and the handler type constructor,
respectively.

Subtyping for computation types is witnessed by coercion form
$\coercion_1~!~\coercion_2$, which combines subtyping proofs of their
components.

Finally, coercion forms $\emptyset_\dirt$ and $\{\op\} \cup \coercion$ are
concerned with dirt subtyping. Form $\emptyset_\dirt$ witnesses that the empty
dirt $\emptyset$ is a subdirt of any dirt $\dirt$. Lastly, coercion form
$\{\op\} \cup \coercion$ witnesses that subtyping between dirts is preserved
under extension with a new operation.

\begin{myfigure}[t]
\vspace{-3mm}
\begin{mathpar}
\inferrule*[right=]
           { (x : \vty) \in \tyEnv }
           { \tcTrgVal{\tyEnv}{x}{\vty} }

\inferrule*[right=]
           { }
           { \tcTrgVal{\tyEnv}{\tmUnit}{\tyUnit} }

\inferrule*[right=]
           { \tcTrgComp{\tyEnv, x : \vty}{c}{\cty} \\
             \tcTrgVty{\tyEnv}{\vty}{\ety}
           } 
           { \tcTrgVal{\tyEnv}{(\fun{x:\vty}{c})}{\vty \to \cty} }

\inferrule*[right=]
           { \tcTrgVal{\tyEnv}{v}{\vty_1} \\
             \tcTrgCo{\tyEnv}{\coercion}{\vty_1 \le \vty_2}
           } 
           { \tcTrgVal{\tyEnv}{\cast{v}{\coercion}}{\vty_2} }

\inferrule*[right=]
           { \tcTrgVal{\tyEnv, \evar}{v}{\vty} }
           { \tcTrgVal{\tyEnv}{\Lambda \evar. v}{\forall \evar.\vty} }

\inferrule*[right=]
           { \tcTrgVal{\tyEnv, \tyvar : \ety}{v}{\vty} }
           { \tcTrgVal{\tyEnv}{\Lambda \tyvar : \ety. v}{\forall \tyvar : \ety.\vty} }

\inferrule*[right=]
           { \tcTrgVal{\tyEnv, \dirtvar}{v}{\vty} }
           { \tcTrgVal{\tyEnv}{\Lambda \dirtvar. v}{\forall \dirtvar. \vty} }

\inferrule*[right=]
           { \tcTrgVal{\tyEnv, \covar : \trgPartialCt}{v}{\vty} \\
             \wfTrgCt{\tyEnv}{\trgPartialCt}
           } 
           { \tcTrgVal{\tyEnv}{\Lambda (\covar : \trgPartialCt). v}{\trgPartialCt \Rightarrow \vty} }

\inferrule*[right=]
           { \tcTrgVal{\tyEnv}{v}{\trgPartialCt \Rightarrow \vty} \\
             \tcTrgCo{\tyEnv}{\coercion}{\trgPartialCt}
           } 
           { \tcTrgVal{\tyEnv}{v~\coercion}{\vty} }

\inferrule*[right=]
           { \tcTrgComp{\tyEnv, x : \vty_x}{c_r}{\vty\,!\,\dirt} \\
             \left[
               (\op : \vty_1 \to \vty_2) \in \sig \qquad
               \tcTrgComp{\tyEnv, x : \vty_1, k : \vty_2 \to \vty\,!\,\dirt}{c_\op}{\vty\,!\,\dirt}
             \right]_{\op \in \ops}
           } 
           { \tcTrgVal{\tyEnv}{\trgShorthand}{\vty_x~!~\dirt \cup \ops \hto \vty~!~\dirt} }

\inferrule*[right=]
           { \tcTrgVal{\tyEnv}{v}{\forall \evar. \vty} \\\\
             \tcSkeleton{\tyEnv}{\ety}
           } %
           { \tcTrgVal{\tyEnv}{v~\ety}{\vty[\ety/\evar]} }

\inferrule*[right=]
           { \tcTrgVal{\tyEnv}{v}{\forall \tyvar : \ety. \vty_1} \\\\
             \tcTrgVty{\tyEnv}{\vty_2}{\ety}
           } %
           { \tcTrgVal{\tyEnv}{v~\vty_2}{\vty_1[\vty_2/\tyvar]} }

\inferrule*[right=]
           { \tcTrgVal{\tyEnv}{v}{\forall \dirtvar. \vty} \\\\
             \wfDirt{\tyEnv}{\dirt}
           } 
           { \tcTrgVal{\tyEnv}{v~\dirt}{\vty[\dirt/\dirtvar]} }
\end{mathpar}

\vspace{-5mm}
\caption{\target Value Typing}
\label{fig:target-typing-values}
\end{myfigure}

\begin{myfigure}[t]
\vspace{-3mm}
\begin{mathpar}
\inferrule*[right=]
           { \tcTrgVal{\tyEnv}{v_1}{\vty \to \cty} \\
             \tcTrgVal{\tyEnv}{v_2}{\vty}
           } 
           { \tcTrgComp{\tyEnv}{v_1~v_2}{\cty} }

\inferrule*[right=]
           { \tcTrgVal{\tyEnv}{v}{\vty} \\
             \tcTrgComp{\tyEnv, x : \vty}{c}{\cty}
           } 
           { \tcTrgComp{\tyEnv}{\letval{x}{v}{c}}{\cty} }

\inferrule*[right=]
           { \tcTrgComp{\tyEnv}{v}{\vty} }
           { \tcTrgComp{\tyEnv}{\return{v}}{\vty~!~\emptyset} }

\inferrule*[right=]
           { \tcTrgComp{\tyEnv}{c_1}{\vty_1~!~\dirt} \\
             \tcTrgComp{\tyEnv, x : \vty_1}{c_2}{\vty_2~!~\dirt}
           } 
           { \tcTrgComp{\tyEnv}{\doin{x}{c_1}{c_2}}{\vty_2~!~\dirt} }

\inferrule*[right=]
           { (\op : \vty_1 \to \vty_2) \in \sig \\
             \tcTrgVal{\tyEnv}{v}{\vty_1} \\
             \tcTrgComp{\tyEnv, y : \vty_2}{c}{\vty~!~\dirt} \\
             \op \in \dirt
           } 
           { \tcTrgComp{\tyEnv}{\operation{v}{y : \vty_2}{c}}{\vty~!~\dirt} }

\inferrule*[right=]
           { \tcTrgVal{\tyEnv}{v}{\cty_1 \hto \cty_2} \\
             \tcTrgComp{\tyEnv}{c}{\cty_1}
           } 
           { \tcTrgComp{\tyEnv}{\withhandle{v}{c}}{\cty_2} }

\inferrule*[right=]
           { \tcTrgComp{\tyEnv}{c}{\cty_1} \\
             \tcTrgCo{\tyEnv}{\coercion}{\cty_1 \le \cty_2}
           } 
           { \tcTrgComp{\tyEnv}{\cast{c}{\coercion}}{\cty_2} }
\end{mathpar}

\vspace{-5mm}
\caption{\target Computation Typing}
\label{fig:target-typing-computations}
\end{myfigure}

\paragraph{\bf A Note on Reflexivity of Arbitrary Types}

In contrast to our earlier work~\cite{esop2018}, \target (and the other calculi
we present in the remainder of this paper) does not syntactically allow for
reflexivity of arbitrary types. Nevertheless, we avoid notational burden and
throughout the paper write $\refl{\vty}$ to denote the coercion that witnesses
$\vty \le \vty$; such a coercion can be built by traversing the structure of
$\vty$ (see Appendix~\ref{appendix:target-additional}). A similar situation
arises when applying a type substitution on a coercion, but it can be remedied
in exactly the same way.

One of the problems with reflexivity of arbitrary types is that it allows for
many trivially different proofs for the same constraint. The same is also true
for \emph{inversion coercions}, which are coercion formers that allow for
decomposition of coercion types. For example, our earlier
work~\cite{esop2018} included a coercion former $\leftinv{\coercion}$ which is
a proof of $\vty_2 \le \vty_1$, if $\coercion$ is a proof of $\vty_1 \to \cty_1
\le \vty_2 \to \cty_2$.

By removing both, we have managed to greatly simplify the proofs of
the metatheoretical properties of our calculi, since now there are much less
proofs for any type inequality. Additionally, as we show in
Section~\ref{subsec:exeff-operational-semantics}, \target's operational
semantics inspect the coercions so having uniqueness of proofs (coercions) is
essential.

The situation is quite different when it comes to dirts. Dirts can take much
less forms than types do (and so do coercions about them), and coercions
regarding dirts need never be inspected during evaluation. Hence, we do not
require unique coercion forms for dirt inequalities and can allow the simpler
and more conventional reflexivity coercions~$\refl{\dirt}$ for arbitrary dirts~$\dirt$.

\subsection{Typing}

\paragraph{\bf Value \& Computation Typing}

Typing for \target values and computations is presented in
Figures~\ref{fig:target-typing-values} and~\ref{fig:target-typing-computations}
and is given by two mutually recursive relations of the form
$\tcTrgVal{\tyEnv}{v}{\vty}$ (values) and $\tcTrgComp{\tyEnv}{c}{\cty}$
(computations). \target typing environments $\tyEnv$ contain bindings for
variables of all sorts:
\[
\begin{array}{r@{~}c@{~}l}
  \ctx & ::= & \epsilon                             \mid
               \tyEnv, \evar                        \mid
               \tyEnv, \tyvar : \ety                \mid
               \tyEnv, \dirtvar                     \mid
               \tyEnv, x : \vty                     \mid
               \tyEnv, \covar : \trgPartialCt       \\
\end{array}
\]
Typing is entirely syntax-directed. Apart from the typing rules for skeleton,
type, dirt, and coercion abstraction (and, subsequently, skeleton, type, dirt,
and coercion
application), the main difference between typing for \source and \target lies
in the explicit cast forms, $(\cast{v}{\coercion})$ and
$(\cast{c}{\coercion})$.
Given that a value $v$ has type $\vty_1$ and that $\coercion$ is a proof that
$\vty_1$ is a subtype of $\vty_2$, we can upcast $v$ with an explicit
cast operation $(\cast{v}{\coercion})$. Upcasting for computations works
analogously.

\paragraph{\bf Well-formedness of Types, Constraints, Dirts \& Skeletons}

The definitions of the judgements that check the well-formedness of \target
value types ($\tcTrgVty{\tyEnv}{\vty}{\ety}$), computation types
($\tcTrgCty{\tyEnv}{\cty}{\ety}$), dirts ($\wfDirt{\tyEnv}{\dirt}$), and
skeletons ($\tcSkeleton{\tyEnv}{\ety}$) are equally straightforward as those
for \source and can be found in Appendix~\ref{appendix:target-additional}.

\paragraph{\bf Coercion Typing}

Coercion typing formalizes the intuitive interpretation of coercions we gave in
Section~\ref{subsec:target-syntax} and takes the form
$\tcTrgCo{\tyEnv}{\coercion}{\trgFullCt}$, defined in Appendix~\ref{appendix:target-additional}. It is essentially an
extension of the constraint entailment relation of
Figure~\ref{fig:source-coercion-typing}.

\subsection{Operational Semantics}
\label{subsec:exeff-operational-semantics}

\begin{myfigure}[t]
\begin{center}
\end{center}
$\ruleform{\smallStepVal{v}{v'}}$ \textbf{Values}
\begin{mathpar}
\inferrule*[right=VCast]
           { \smallStepVal{v}{v'} }
           { \smallStepVal{\cast{v}{\coercion}}{\cast{v'}{\coercion}} }

\inferrule*[lab=VPushUnit]
           {}
           { \smallStepVal{\cast{\valRes}{\trgUnitRefl}}{\valRes} }

\inferrule*[lab=VPushSkel]
           {}
           { \smallStepVal{(\cast{\valRes}{(\forall \evar. \coercion)})~\ety}{\cast{\valRes~\ety}{\coercion[\ety/\evar]}} }

\inferrule*[lab=VPushTy]
           {}
           { \smallStepVal{(\cast{\valRes}{(\forall (\tyvar : \ety). \coercion)})~\vty}{\cast{\valRes~\vty}{\coercion[\vty/\tyvar]}} }

\inferrule*[lab=VPushDirt]
           {}
           { \smallStepVal{(\cast{\valRes}{(\forall \dirtvar. \coercion)})~\dirt}{\cast{\valRes~\dirt}{\coercion[\dirt/\dirtvar]}} }

\inferrule*[lab=VPushQual]
           {}
           { \smallStepVal{(\cast{\valRes}{(\trgPartialCt \Rightarrow \coercion_1)})~\coercion_2}{\cast{\valRes~\coercion_2}{\coercion_1}} }
\end{mathpar}

$\ruleform{\smallStepComp{c}{c'}}$ \textbf{Computations}
\begin{mathpar}
\inferrule*[right=CCast]
           { \smallStepComp{c}{c'} }
           { \smallStepComp{\cast{c}{\coercion}}{\cast{c'}{\coercion}} }

\inferrule*[lab=CPushApp]
           {}
           { \smallStepComp{(\cast{\valRes}{(\coercion_1 \to \coercion_2)})~v}
                           {\cast{(\valRes~(\cast{v}{\coercion_1}))}{\coercion_2}}
           } 

\inferrule*[lab=CPushOp]
           {}
           { \smallStepComp{\cast{(\operation{\valRes}{x : \vty}{c})}{\coercion}}
                           {\operation{\valRes}{x : \vty}{(\cast{c}{\coercion})}}
           } 

\inferrule*[lab=CDoRet]
           {} 
           { \smallStepComp{\doin{x}
                                 {(\cast{\cast{\cast{(\return{\valRes})}{(\coercion_1~!~\coercion_1')}}{\ldots}}{(\coercion_n~!~\coercion_n')})}
                                 {c_2}}
                           {c_2[(\cast{\cast{\cast{\valRes}{\coercion_1}}{\ldots}}{\coercion_n})/x]} }

\inferrule*[lab=CDoOp]
  {}
  { \smallStepComp
       {\doin{x}{\operation{\valRes}{y : \vty}{c_1}}{c_2}}
       {\operation{\valRes}{y : \vty}{\doin{x}{c_1}{c_2}}}
  } 

\inferrule*[lab=CPushHandle]
           {}
           { \smallStepComp{\withhandle{(\cast{\valRes}{(\coercion_1 \hto \coercion_2)})}{c}}
                           {\cast{(\withhandle{\valRes}{(\cast{c}{\coercion_1})})}{\coercion_2}}
           } 

\inferrule*[lab=CHandleRet]
  {(\ret x \mapsto c_r) \in h } 
  { \smallStepComp{\withhandle{h}{( \cast{\cast{\cast{(\return{\valRes})}{(\coercion_1~!~\coercion_1')}}{\ldots}}{(\coercion_n~!~\coercion_n')})}}
                   {c_r[(\cast{\cast{\cast{\valRes}{\coercion_1}}{\ldots}}{\coercion_n})/ x]}
  } 

\inferrule*[right=CHandleOp1]
  { (\call{\op}{x}{k} \mapsto c_{\op}) \in h }
  { \smallStepComp{
      \withhandle{h}{(\operation{\valRes}{y : \vty}{c})}
    }{
      c_{\op}[\valRes / x, (\fun{(y : \vty)}{\withhandle{h}{c}}) / k]
    }
  }

\inferrule*[right=CHandleOp2]
  { (\call{\op}{x}{k} \mapsto c_{\op}) \notin h }
  { \smallStepComp{
      \withhandle{h}{(\operation{\valRes}{y : \vty}{c})}
    }{
      \operation{\valRes}{y : \vty}{\withhandle{h}{c}}
    }
  }
\end{mathpar}

\vspace{-7mm}
\caption{\target Operational Semantics (Selected Rules)}
\label{fig:target-opsem-partial}
\end{myfigure}

Figure~\ref{fig:target-opsem-partial} presents selected rules of \target's
small-step, call-by-value operational semantics. For lack of
space, we omit $\beta$-rules and other common rules and focus only on
cases of interest. The complete operational semantics can be found in
Appendix~\ref{appendix:target-additional}.

Firstly, one of the non-conventional features of our system lies in the
stratification of results in plain results and cast results:
\begin{center}\begin{shaded}\begin{minipage}{\columnwidth}
\small
\vspace{-1mm}
\[
\begin{array}{r@{~}c@{~}l}
  \text{terminal value}~\termVal & ::= &%
    \tmUnit                                   \mid
    \fun{x:\vty}{c}                           \mid
    h                                         \mid
    \Lambda \evar. v                          \mid
    \Lambda (\tyvar : \ety). v                \mid
    \Lambda \dirtvar. v                       \mid
    \lambda (\covar  : \trgPartialCt). v      \\

  \text{value result}~\valRes & ::= &%
    \termVal                                              \mid
    \cast{\valRes}{(\coercion_1 \to  \coercion_2)}        \mid
    \cast{\valRes}{(\coercion_1 \hto \coercion_2)}        \mid
    \cast{\valRes}{(\forall \evar. \coercion)}            \\
  & \mid &%
    \cast{\valRes}{(\forall (\tyvar : \ety). \coercion)}  \mid
    \cast{\valRes}{(\forall \dirtvar. \coercion)}         \mid
    \cast{\valRes}{(\trgPartialCt \Rightarrow \coercion)} \\

  \text{terminal computation}~\termComp & := &%
    \return{\valRes}                              \mid
    \cast{\termComp}{(\coercion_1~!~\coercion_2)} \\

  \text{computation result}~\compRes & ::= &%
    \termComp                                 \mid
    \operation{\valRes}{y : \vty}{c}          \\
\end{array}
\]
\end{minipage}\end{shaded}\end{center}
Terminal values $\termVal$ represent conventional values, and value results
$\valRes$ can either be plain terminal values $\termVal$ or cast value results,
where we exclude reflexivity coercions, as those can be further reduced.
This stratification can also be found in Henglein's coercion
calculus~\citep{CoercionCalculus}, Crary's coercion calculus for inclusive
subtyping~\citep{crary}, and, more recently, in \SystemFC~\citep{systemfc}.

Computations evaluate either to a returned value or an operation call. Both can be further cast, though we are able to delegate any coercion on the operation call to its continuation, leading to a slightly different stratification than in values. The same is not true for returned values. Consider for example the expression
$(\cast{\return{5}}{\refl{\tyInt}\,!\,\emptyset_{\{\op\}}})$, of type
$\tyInt\,!\,\{\op\}$. We can not reduce the expression further without
losing effect information; removing the cast would result in computation
$(\return{5})$, of type $\tyInt\,!\,\emptyset$. Even if we consider
type preservation only up to subtyping, the redex may still occur as a subterm
in a context that expects solely the larger type.

Secondly, we need to make sure that casts do not stand in the way of
evaluation. This is captured in the so-called ``push'' rules, all of which
appear in Figure~\ref{fig:target-opsem-partial}.

In relation $\smallStepVal{v}{v'}$, Rule~\textsc{VCast} evaluates under the coercion, while the
rest are push rules: whenever a redex is ``blocked'' due to a cast, we take
the coercion apart and redistribute it (in a type-preserving manner) over the
subterms, so that evaluation can progress.

\begin{example}
Consider the evaluation of $((\cast{(\Lambda \tyvar. v)}{(\forall \tyvar.
\coercion)})~\vty)$ (we elide skeleton annotations for clarity; they are
orthogonal to the task at hand). The evaluation is ``blocked'' because of the
type cast; in order to expose the redex $((\Lambda \tyvar. v)~\vty)$ we need to
\emph{push} the coercion outside the redex, which we achieve using
Rule~\textsc{VPushTy}:
\[
(\cast{(\Lambda \tyvar. v)}{(\forall \tyvar. \coercion)})~\vty \leadsto_\mathrm{v} \cast{((\Lambda \tyvar. v)~\vty)}{\coercion[\vty/\tyvar]}
\]
Since the type cast now happens after the instantiation, we change the coercion
accordingly (to $\coercion[\vty/\tyvar]$), to ensure that the type of the
expression remains the same as before (preservation).
Now using Rule~\textsc{CCast} we can continue with the evaluation of the redex
under the cast, thus obtaining:
\[
\cast{((\Lambda \tyvar. v)~\vty)}{\coercion[\vty/\tyvar]} \leadsto_\mathrm{v} \cast{v}{\coercion[\vty/\tyvar]}
\]
The rest of the push rules behave similarly.
\end{example}

The situation in relation $\smallStepComp{c}{c'}$ is quite similar.
Rule~\textsc{CCast} continues evaluating the computation under the coercion.
Rule~\textsc{CPushApp} is a push rule for function application.
Rule~\textsc{CPushOp} pushes a coercion inside an operation-computation, illustrating
why the syntax for $\compRes$ does not require casts on operation-computations;
we can always push the casts inside the continuation.
Rule~\textsc{CDoRet} is a $\beta$-reduction for sequencing and performs two
tasks at once. Since we know that the computation bound to $x$ calls no
operations, we
\begin{inparaenum}[(a)]
\item safely ``drop'' the impure part of coercions, and
\item substitute $x$ with $\valRes$, cast with the pure part of coercions
  (so that types are preserved).
\end{inparaenum}
Rule~\textsc{CDoOp} handles operation calls in sequencing computations. If an
operation is called in a sequencing computation, evaluation is suspended and
the rest of the computation is captured in the continuation.

The last four rules are concerned with effect handling.
Rule~\textsc{CPushHandle} pushes a coercion on the handler ``outwards'', such that the
handler can be exposed and evaluation is not stuck (similarly to the push rule
for term application).
Rule~\textsc{CHandleRet} behaves similarly to the push/beta rule for sequencing computations.
Finally, the last two rules are concerned with handling of operations. Rule~\textsc{CHandleOp1}
captures cases where the called operation is handled by the
handler, in which case the respective clause of the handler is called. As
illustrated by the rule, like~\citet{Pretnar}{DBLP:journals/corr/Pretnar13}, \target features
{\em deep handlers}: the continuation is also wrapped within a
$\mathtt{with}$-$\mathtt{handle}$ construct.
Rule~\textsc{CHandleOp2} captures cases where the operation is not covered by the
handler and thus remains unhandled.

We have shown that \target is type safe:
\begin{theorem}[Type Safety]
\label{thm:safety}
\vspace{-1mm}
\begin{itemize}
\item
  If $\tcTrgVal{\tyEnv}{v}{\vty}$ then either $v$ is a result value or
  $\smallStepVal{v}{v'}$ and $\tcTrgVal{\tyEnv}{v'}{\vty}$.
\item
  If $\tcTrgComp{\tyEnv}{c}{\cty}$ then either $c$ is a result computation or
  $\smallStepComp{c}{c'}$ and $\tcTrgComp{\tyEnv}{c'}{\cty}$.
\end{itemize}
\end{theorem}

\section{Type Inference \& Elaboration}\label{sec:inference}
This section presents the typing-directed elaboration of \source into \target.
This elaboration makes all the implicit type and effect information explicit,
and introduces explicit term-level coercions to witness the use of subtyping.

After covering the declarative specification of this elaboration, we present
a constraint-based algorithm to infer \source types and at the same time elaborate into \target.
This algorithm alternates between two phases: 1) the syntax-directed generation of constraints
from the \source term, and 2) solving these constraints.

\subsection{Elaboration of \source into \target}

The greyed parts of Figure~\ref{fig:source-typing} augment the typing rules for
\source value and computation terms with typing-directed elaboration to
corresponding \target terms. The elaboration is mostly straightforward, mapping
every \source construct onto its corresponding \target construct while adding
explicit type annotations to binders in Rules \textsc{TmTmAbs},
\textsc{TmHandler} and \textsc{TmOp}. Implicit appeals to subtyping are turned
into explicit casts with coercions in Rules \textsc{TmCastV} and
\textsc{TmCastC}. Rule \textsc{TmLet} introduces explicit binders for skeleton,
type, and dirt variables, as well as for constraints. These last also introduce
coercion variables $\covar$ that can be used in casts.

Binders introduced by Rule \textsc{TmLet} are
eliminated in Rule \textsc{TmVar} by means of explicit application with
skeletons, types, dirts and coercions. The coercions are produced by the auxiliary
judgement $\proveCt{\tyEnv}{\coercion}{\srcPartialCt}$, defined in
Figure~\ref{fig:source-coercion-typing}, which provides a coercion witness for
every subtyping proof.

As a sanity check, we have shown that elaboration preserves types.
\begin{theorem}[Type Preservation]
\label{thm:type_preservation}
\vspace{-1mm}
\begin{itemize}
\item
If $\tcElabVal{\tyEnv}{v}{\aVty}{v'}$
then $\tcTrgVal{\elabTyEnv{\tyEnv}}{v'}{\elabVty{\aVty}}$.
\item
If $\tcElabComp{\tyEnv}{c}{\cCty}{c'}$
then $\tcTrgComp{\elabTyEnv{\tyEnv}}{c'}{\elabCty{\cCty}}$.
\end{itemize}
\end{theorem}
Here $\elabTyEnv{\tyEnv}$, $\elabVty{\aVty}$ and $\elabCty{\cCty}$ convert
\source environments and types into \target environments and types; they
are defined in Appendix~\ref{appendix:inference-additional}.

\begin{example}
\label{exa:running-target}
  A valid elaboration of the polymorphic expression
  \[
  \keylet~\mathit{f} = (\fun{g}{g~\tmUnit})~\keyin~\ldots
  \]
  from Example~\ref{exa:running-source} can be
  \[
  \begin{array}{l@{\hspace{1mm}}c@{\hspace{1mm}}l}
    \keylet~~\mathit{f} & : & (\tyUnit \to \tyUnit~!~\emptyset) \to \tyUnit~!~\emptyset \\
                        & = & \fun{(g : \tyUnit \to \tyUnit~!~\emptyset)}{g~\tmUnit} \\
    \keyin~\ldots \\
  \end{array}
  \]
  if the simple monomorphic typing is used (we have included the signature of $\mathit{f}$ for clarity).
  For the polymorphic variant, the elaboration features both type-level abstractions and explicit casts:
  \[
  \begin{array}{l@{\hspace{1mm}}c@{\hspace{1mm}}l}
    \keylet~~\mathit{f} & : & \forall \evar. \forall \tyvar : \evar. \forall \tyvar' : \evar. \forall \dirtvar. \forall \dirtvar'.
                                (\tyvar \le \tyvar') \Rightarrow (\dirtvar \le \dirtvar') \Rightarrow (\tyUnit \to \tyvar~!~\dirtvar) \to \tyvar' ~!~ \dirtvar' \\
                        & = & \Lambda \evar. \Lambda (\tyvar : \evar). \Lambda (\tyvar' : \evar). \Lambda \dirtvar. \Lambda \dirtvar'. \Lambda (\covar : \tyvar \le \tyvar'). \Lambda (\covar' : \dirtvar \le \dirtvar'). \\
                        &   & \quad \fun{(g : \tyUnit \to \tyvar \,!\, \dirtvar)}{(\cast{(g~\tmUnit)}{(\covar \,!\, \covar')})} \\
    \keyin~\ldots \\
  \end{array}
  \]
  Here, coercion variables $\covar$ and $\covar'$ are utilized by the body of $\mathit{f}$ for
  upcasting $(g~\tmUnit)$ to have type $\tyvar' ~!~ \dirtvar'$.
  
  Similarly, applications of the latter variant need to include explicit type-level applications and coercion witnesses.
  Elaborating the application of $f$ to the pure function $\mathit{id}$ we get
  \[
    f~\tyUnit~\tyUnit~\tyUnit~\emptyset~\emptyset~\trgUnitRefl~\emptyset_\emptyset~(\fun{(x:\tyUnit)}{\return{x}})
  \]
  whereas for the impure $\mathit{tick}$ at a type $\tyUnit~!~\{\texttt{Tick}, \texttt{Tock}\}$ we get
  \begin{align*}
    &f~\tyUnit~\tyUnit~\tyUnit~\{\texttt{Tick}\}~\{\texttt{Tick}, \texttt{Tock}\}~\trgUnitRefl~(\{\texttt{Tick}\} \cup \emptyset_{\{\texttt{Tock}\}}) \\
    &\quad (\fun{x : \tyUnit}{\operation[\mathtt{Tick}]{x}{y : \tyUnit}{(\cast{(\return{y})}{\trgUnitRefl~!~\emptyset_{\{\texttt{Tick}\}}})}})
  \end{align*}
  where $\keyreturn$ had to be coerced in order to match the dirt of the operation call.
\end{example}

\subsection{Constraint Generation \& Elaboration}\label{subsec:constraint-generation}

Constraint generation with elaboration into \target is presented in
Figures~\ref{fig:constraint-generation-values} (values)
and~\ref{fig:constraint-generation-computations} (computations).
Before going into the details of each, we first introduce the three auxiliary
constructs they use.
\begin{center}\begin{shaded}\begin{minipage}{\columnwidth}
\small
\[
\begin{array}{r@{~}c@{~}l}
  \text{constraint set}~\cstr, \queue & ::= & \bullet                                   \mid
                                              \ety_1 = \ety_2,                    \cstr \mid
                                              \tyvar : \ety,                      \cstr \mid
                                              \highlight{\covar :} \srcPartialCt, \cstr \\
  \text{typing environment}~\ctx      & ::= & \epsilon                                  \mid
                                              \tyEnv, x : \polyTy                       \\
  \text{substitution}~\sol            & ::= & \bullet                                   \mid
                                              \sol \cdot [\ety/\evar]               \mid
                                              \sol \cdot [\aVty/\tyvar]             \mid
                                              \sol \cdot [\dirt/\dirtvar]           \mid
                                              \sol \cdot \highlight{[\coercion/\covar]}         \\
\end{array}
\]
\end{minipage}\end{shaded}\end{center}
At the heart of our algorithm are sets $\cstr$,
containing three different kinds of constraints:
\begin{inparaenum}[(a)]
  \item skeleton equalities of the form $\ety_1 = \ety_2$,
  \item skeleton constraints of the form $\tyvar : \ety$, and
  \item wanted subtyping constraints of the form $\covar : \srcPartialCt$.
\end{inparaenum}
The purpose of the first two becomes clear when we discuss constraint solving,
in Section~\ref{subsec:constraint-solving}.
Next, typing environments $\tyEnv$ only contain term variable bindings, while
other variables represent unknowns of their sort and may end up being
instantiated after constraint solving.
Finally, during type inference we compute substitutions $\sol$, for refining as
of yet unknown skeletons, types, dirts, and coercions. The last one is
essential, since our algorithm simultaneously performs type inference and
elaboration into \target.

\paragraph{\bf Values.}

\begin{myfigure}[t!]
$\ruleform{\inferStVal{\constraints}{\tyEnv}{v}{\aVty}{\constraints'}{\sol}{v'}}$ \textbf{Values}
\begin{mathpar}
\inferrule*[right=TmVar]
           { (x : \forall \bar{\evar}. \overline{\tyvar:\ety}. \forall \bar{\dirtvar}. \bar{\srcPartialCt} \Rightarrow \aVty) \in \tyEnv \\
             \sol = [\overline{\evar'/\evar}, \overline{\tyvar'/\tyvar}, \overline{\dirtvar'/\dirtvar}]
           } 
           { \inferStVal{\constraints}
                        {\tyEnv}
                        {x}
                        {\sol(\aVty)}
                        {\overline{\highlight{\covar :} \sol{(\srcPartialCt)}}, \overline{\tyvar':\sol{(\ety)}}, \constraints}
                        {\bullet}
                        {x~\bar{\evar'}~\bar{\tyvar'}~\bar{\dirtvar'}~\bar{\covar}}
           } 

\inferrule*[right=TmUnit]
           { }
           { \inferStVal{\constraints}{\tyEnv}{\tmUnit}{\tyUnit}{\constraints}{\bullet}{\tmUnit} }

\inferrule*[right=TmAbs]
           { \inferStComp{\tyvar : \evar, \constraints}{\tyEnv, x : \tyvar}{c}{\cCty}{\constraints'}{\sol}{c'} }
           { \inferStVal{\constraints}{\tyEnv}{(\fun{x}{c})}{\sol(\tyvar) \to \cCty}{\constraints'}{\sol}{\fun{x : \sol(\tyvar)}{c'}} }

\inferrule*[right=TmHand]
           { \inferStComp{\tyvar_r : \evar_r, \constraints}{\tyEnv, x : \tyvar_r}{c_r}{\bVty_r~!~\dirt_r}{\constraints_0}{\sol_r}{c_r'} \\
             \sol^i = \sol_i \cdot \sol_{i-1} \cdot \ldots \cdot \sol_1 \\\\
             \mbox{$
               \begin{array}{@{\hspace{0mm}}l@{\hspace{0mm}}}
                 \op_i \in \ops: \\
                 \quad (\op_i : \aVty_i \to \bVty_i) \in \sig \\
                 \quad \inferStComp{\tyvar_i : \evar_i, \constraints_{i-1}}
                                   {\sol^{i-1}(\sol_r(\tyEnv)), x : \aVty_i, k : \bVty_i \to \tyvar_i\,!\,\dirtvar_i}
                                   {c_{\op_i}}
                                   {\bVty_{\op_i}\,!\,\dirt_{\op_i}}
                                   {\constraints_i}
                                   {\sol_i}
                                   {c_{\op_i}'} \\
               \constraints' = \constraints_n,
                               \tyvar_\mathit{in}   : \evar_\mathit{in},
                               \tyvar_\mathit{out}  : \evar_\mathit{out}, \\
               \begin{array}{l@{\qquad}l@{\hspace{0mm}}}
                 ~~~~~~~ \highlight{\covar_1 :} \sol^n(\bVty_r) \le \tyvar_\mathit{out}, \\
                 ~~~~~~~ \highlight{\covar_2 :} \sol^n(\dirt_r) \le \dirtvar_\mathit{out}, \\
                 ~~~~~~~ \highlight{\covar_{3_i} :} \sol^n(\bVty_{\op_i}) \le \tyvar_\mathit{out},   & (\forall i \in [1\dots n]) \\
                 ~~~~~~~ \highlight{\covar_{4_i} :} \sol^n(\dirt_{\op_i}) \le \dirtvar_\mathit{out}, & (\forall i \in [1\dots n]) \\
                 ~~~~~~~ \highlight{\covar_{5_i} :}
                   \bVty_i \to \tyvar_\mathit{out}\,!\,\dirtvar_\mathit{out} \le
                   \bVty_i \to \sol^n(\tyvar_i\,!\,\dirtvar_i), & (\forall i \in [1\dots n]) \\
               \end{array} \\
               ~~~~~~~~~ \highlight{\covar_6 :} \tyvar_\mathit{in} \le \sol^n(\sol_r(\tyvar_r)), \\
               ~~~~~~~~~ \highlight{\covar_7 :} \dirtvar_\mathit{in} \le \dirtvar_\mathit{out} \cup \ops \\
               \end{array}
             $} \\\\
             \highlight{
               \mbox{$
                 \begin{array}{@{\hspace{0mm}}l@{~}l@{\hspace{0mm}}}
                   c_\mathit{res} = &\{\,\return{y : \sol^n(\sol_r(\tyvar_r))}
                                           \mapsto \cast{\sol^n(c_r')[\cast{y}{\covar_6}/x]}{\covar_1\,!\,\covar_2} \\
                                    & , \left[
                                      \op_i~x~l \mapsto \cast{\sol^n(c_{\op_i}')[\cast{l}{\covar_{5_i}}/k]}{\covar_{3_i}\,!\,\covar_{4_i}}
                                      \right]_{\op_i \in \ops}
                                    \} \vartriangleright (\refl{\tyvar_\mathit{in}}\,!\,\covar_7 \hto \refl{\tyvar_\mathit{out}}\,!\,\refl{\dirtvar_\mathit{out}})
                 \end{array}
               $}
             }
           } 
           { \inferStVal{\constraints}
                        {\tyEnv}
                        {\shorthand}
                        {\tyvar_\mathit{in}\,!\,\dirtvar_\mathit{in} \hto \tyvar_\mathit{out}\,!\,\dirtvar_\mathit{out}}
                        {\constraints'}
                        {(\sol^n \cdot \sol_r)}
                        {c_\mathit{res}}
           } 
\end{mathpar}

\vspace{-7mm}
\caption{Constraint Generation with Elaboration (Values)}
\label{fig:constraint-generation-values}
\end{myfigure}

Constraint generation for values takes the form
$\inferStVal{\constraints}{\tyEnv}{v}{\aVty}{\constraints'}{\sol}{v'}$. It takes
as inputs a set of {\em} wanted constraints $\constraints$, a typing environment
$\tyEnv$, and a \source value $v$, and produces a value type $\aVty$, a new set of
wanted constraints $\constraints'$, a substitution $\sol$, and a \target
value $v'$.

In order to support let generalization, our inference algorithm does not keep constraint generation and solving
separate. Instead, the two are interleaved, as
indicated by the additional arguments of our relation:
\begin{inparaenum}[(a)]
\item constraints $\constraints$ are passed around in a stateful manner (i.e., they are input
  and output), and
\item substitutions $\sol$ generated from constraint solving constitute
  part of the relation output.
\end{inparaenum}

The rules are syntax-directed on the input \source value.
Rule~\textsc{TmVar} handles
term variables $x$: as usual for constraint-based type inference the rule
instantiates the polymorphic type $(\forall \bar{\evar}.
\overline{\tyvar:\ety}. \forall \bar{\dirtvar}. \bar{\srcPartialCt} \Rightarrow
\aVty)$ of $x$ with fresh variables; these are placeholders that are determined during constraint
solving. Moreover, the rule
extends the wanted constraints $\cstr$ with
$\bar{\srcPartialCt}$, appropriately instantiated. In \target, this corresponds
to explicit skeleton, type, dirt, and coercion applications.

More interesting is Rule~\textsc{TmAbs}, which handles term abstractions. Like in
standard Hindley-Damas-Milner~\citep{DamasMilner}, it generates a fresh type
variable $\tyvar$ for the type of the abstracted term variable $x$. In
addition, it generates a fresh skeleton variable $\evar$, to capture the (yet
unknown) shape of $\tyvar$.

As explained in detail in Section~\ref{subsec:constraint-solving}, the
constraint solver instantiates type variables only through their skeletons
annotations. Because we want to allow local constraint solving for the body
$c$ of the term abstraction the opportunity to
produce a substitution $\sol$ that instantiates $\tyvar$, we have to pass in the annotation constraint
$\tyvar : \evar$, which hints at why we
need to pass constraints in a stateful manner. We apply the resulting
substitution $\sol$ to the result type $\sol(\tyvar) \to
\cCty$ (though $\sol$ refers to \source types, we abuse notation to
save clutter and apply it directly to \target entities too).

Finally, Rule~\textsc{TmHand} is concerned with
handlers. Since it is the most complex of the rules, we discuss each of its
premises separately:

Firstly, we infer a type $\bVty_r\,!\,\dirt_r$ for the right hand side of the
$\mathtt{return}$-clause. Since $\tyvar_r$ is a fresh unification variable,
just like for term abstraction we require $\tyvar_r : \evar_r$, for a fresh
skeleton variable $\evar_r$.

Secondly, we check every operation clause in $\ops$ in order.
For each clause, we generate fresh skeleton, type, and dirt variables
($\evar_i$, $\tyvar_i$, and $\dirtvar_i$), to account for the (yet unknown)
result type $\tyvar_i\,!\,\dirtvar_i$ of the continuation $k$, while inferring
type $\bVty_{\op_i}\,!\,\dirt_{\op_i}$ for the right-hand-side $c_{\op_i}$.

More interesting is the (final) set of wanted constraints $\constraints'$.
First, we assign to the handler the overall type
\[
  \tyvar_\mathit{in}\,!\,\dirtvar_\mathit{in} \hto \tyvar_\mathit{out}\,!\,\dirtvar_\mathit{out}
\]
where $\evar_\mathit{in}, \tyvar_\mathit{in}, \dirtvar_\mathit{in},
\evar_\mathit{out}, \tyvar_\mathit{out}, \dirtvar_\mathit{out}$ are fresh
variables of the respective sorts.
In turn, we require that
\begin{inparaenum}[(a)]
\item the type of the return clause is a subtype of
  $\tyvar_\mathit{out}\,!\,\dirtvar_\mathit{out}$ (given by the combination of
  $\covar_1$ and $\covar_2$),
\item the right-hand-side type of each operation clause is a subtype of the
  overall result type: $\sol^n(\bVty_{\op_i}\,!\, \dirt_{\op_i}) \le
  \tyvar_\mathit{out}\,!\,\dirtvar_\mathit{out}$ (witnessed by
  $\covar_{3_i}\,!\,\covar_{4_i}$),
\item the actual types of the continuations $\bVty_i \to \tyvar_\mathit{out}\,!\,\dirtvar_\mathit{out}$ in the operation
  clauses should be subtypes of their assumed types $\bVty_i \to \sol^n(\tyvar_i\,!\,\dirtvar_i)$ (witnessed by $\covar_{5_i}$).
\item the overall argument type $\tyvar_\mathit{in}$ is a subtype of the
  assumed type of $x$:~$\sol^n(\sol_r(\tyvar_r))$ (witnessed by
  $\covar_6$), and
\item
  the input dirt set $\dirtvar_\mathit{in}$ is a subtype of the resulting dirt
  set $\dirtvar_\mathit{out}$, extended with the handled operations $\ops$
  (witnessed by $\covar_7$).
\end{inparaenum}

All the aforementioned implicit subtyping relations become explicit in the
elaborated term $c_\mathit{res}$, via explicit casts.

\paragraph{\bf Computations.}

\begin{myfigure}[t!]
$\ruleform{\inferStComp{\constraints}{\tyEnv}{c}{\cCty}{\constraints'}{\sol}{c'}}$ \textbf{Computations}
\begin{mathpar}
\inferrule*[right=TmApp]
           { \inferStVal{\constraints}{\tyEnv}{v_1}{\aVty_1}{\constraints_1}{\sol_1}{v_1'} \\
             \inferStVal{\constraints_1}{\sol_1(\tyEnv)}{v_2}{\aVty_2}{\constraints_2}{\sol_2}{v_2'}
           } 
           { \inferStComp{\constraints}
                         {\tyEnv}
                         {v_1~v_2}
                         {\tyvar\,!\,\dirtvar}
                         {\tyvar : \evar, \highlight{\covar :} \sol_2(\aVty_1) \le \aVty_2 \to \tyvar\,!\,\dirtvar, \constraints_2}
                         {(\sol_2 \cdot \sol_1)}
                         {(\cast{\sol_2(v_1')}{\covar})~v_2'}
           } 

\inferrule*[right=TmReturn]
           { \inferStVal{\constraints}{\tyEnv}{v}{\aVty}{\constraints'}{\sol}{v'} }
           { \inferStComp{\constraints}{\tyEnv}{\return{v}}{\aVty~!~\emptyset}{\constraints'}{\sol}{\return{v'}} }

\inferrule*[right=TmLet]
           { \inferStVal{\constraints}{\tyEnv}{v}{\aVty}{\constraints_v}{\sol_1}{v'} \\
             \unify(\bullet;\, \bullet;\, \constraints_v) = (\sol_1', \constraints_v') \\
             \splash{ \sol_1'(\sol_1(\tyEnv ))}{\constraints_v'}{\sol_1'(\aVty)}
               = \langle \bar{\evar}, \overline{\tyvar:\ety}, \bar{\dirtvar}, \overline{\highlight{\covar :} \srcPartialCt}, \constraints_1 \rangle \\
             \inferStComp{\constraints_1}
                         {\sol_1'(\sol_1(\tyEnv)),
                          x : \forall \bar{\evar}. \forall \overline{\tyvar:\ety}. \forall \bar{\dirtvar}. \overline{\srcPartialCt} \Rightarrow \sol_1'(\aVty)}
                         {c}{\cCty}{\constraints_2}{\sol_2}{c'} \\
             \highlight{c_\mathit{res} = \letval{x}{\sol_2(\Lambda \bar{\evar}. \Lambda \overline{\tyvar:\ety}. \Lambda \bar{\dirtvar}. \Lambda \overline{(\covar : \elabConstraint{\srcPartialCt})}. v')}{c'}}
           } 
           { \inferStComp{\constraints}
                         {\tyEnv}
                         {\letval{x}{v}{c}}
                         {\cCty}
                         {\constraints_2}
                         {(\sol_2 \cdot \sol_1' \cdot \sol_1)}
                         {c_\mathit{res}}
           } 

\inferrule*[right=TmOp]
           { \inferStVal{\constraints}{\tyEnv}{v}{\aVty_1}{\constraints_1}{\sol_1}{v'} \\
             \inferStComp{\constraints_1}{\sol_1(\tyEnv), y : \bVty_\op}{c}{\aVty_2~!~\dirt_2}{\constraints_2}{\sol_2}{c'} \\
             (\op : \aVty_\op \to \bVty_\op) \in \sig \\
             \highlight{c_\mathit{res} = \operation{(\cast{\sol_2(v')}{\covar})}{y : \elabVty{\bVty_\op}}{c'}}
           } 
           { \inferStComp{\constraints}
                         {\tyEnv}
                         {\operation{v}{y : \bVty_\op}{c}}
                         {\aVty_2~!~\{\op\} \cup \dirt_2}
                         {\highlight{\covar :} \sol_2(\aVty_1) \le \aVty_\op, \constraints_2}
                         {(\sol_2 \cdot \sol_1)}
                         {c_\mathit{res}}
           } 

\inferrule*[right=TmDo]
           { \inferStComp{\constraints}{\tyEnv}{c_1}{\aVty_1\,!\,\dirt_1}{\constraints_1}{\sol_1}{c_1'} \\
             \inferStComp{\constraints_1}{\sol_1(\tyEnv), x : \aVty_1}{c_2}{\aVty_2\,!\,\dirt_2}{\constraints_2}{\sol_2}{c_2'} \\
             \highlight{c_\mathit{res} = \doin{x}{(\cast{\sol_2(c_1')}{\refl{\sol_2(\aVty_1)}\,!\,\covar_1})}{(\cast{c_2'}{\refl{\aVty_2}\,!\,\covar_2})}}
           } 
           { \inferStComp{\constraints}
                         {\tyEnv}
                         {\doin{x}{c_1}{c_2}}
                         {\aVty_2~!~\dirtvar}
                         {\highlight{\covar_1 :} \sol_2(\dirt_1) \le \dirtvar, \highlight{\covar_2 :} \dirt_2 \le \dirtvar, \constraints_2}
                         {(\sol_2 \cdot \sol_1)}
                         {c_\mathit{res}}
           }

\inferrule*[right=TmHandle]
           { \inferStVal{\constraints}{\tyEnv}{v}{\aVty_1}{\constraints_1}{\sol_1}{v'} \\
             \inferStComp{\constraints_1}{\sol_1(\tyEnv)}{c}{\aVty_2\,!\,\dirt_2}{\constraints_2}{\sol_2}{c'} \\\\
             \constraints' = \tyvar_1 : \evar_1,
                             \tyvar_2 : \evar_2,
                             \highlight{\covar_1 :} \sol_2(\aVty_1) \le (\tyvar_1\,!\,\dirtvar_1 \hto \tyvar_2\,!\,\dirtvar_2),
                             \highlight{\covar_2 :} \aVty_2 \le \tyvar_1,
                             \highlight{\covar_3 :} \dirt_2 \le \dirtvar_1,
                             \constraints_2 \\
             \highlight{c_\mathit{res} = \withhandle{(\cast{\sol_2(v')}{\covar_1})}{(\cast{c'}{(\covar_2~!~\covar_3)})}}
           } 
           { \inferStComp{\constraints}
                         {\tyEnv}
                         {\withhandle{v}{c}}
                         {\tyvar_2~!~\dirt_2}
                         {\constraints'}
                         {(\sol_2 \cdot \sol_1)}
                         {c_\mathit{res}}
           } 
\end{mathpar}

\vspace{-7mm}
\caption{Constraint Generation with Elaboration (Computations)}
\label{fig:constraint-generation-computations}
\end{myfigure}

The judgement
$\inferStComp{\constraints}{\tyEnv}{c}{\cCty}{\constraints'}{\sol}{c'}$
generates constraints for computations.

Rule~\textsc{TmApp} handles term applications of the form $v_1\,v_2$. After
inferring a type for each subterm ($\aVty_1$ for $v_1$ and $\aVty_2$ for
$v_2$), we generate the wanted constraint $\sol_2(\aVty_1) \le \aVty_2 \to
\tyvar\,!\,\dirtvar$, with fresh type and dirt variables $\tyvar$ and
$\dirtvar$, respectively. Associated coercion variable $\covar$ is then used in
the elaborated term to explicitly (up)cast $v_1'$ to the expected type $\aVty_2
\to \tyvar\,!\,\dirtvar$.

Rule~\textsc{TmReturn} handles \texttt{return}-computations and is entirely
straightforward.

Rule~\textsc{TmLet} handles polymorphic let-bindings. First, we infer a type $\aVty$
for $v$, as well as wanted constraints $\constraints_v$. Then, we simplify
wanted constraints $\constraints_v$ by means of function $\unify$ (which we
explain in detail in Section~\ref{subsec:constraint-solving} below), obtaining
a substitution $\sol_1'$ and a set of {\em residual constraints}
$\constraints_v'$.

Generalization of $x$'s type is performed by the auxiliary function $\splashName$,
given by the following clause:
\[
\inferrule*[right=]
           { \bar{\evar}    = \{ \evar \mid (\tyvar : \evar) \in \constraints,
                                            \nexists \tyvar'. \tyvar' \notin \bar{\tyvar} \land (\tyvar' : \evar) \in \constraints \}
             \\
             \bar{\tyvar}   = \ftv{\constraints} \cup \ftv{\aVty}~\backslash~\ftv{\tyEnv}
             \\
             \constraints_1 = \{ (\covar : \srcPartialCt) \mid
               (\covar : \srcPartialCt) \in \constraints, \fv{\srcPartialCt} \not\subseteq \fv{\tyEnv} \}
             \\
             \bar{\dirtvar} = \fdv{\constraints} \cup \fdv{\aVty}~\backslash~\fdv{\tyEnv} ~~
             \\
             \constraints_2 = \constraints - \constraints_1
           } 
           { \splash{\tyEnv}{\constraints}{\aVty} = \langle \bar{\evar}, \overline{\tyvar : \ety}, \bar{\dirtvar}, \constraints_1, \constraints_2 \rangle }
\]
In essence, $\splashName$ generates the type (scheme) of $x$ in parts. Additionally, it
computes the subset $\constraints_2$ of the input constraints $\constraints$
that do not depend on locally-bound variables. Such constraints can be floated
``upwards'', and are passed as input when inferring a type for $c$. The remainder
of the rule is self-explanatory.

Rule~\textsc{TmOp} handles operation calls. Observe that in the elaborated term,
we upcast the inferred type to match the expected type in the signature.

Rule~\textsc{TmDo} handles sequences.
The requirement that all computations in a $\mathtt{do}$-construct have the
same dirt set is expressed in the wanted constraints $\sol_2(\dirt_1)
\le \dirtvar$ and
$\dirt_2 \le \dirtvar$ (where $\dirtvar$ is a fresh dirt variable; the
resulting dirt set), witnessed by coercion variables $\covar_1$
and $\covar_2$. Both coercion variables are used in the
elaborated term to upcast $c_1$ and $c_2$, such that both draw effects from the
same dirt set $\dirtvar$.

Finally, Rule~\textsc{TmHandle} is concerned with effect handling.
After inferring type $\aVty_1$ for the handler $v$, we require that it takes
the form of a handler type, witnessed by coercion variable $\covar_1 :
\sol_2(\aVty_1) \le (\tyvar_1\,!\,\dirtvar_1 \hto
\tyvar_2\,!\,\dirtvar_2)$, for fresh $\tyvar_1,
\tyvar_2, \dirtvar_1, \dirtvar_2$. To ensure that the type
$\aVty_2\,!\,\dirt_2$ of $c$ matches the expected type,
we require that $\aVty_2\,!\,\dirt_2 \le \tyvar_1\,!\,\dirtvar_1$.
Our syntax does not include coercion variables for computation subtyping;
we achieve the same effect by combining
$\covar_2 : \aVty_2 \le \tyvar_1$ and $\covar_3 : \dirt_2 \le \dirtvar_1$.

In the following, notation $\sol \models \constraints$ denotes that the
substitution $\sol$ is a \emph{solution} of the constraint set $\constraints$,
i.e., when after applying $\sol$ to all constraints in $\constraints$,
we get derivable judgements according to rules of Figure~\ref{fig:source-coercion-typing}.

\begin{theorem}[Soundness of Inference]
If $\inferStVal{\bullet}{\tyEnv}{v}{\aVty}{\constraints}{\sol}{v'}$ then for any $\sol' \models \constraints$, we have
$\tcElabVal{(\sol' \cdot \sol)(\tyEnv)}{v}{\sol'(\aVty)}{\sol'(v')}$, and analogously for computations.
\end{theorem}

\begin{theorem}[Completeness of Inference]
If $\tcElabVal{\tyEnv}{v}{\aVty}{v'}$ then we have
$\inferStVal{\bullet}{\tyEnv}{v}{\aVty'}{\constraints}{\sol}{v''}$ and there exists $\sol' \models \constraints$ and $\coercion$, such that
$\sol'(v'') = v'$ and $\proveCt{\sol(\tyEnv)}{\coercion}{\sol'(\aVty') \le \aVty}$. An analogous statement holds for computations.
\end{theorem}

\subsection{Constraint Solving}\label{subsec:constraint-solving}

The second phase of our inference-and-elaboration algorithm is the constraint
solver. It is defined by the $\unify$ function signature:
\[
  \ruleform{ \unify(\sol;\, \cstr;\, \queue) = (\sol',\, \cstr') }
\]
It takes three inputs: the substitution $\sol$ accumulated so far, a list of
already processed constraints $\cstr$, and a queue of still to be processed
constraints $\queue$. There are two outputs: the substitution $\sol'$ that
solves the constraints and the residual constraints $\cstr'$. The substitutions
$\sol$ and $\sol'$ contain four kinds of mappings: $\evar \mapsto \ety$, $\tyvar \mapsto \aVty$, $\dirtvar \mapsto \dirt$ and
$\covar \to \coercion$ which respectively instantiate skeleton variables,
type variables, dirt variables and coercion variables.

\begin{theorem}[Correctness of Solving]
For any set $\queue$, the call $\unify(\bullet; \bullet; \queue)$ either results in a failure,
in which case $\queue$ has no solutions, or returns $(\sol, \cstr)$ such that for any $\sol' \models \queue$, there exists
$\sol'' \models \cstr$ such that $\sol' = \sol'' \cdot \sol$.
\end{theorem}

The solver is invoked with $\unify(\bullet;\, \bullet;\, \constraints)$, to
process the constraints $\constraints$ generated in the first phase of the
algorithm, i.e., with an empty substitution and no processed constraints.
The $\unify$ function is defined by case analysis on the queue.

\paragraph{\bf Empty Queue} When the queue is empty, all constraints
have been processed. What remains are the residual constraints
and the solving substitution $\sol$, which are both returned as the result
of the solver.

\begin{center}\begin{shaded}\begin{minipage}{\columnwidth}
\small
\vspace{-3mm}
\[
\begin{array}{l}
  \unify(\sol;\, \cstr;\, \bullet) = (\sol,\, \cstr) \\
\end{array}
\]
\vspace{-5mm}
\end{minipage}\end{shaded}\end{center}

\paragraph{\bf Skeleton Equalities}
The next set of cases we consider are those where the queue is non-empty
and its first element is an equality between skeletons $\ety_1 = \ety_2$.
We consider seven possible cases based on the structure of $\ety_1$ and $\ety_2$
that together essentially implement conventional unification as used in
Hindley-Milner type inference~\citep{DamasMilner}.

\begin{center}\begin{shaded}\begin{minipage}{\columnwidth}
\small
\vspace{-2mm}
\[
\begin{array}{l}
  \unify(\sol;\, \cstr; \ety_1 = \ety_2, \queue) = \kpre{match} \ety_1 = \ety_2 \kpost{with} \\
  \quad \case \evar = \evar     \mapsto \unify(\sol;\, \cstr;\, \queue) \\[0.5mm]
  \quad \case \evar = \ety      \mapsto \kpre{if} \evar \notin \fsv{\ety}
    \kop{then} \letin{\sol' = [\ety/\evar]} \unify(\sol' \cdot \sol;\, \bullet; \sol'(\queue, \cstr))
    \kop{else} \mathtt{fail} \\[0.5mm]
  \quad \case \ety = \evar      \mapsto \kpre{if} \evar \notin \fsv{\ety}
    \kop{then} \letin{\sol' = [\ety/\evar]} \unify(\sol' \cdot \sol;\, \bullet; \sol'(\queue, \cstr))
    \kop{else} \mathtt{fail} \\[0.5mm]
  \quad \case \tyUnit = \tyUnit \mapsto \unify(\sol;\, \cstr;\, \queue) \\[0.5mm]
  \quad \case (\ety_1 \to \ety_2) = (\ety_3 \to \ety_4) \mapsto
    \unify(\sol;\, \cstr;\, \ety_1 = \ety_3, \ety_2 = \ety_4, \queue) \\[0.5mm]
  \quad \case (\ety_1 \hto \ety_2) = (\ety_3 \hto \ety_4) \mapsto
    \unify(\sol;\, \cstr;\, \ety_1 = \ety_3, \ety_2 = \ety_4, \queue) \\[0.5mm]
  \quad \case \mathtt{otherwise} \mapsto \mathtt{fail} \\
\end{array}
\]
\vspace{-2mm}
\end{minipage}\end{shaded}\end{center}

The first case applies when both skeletons are the same type variable $\evar$.
Then the equality trivially holds. Hence we drop it and proceed with solving
the remaining constraints.
The next two cases apply when either $\ety_1$ or $\ety_2$ is a skeleton
variable $\evar$. If the occurs check fails, there is no finite solution and
the algorithm signals failure. Otherwise, the constraint is solved by
instantiating the $\evar$. This additional substitution
is accumulated and applied to all other constraints $\cstr,\queue$.  Because
the substitution might have modified some of the already processed constraints
$\cstr$, we have to revisit them. Hence, they are all pushed back onto the
queue, which is processed recursively.

The next three cases consider three different ways in which the two skeletons
can have the same instantiated top-level structure. In those cases
the equality is decomposed into equalities on the subterms, which are pushed
onto the queue and processed recursively.

The last catch-all case deals with all ways in which the two skeletons can
be instantiated to different structures. Then there is no solution.

\paragraph{\bf Skeleton Annotations}
The next four cases consider a skeleton annotation $\tyvar : \ety$ at the head
of the queue, and propagate the skeleton instantiation to the type variable.
The first case, where the skeleton is a variable $\evar$, has nothing to do,
moves the annotation to the processed constraints and proceeds with the remainder
of the queue. In the other three cases, the skeleton is instantiated and
the solver instantiates the type variable with the corresponding structure, introducing
fresh variables for any subterms, where implicitly annotate every type variable with its
skeleton: $\tyvar^{\ety}$. The instantiating substitution is accumulated and applied
to the remaining constraints, which are processed recursively.

\begin{center}\begin{shaded}\begin{minipage}{\columnwidth}
\small
\vspace{-2mm}
\[
\begin{array}{l@{~}c@{~}l}
 \multicolumn{3}{l}{\unify(\sol;\, \cstr;\, \tyvar : \ety, \queue) = \kpre{match} \ety \kpost{with}} \\
  \quad \case \evar & \mapsto & \unify(\sol;\, \cstr, \tyvar : \ety;\, \queue) \\[0.5mm]
  \quad \case \tyUnit & \mapsto &
    \letin{\sol' = [\tyUnit/\tyvar]}
    \unify(\sol' \cdot \sol;\, \bullet;\, \sol'(\queue, \cstr)) \\[0.5mm]
  \quad \case \ety_1 \to \ety_2 & \mapsto &
    \letin{\sol' = [(\tyvar_1^{\ety_1} \to \tyvar_2^{\ety_2}\,!\,\dirtvar)/\tyvar]}
    \unify(\sol' \!\cdot\! \sol;\, \bullet;\, \tyvar_1 : \ety_1, \tyvar_2 : \ety_2, \sol'(\queue, \cstr)) \\[0.5mm]
  \quad \case \ety_1 \hto \ety_2 & \mapsto &
    \letin{\sol' = [(\tyvar_1^{\ety_1}\,!\,\dirtvar_1 \hto \tyvar_2^{\ety_2}\,!\,\dirtvar_2)/\tyvar]}
    \unify(\sol'\! \cdot \! \sol;\, \bullet;\, \tyvar_1 : \ety_1, \tyvar_2 : \ety_2, \sol'(\queue, \cstr)) \\
\end{array}
\]
\vspace{-2mm}
\end{minipage}\end{shaded}\end{center}

\paragraph{\bf Value Type Subtyping}
Next are the cases where a subtyping constraint between two value types
$\aVty_1 \le \aVty_2$---evidenced by coercion variable $\covar$---is at
the head of the queue.
We consider six different situations.
\begin{center}\begin{shaded}\begin{minipage}{\columnwidth}
\small
\vspace{-2mm}
\[
\begin{array}{l}
  \unify(\sol;\, \cstr;\, \covar : \aVty_1 \le \aVty_2, \queue) = \kpre{match} \aVty_1 \le \aVty_2 \kpost{with} \\[0.5mm]
  \quad \case \aVty \le \aVty \mapsto
    \letin{\vty = \elabVty{\aVty}} \unify([\refl{\vty}/\covar] \cdot \sol;\, \cstr;\, \queue) \\[0.5mm]
  \quad \case \tyvar^{\ety_1} \le \aVty \mapsto
     \letin{\ety_2 = \skeletonOf{\aVty}} 
    \unify(\sol;\, \cstr, \covar : \tyvar^{\ety_1} \le \aVty;\, \ety_1 = \ety_2, \queue) \\[0.5mm]
  \quad \case \aVty \le \tyvar^{\ety_1} \mapsto
     \letin{\ety_2 = \skeletonOf{\aVty}} 
    \unify(\sol;\, \cstr, \covar : \aVty \le \tyvar^{\ety_1};\, \ety_2 = \ety_1, \queue) \\[0.5mm]
  \quad \case (\aVty_1 \to \bVty_1\,!\,\dirt_1) \le (\aVty_2 \to \bVty_2\,!\,\dirt_2) \mapsto
    \letin{\sol' = [(\covar_1 \to \covar_2\,!\,\covar_3)/\covar]} \\[0.5mm]
    \qquad\qquad
    \unify(\sol' \cdot \sol;\, \cstr;\, \covar_1 : \aVty_2 \le \aVty_1, \covar_2 : \bVty_1 \le \bVty_2, \covar_3 : \dirt_1 \le \dirt_2, \queue) \\[0.5mm]
  \quad \case (\aVty_1\,!\,\dirt_1 \hto \aVty_2\,!\,\dirt_2) \le (\aVty_3\,!\,\dirt_3 \hto \aVty_4\,!\,\dirt_4) \mapsto
    \letin{\sol' = [(\covar_1\,!\,\covar_2 \hto \covar_3\,!\,\covar_4)/\covar]} \\[0.5mm]
    \qquad\qquad
    \unify(\sol' \cdot \sol;\, \cstr;\, \covar_1 : \aVty_3 \le \aVty_1, \covar_2 : \dirt_3 \le \dirt_1, \covar_3 : \aVty_2 \le \aVty_4, \covar_4 : \dirt_2 \le \dirt_4, \queue) \\[0.5mm]
  \quad \case \mathtt{otherwise} \mapsto \mathtt{fail} \\
\end{array}
\]
\vspace{-2mm}
\end{minipage}\end{shaded}\end{center}
If the two types are equal, the subtyping
holds trivially through reflexivity. The solver thus drops the constraint and
instantiates $\covar$ with the reflexivity coercion $\refl{\vty}$. Note that
each coercion variable only appears in one constraint.  So we only accumulate
the substitution and do not have to apply it to the other constraints.
In the next two cases, one of the two types is a type variable $\tyvar$. Then
we move the constraint to the processed set.
We also add an equality constraint between the skeletons to the queue, thus enforcing the
invariant that only types with the same skeleton are compared.
Through the skeleton equality the type structure (if any) from the type is also
transferred to the type variable.
The next two cases concern two types with the same top-level instantiation. In these cases the solver
decomposes the constraint into constraints on the corresponding subterms and
appropriately relates the evidence of the old constraint to the new ones.
The final case catches all situations where the two types are instantiated
with a different structure and thus there is no solution.

\noindent
Auxiliary function $\skeletonOf{\aVty}$, defined in Appendix~\ref{appendix:inference-additional}, computes the skeleton of $\aVty$.

\paragraph{\bf Dirt Subtyping}

The final six cases deal with subtyping constraints between dirts.

\begin{center}\begin{shaded}\begin{minipage}{\columnwidth}
\small
\vspace{-2mm}
\[
\begin{array}{@{\hspace{0mm}}l}
  \unify(\sol;\ \cstr; \covar :\dirt \le \dirt', \queue) = \kpre{match} \dirt \le \dirt' \kpost{with} \\[0.5mm]
   \quad \case \ops \cup \dirtvar \le \ops' \cup \dirtvar' \mapsto
     \kpre{if} \ops \neq \emptyset \kop{then}
     \letin{\sol' = [((\ops \backslash \ops') \cup \dirtvar'')/\dirtvar',\ops \cup \covar'/\covar]} \\[0.5mm]
     ~~~~~~~~~~~~~~~~~~~~~~~~~~~~~~~~~~~~~~~~~~~~~~~
     \unify(\sol' \cdot \sol;\ \bullet; (\covar': \dirtvar \leq \sol'(\dirt')) , \sol'(\queue, \cstr) ) \\[0.5mm]
     ~~~~~~~~~~~~~~~~~~~~~~~~~~~~~~~~~~~~~~~
     \kop{else} \unify(\sol;\ \cstr, (\covar :\dirt \le \dirt') ;\ \queue) \\[0.5mm]
   \quad \case \emptyset \le \dirt' \mapsto \unify([\emptyset_{\dirt'}/\covar] \cdot \sol;\ \cstr ;\ \queue ) \\[0.5mm]
   \quad \case \dirtvar \le \emptyset \mapsto \letin{\sol' = [\emptyset/\dirtvar ;\ \emptyset_\emptyset / \covar]}
      \unify( \sol' \cdot \sol ;\  \bullet ;\ \sol'(\queue, \cstr) ) \\[0.5mm]
    \quad \case \ops \cup \dirtvar \le \ops' \mapsto \\[0.5mm]
      ~~~~~
      \kpre{if} \ops \subseteq \ops' \kop{ then } \letin{\sol' =  [\ops \cup \covar'/\covar]} \unify(\sol' \cdot \sol;\ \cstr, (\covar' :\dirtvar \le \ops') ;\ \queue) \kop{else} \mathtt{fail } \\[0.5mm]
   \quad \case \ops \le \ops' \mapsto \kpre{if} \ops \subseteq \ops'
     \kop{then} \letin{\sol' =  [\ops \cup \emptyset_{\ops' \backslash \ops}/\covar]} \unify(\sol' \cdot \sol;\ \cstr ;\ \queue)
     \kop{else} \mathtt{fail} \\[0.5mm]
   \quad \case \ops \le \ops' \cup \dirtvar' \mapsto
     \letin{\sol' =  [ (\ops \backslash \ops' ) \cup \dirtvar'' /\dirtvar' ;\ \ops' \cup \emptyset_{(\ops' \backslash \ops) \cup \dirtvar''}/\covar]}
     \unify(\sol' \cdot \sol;\ \bullet ;\ \sol'(\queue, \cstr)) \\
\end{array}
\]
\vspace{-2mm}
\end{minipage}\end{shaded}\end{center}

If the two dirts are of the general form $\ops \cup \dirtvar$ and $\ops' \cup
\dirtvar'$, we distinguish two subcases. Firstly, if $\ops$ is empty, there
is nothing to be done and we move the constraint to the processed set. Secondly,
if $\ops$ is non-empty, we partially instantiate $\delta'$ with any of the operations
that appear in $\ops$ but not in $\ops'$. We then drop $\ops$ from the
constraint, and, after substitution, proceed with processing all constraints.
For instance, for $\{ \op_1 \} \cup \dirtvar \le \{ \op_2 \} \cup \dirtvar'$,
we instantiate $\dirtvar'$ to $\{ \op_1 \} \cup \dirtvar''$---where $\dirtvar''$ is a fresh dirt variable---and proceed with
the simplified constraint $\dirtvar \le \{ \op_1, \op_2 \} \cup \dirtvar''$.
Note that due to the set semantics of dirts, it is not valid to simplify the
above constraint to $\dirtvar \le \{ \op_2 \} \cup \dirtvar''$. After all
the substitution $[\dirtvar \mapsto \{ \op_1 \}, \dirtvar'' \mapsto \emptyset]$ solves the
former and the original constraint, but not the latter.

The second case, $\emptyset \le \dirt'$, always holds and is discharged by
instantiating $\covar$ to $\emptyset_{\dirt'}$.
The third case, $\dirtvar \le \emptyset$, has only one solution: $\dirtvar \mapsto \emptyset$
with coercion $\emptyset_\emptyset$.
The fourth case, $\ops \cup \dirtvar \le \ops'$, has as many solutions as there are
subsets of $\ops'$, provided that $\ops \subseteq \ops'$. We then simplify the constraint
to $\dirtvar \le \ops'$, which we move to the set of processed constraints.
The fifth case, $\ops \le \ops'$, holds iff $\ops \subseteq \ops'$.
The last case, $\ops \le \ops' \cup \dirtvar'$, is like the first, but
without a dirt variable in the left-hand side. We can satisfy it in a similar
fashion, by partially instantiating $\dirtvar'$ with $(\ops \setminus \ops')
\cup \dirtvar''$---where $\dirtvar''$ is a fresh dirt variable. Now the constraint is satisfied and can be
discarded.

\section{Erasure of Effect Information from \target}\label{sec:erasure}
Our first backend for \target is \erased, which is essentially a copy of \target
from which all effect information $\dirt$, type information $\vty$ and
coercions $\coercion$ have been erased. Instead, skeletons $\ety$ play the role
of plain types. Thus, \erased is essentially System F extended with term-level
(but not type-level) support for algebraic effects.

The main point of \erased is to show that we can erase
the effects and subtyping from \target to obtain types that are compatible with
a System F-like language. At the term-level \erased also resembles a subset of
Multicore OCaml~\cite{ocaml}, which provides native support for algebraic effects
and handlers but features no explicit polymorphism.

\subsection{The \erased Language}

Figure \ref{fig:erased-syntax} defines the syntax of \erased. The type system
and operational semantics of \erased follow from those of \target, and can be
found in Appendix~\ref{appendix:erased-additional}.

\begin{myfigure}[h]
\textbf{Terms}
\vspace{-3mm}
\[
\begin{array}{r@{~}c@{~}l}
    \text{value}~v & ::= & x                                         \mid
                           \tmUnit                                   \mid
                           h                                         \mid
                           \fun{(x:\ety)}{c}                         \mid
                           \Lambda \evar. v                         \mid
                           v~\ety                           \\

    \text{handler}~h & ::= & \{ \return{(x : \ety)} \mapsto c_r, \longcases \} \\

    \text{computation}~c & ::=  & v_1~v_2                          \mid
                                  \letval{x}{v}{c}                 \mid
                                  \return{v}                       \mid
                                  \operation{v}{y:\ety}{c}         \\
                         & \mid & \doin{x}{c_1}{c_2}               \mid
                                  \withhandle{v}{c}                \\
\end{array}
\]
\textbf{Types}
\vspace{-3mm}
\[
\begin{array}{r@{~}c@{~}l}
  \text{type}~\ety & ::= & \evar                            \mid
                           \ety_1 \to \ety_2                \mid
                           \ety_1 \hto \ety_2               \mid
                           \tyUnit                          \mid
                           \forall \evar. \ety              \\
\end{array}
\]

\vspace{-5mm}
\caption{\erased Syntax}
\label{fig:erased-syntax}
\end{myfigure}


\subsection{Erasure}

Figure \ref{fig:erasure} defines erasure functions $\eraseV{v}$, $\eraseC{c}$,
$\eraseVT{\vty}$, $\eraseCT{\cty}$ and $\eraseEnv{\tyEnv}$ for values,
computations, value types, computation types, and type environments
respectively. All five functions take a substitution $\sigma$ from the free
type variables $\tyvar$ to their skeleton $\ety$ as an additional parameter.

\begin{myfigure}[h]
\vspace{-3mm}
\[
\begin{array}{@{\hspace{0mm}}c@{\hspace{10mm}}c@{\hspace{0mm}}}
  \begin{array}{@{\hspace{0mm}}r@{~}c@{~}l@{\hspace{0mm}}}
    \eraseV{x}                        & = & x \\
    \eraseV{\tmUnit}                  & = & \tmUnit \\
    \eraseV{\cast{v}{\coercion}}      & = & \eraseV{v} \\
    \eraseV{\fun{(x:\vty)}{c}}        & = & \fun{(x:\eraseVT{\vty})}{\eraseC{c}} \\
    \eraseV{\Lambda \evar. v}         & = & \Lambda \evar. \eraseV{v} \\
    \eraseV{\Lambda (\tyvar:\ety). v} & = & \eraseV[\sigma \cdot \{\tyvar \mapsto \ety\}]{v} \\
  \end{array}
&
  \begin{array}{@{\hspace{0mm}}r@{~}c@{~}l@{\hspace{0mm}}}
    \eraseV{\Lambda \dirtvar. v}                 & = & \eraseV{v} \\
    \eraseV{\Lambda (\covar : \trgPartialCt). v} & = & \eraseV{v} \\
    \eraseV{v~\ety}                              & = & \eraseV{v}~\ety \\
    \eraseV{v~\vty}                              & = & \eraseV{v} \\
    \eraseV{v~\dirt}                             & = & \eraseV{v} \\
    \eraseV{v~\coercion}                         & = & \eraseV{v} \\
  \end{array}
\end{array}
\]
\[
\begin{array}{@{\hspace{0mm}}l@{\hspace{0mm}}}
  \eraseV{\{ \return{(x : \vty)} \mapsto c_r, \shortcases \}} = \\
  \quad~ \{ \return{(x : \eraseVT{\vty})} \mapsto \eraseC{c_r}, \erasedshortcases \} \\
\end{array}
\]

\[
\begin{array}{@{\hspace{0mm}}r@{~}c@{~}l@{\hspace{0mm}}}
  \eraseC{v_1 \ v_2}                & = & \eraseV{v_1} \ \eraseV{v_2} \\
  \eraseC{\letval{x}{v}{c}}         & = & \letval{x}{\eraseV{v}}{\eraseC{c}} \\
  \eraseC{\return{v}}               & = & \return{(\eraseV{v})} \\
  \eraseC{\operation{v}{y:\vty}{c}} & = & \operation{(\eraseV{v})}{y:\eraseVT{\vty}}{\eraseC{c}} \\
  \eraseC{\doin{x}{c_1}{c_2}}       & = & \doin{x}{\eraseC{c_1}}{\eraseC{c_2}} \\
  \eraseC{\withhandle{v}{c}}        & = & \withhandle{\eraseV{v}}{\eraseC{c}} \\
  \eraseC{\cast{c}{\coercion}}      & = & \eraseC{c} \\\\
\end{array}
\]
\[
\begin{array}{@{\hspace{0mm}}c@{\hspace{10mm}}c@{\hspace{0mm}}}
  \begin{array}{@{\hspace{0mm}}r@{~}c@{~}l@{\hspace{0mm}}}
    \eraseVT{\tyvar}                         & = & \sigma(\tyvar) \\
    \eraseVT{\vty \to \cty}                  & = & \eraseVT{\vty} \to \eraseCT{\cty} \\
    \eraseVT{\cty_1 \hto \cty_2}             & = & \eraseCT{\cty_1} \hto \eraseCT{\cty_2} \\
    \eraseVT{\tyUnit}                        & = & \tyUnit \\
    \eraseVT{\trgPartialCt \Rightarrow \vty} & = & \eraseVT{\vty} \\
    \eraseVT{\forall \evar. \vty}            & = & \forall \evar. \eraseVT{\vty} \\
    \eraseVT{\forall (\tyvar:\ety). \vty}    & = & \eraseVT[\sigma \cdot \{ \tyvar \mapsto \ety \}]{\vty} \\
    \eraseVT{\forall \dirtvar. \vty}         & = & \eraseVT{\vty} \\
  \end{array}
&
  \begin{array}{@{\hspace{0mm}}r@{~}c@{~}l@{\hspace{0mm}}}
    \eraseCT{\vty~!~\dirt}                  & = & \eraseVT{\vty} \\
    \\
    \eraseEnv{\epsilon}                     & = & \epsilon \\
    \eraseEnv{\tyEnv, \evar}                & = & \eraseEnv{\tyEnv}, \evar \\
    \eraseEnv{\tyEnv, \tyvar:\ety}          & = & \eraseEnv[\sigma \cdot \{ \tyvar \mapsto \ety \}]{\tyEnv} \\
    \eraseEnv{\tyEnv, \dirtvar}             & = & \eraseEnv{\tyEnv} \\
    \eraseEnv{\tyEnv, x : \vty}             & = & \eraseEnv{\tyEnv}, x : \eraseVT{\vty} \\
    \eraseEnv{\tyEnv, \covar : \constraint} & = & \eraseEnv{\tyEnv} \\
  \end{array}
\end{array}
\]

\vspace{-3mm}
\caption{Definition of type erasure.}
\label{fig:erasure}
\end{myfigure}

Thanks to the skeleton-based design of \target, erasure is straightforward.
All types are erased to their skeletons,
dropping quantifiers for type variables and all occurrences of dirt sets.
Moreover, coercions are dropped from values and computations. Finally,
all binders and elimination forms for type variables, dirt set variables and coercions are dropped
from values and type environments.

\begin{example}
  \label{exa:running-erasure}
  Continuing the Example~\ref{exa:running-target}, a monomorphic function
  \[
  \begin{array}{l@{\hspace{1mm}}c@{\hspace{1mm}}l}
    \keylet~~\mathit{f} & : & (\tyUnit \to \tyUnit~!~\emptyset) \to \tyUnit~!~\emptyset \\
                        & = & \fun{(g : \tyUnit \to \tyUnit~!~\emptyset)}{g~\tmUnit} \\
    \keyin~\ldots \\
  \end{array}
  \]
  is erased to
    \[
    \begin{array}{l@{\hspace{1mm}}c@{\hspace{1mm}}l}
      \keylet~~\mathit{f} & : & (\tyUnit \to \tyUnit) \to \tyUnit \\
                          & = & \fun{(g : \tyUnit \to \tyUnit)}{g~\tmUnit} \\
      \keyin~\ldots \\
    \end{array}
    \]
  while its polymorphic variant
  \[
  \begin{array}{l@{\hspace{1mm}}c@{\hspace{1mm}}l}
    \keylet~~\mathit{f} & : & \forall \evar. \forall \tyvar : \evar. \forall \tyvar' : \evar. \forall \dirtvar. \forall \dirtvar'.
                                (\tyvar \le \tyvar') \Rightarrow (\dirtvar \le \dirtvar') \Rightarrow (\tyUnit \to \tyvar~!~\dirtvar) \to \tyvar' ~!~ \dirtvar' \\
                        & = & \Lambda \evar. \Lambda (\tyvar : \evar). \Lambda (\tyvar' : \evar). \Lambda \dirtvar. \Lambda \dirtvar'. \Lambda (\covar : \tyvar \le \tyvar'). \Lambda (\covar' : \dirtvar \le \dirtvar'). \\
                        &   & \quad \fun{(g : \tyUnit \to \tyvar \,!\, \dirtvar)}{(\cast{(g~\tmUnit)}{(\covar \,!\, \covar')})} \\
    \keyin~\ldots \\
  \end{array}
  \]
  is erased to
  \[
  \begin{array}{l@{\hspace{1mm}}c@{\hspace{1mm}}l}
    \keylet~~\mathit{f} & : & \forall \evar. (\tyUnit \to \evar) \to \evar \\
                        & = & \Lambda \evar. \fun{(g : \tyUnit \to \evar)}{g~\tmUnit} \\
    \keyin~\ldots \\
  \end{array}
  \]
  Note that in addition to removing all effect annotations and coercions, the erasure
  removed type quantifiers and abstractions, and replaced $\tyvar$ and $\tyvar'$ with their skeleton $\evar$.
  
  We proceed similarly in applications, where
  \[
    f~\tyUnit~\tyUnit~\tyUnit~\emptyset~\emptyset~\trgUnitRefl~\emptyset_\emptyset~(\fun{(x:\tyUnit)}{\return{x}})
  \]
  is erased simply to
  \[
    f~\tyUnit~(\fun{(x:\tyUnit)}{\return{x}})
  \]
  where only the skeleton application remains. Similarly
  \begin{align*}
    &f~\tyUnit~\tyUnit~\tyUnit~\{\texttt{Tick}\}~\{\texttt{Tick}, \texttt{Tock}\}~\trgUnitRefl~(\{\texttt{Tick}\} \cup \emptyset_{\{\texttt{Tock}\}}) \\
    &\quad (\fun{x : \tyUnit}{\operation[\mathtt{Tick}]{x}{y : \tyUnit}{(\cast{(\return{y})}{\trgUnitRefl~!~\emptyset_{\{\texttt{Tick}\}}})}})
  \end{align*}
  is erased to
  \begin{align*}
    &f~\tyUnit~(\fun{x : \tyUnit}{\operation[\mathtt{Tick}]{x}{y : \tyUnit}{\return{y}}})
  \end{align*}
  showing that a polymorphic function is applied in exactly the same way to a pure or an impure function.
\end{example}

The expected theorems hold. Firstly, types are preserved by
erasure, where typing for \erased values and computations takes the obvious forms
$\tcErsVal{\tyEnv}{v}{\ety}$ and $\tcErsComp{\tyEnv}{c}{\ety}$.
\begin{theorem}[Type Preservation]
\label{thm:erasure_type_preservation}
If $\tcTrgVal{\tyEnv}{v}{\vty}$ then  $\tcErsVal{\eraseEnv[\emptyset]{\tyEnv}}{\eraseV[\tyEnv]{v}}{\eraseVT[\tyEnv]{\vty}}$.
If $\tcTrgComp{\tyEnv}{c}{\cty}$ then $\tcErsComp{\eraseEnv[\emptyset]{\tyEnv}}{\eraseC[\tyEnv]{c}}{\eraseCT[\tyEnv]{\cty}}$.
\end{theorem}
Here we abuse of notation and use $\tyEnv$ as a substitution
from type variables to skeletons used by the erasure functions.

Finally, we have that erasure preserves the operational semantics.
\begin{theorem}[Semantic Preservation]
\label{thm:erasure_semantic_preservation}
If $\smallStepVal{v}{v'}$ then $\congStepVal{\eraseV{v}}{\eraseV{v'}}$.
If $\smallStepComp{c}{c'}$ then $\congStepComp{\eraseC{c}}{\eraseC{c'}}$.
\end{theorem}
In both cases, $\equiv^{\leadsto}$ denotes the congruence closure of the step relation in \erased, defined in
Appendix~\ref{appendix:erased-additional}. 
The choice of substitution $\sigma$ does not matter as types do not affect
the behaviour. Note that because coercions are dropped during erasure, this 
means that also in \target they do not have an essential runtime impact.

\begin{corollary}[Coercion Irrelevance]
If $\bigStepVal{v}{v_1}$ and $\bigStepVal{\cast{v}{\coercion}}{v_2}$
then $\congStepVal{\eraseV{v_1}}{\eraseV{v_2}}$.
If $\bigStepComp{c}{c_1}$ and $\bigStepComp{\cast{c}{\coercion}}{c_2}$
then $\congStepComp{\eraseV{c_1}}{\eraseV{c_2}}$.
\end{corollary}

\paragraph{\bf Discussion}

The reason we need to use the symmetric congruence
closure of the step relation in our preservation theorem is that the original
\target program and the resulting \erased program do not necessarily operate in
lockstep. Indeed, the erasure of casts with coercions, of type and coercion
binders and of their applications means that the erased program does not have
to step through their reductions. On the other hand, the erasure of type and
coercion binders may expose applications of skeleton binders that the \erased
program has to reduce whereas the original \target program does not.

For example, take the \target term 
\[ c_1 = (\lambda (x:\forall \dirtvar.\tyUnit). \return{(\lambda (y: \tyUnit). \return{(x~\emptyset)})})~(\Lambda \dirtvar. (\Lambda \evar.\tmUnit)~\tyUnit) \]
which $\beta$-reduces to 
\[ c_2 = \return{(\lambda (y: \tyUnit). \return{((\Lambda \dirtvar. (\Lambda \evar.\tmUnit)~\tyUnit)~\emptyset)})} \]
When we erase $c_1$, we get
\[ \eraseC{c_1} = (\lambda (x:\tyUnit). \return{(\lambda (y: \tyUnit). \return{x})})~((\Lambda \evar.\tmUnit)~\tyUnit) \]
The erasure of the $\Lambda \delta$ binder exposes a new redex that has precedence. Hence, $\eraseC{c_1}$ steps to
\[ (\lambda (x:\tyUnit). \return{(\lambda (y: \tyUnit). \return{x})})~\tmUnit \]
which steps to the irreducible computation
\[ \return{(\lambda (y: \tyUnit). \return{\tmUnit})} \]
In contrast, $c_2$ erases to a different irreducible computation
\[ \eraseC{c_2} = \return{(\lambda (y: \tyUnit). \return{((\Lambda \evar.\tmUnit)~\tyUnit)})} \]
These two irreducible computations can be made equal by reducing under the
$\lambda (y: \tyUnit)$ binder in $\eraseC{c_2}$. The congruence
closure of the step relation allows this reduction under binders. Morever, the closure 
is symmetric because an \target step may defer or block a \erased step
that is exposed by the erasure.


Typically, when type information is erased from call-by-value languages,
type binders are erased by replacing them with other (dummy) binders. For instance,
the expected definition of erasure would be:
\begin{equation*}
  \eraseV{\Lambda (\alpha:\ety).v} = \lambda (x:\tyUnit). \eraseV{v}
\end{equation*}
This replacement is motivated by a desire to preserve the behaviour of the typed
terms. By dropping binders, values might be turned into computations that
trigger their side-effects immediately, rather than at the later point where
the original binder was eliminated. However, there is no call for this circumspect
approach in our setting, as our grammatical partition of terms in values (without
side-effects) and computations (with side-effects) guarantees that this problem
cannot happen when we erase values to values and computations to computations.
Nevertheless, when adding recursion to the language, care is needed to preserve
the termination behavior of values under erasure, though we believe this is not a problem as
appropriate recursive constructs are invoked only at the computation level.

\section{Elaboration to a Language Without Effects}\label{sec:elaboration-to-ocaml}
This section considers an alternative backend for \target, called \noeff. In
contrast to \erased, \noeff's types are explicit about whether or not effects
can be used, but implicit about which effects in particular are used. 

Given that \noeff's types track whether effects are used or not, its name may
seem contradictory. Yet, the calculus is intended to model a purely functional
approach to implementing handlers, e.g., in the pure fragment of OCaml or in
Haskell, where there is no native support for algebraic effects (thus the name
\noeff). In such pure languages, algebraic effects are modeled by
means of a user-defined encoding~\cite{kammar,optimization,eff2ocaml,koka2017}
and the type constructors used by these encodings reveal whether effectful or
pure computations are encoded. Here, to keep \noeff small, we encapsulate
the particular encoding details---which could be implemented in a library---and
present the effect functionality as opaque primitives in \noeff.

\subsection{Syntax of \noeff}\label{sec:ml-syntax}

\begin{myfigure}[t]

\textbf{Terms}
\[
  \begin{array}{r@{~}c@{~}l}
    \text{value}~\mlTm                                               & ::= &
      x                                                              \mid
      \tmUnit                                                        \mid

      \fun{x:\mlTyA}{\mlTm}                                          \mid
      \mlTm_1~\mlTm_2                                                \mid

      \Lambda \tyvar. \mlTm                                          \mid
      \mlTm~\mlTyA                                                   \mid

      \Lambda (\covar : \mlCoTy). \mlTm                              \mid
      \mlTm~\mlCoercion                                              \mid

      \cast{\mlTm}{\mlCoercion}                                      \mid
      \return{\mlTm}                                                 \\
    & \mid &
      \mlHandler                                                     \mid
      \letval{x}{\mlTm_1}{\mlTm_2}                                   \mid
      \operation{\mlTm_1}{y : \mlTyB}{\mlTm_2}                       \mid
      \doin{x}{\mlTm_1}{\mlTm_2}                                     \mid
      \withhandle{\mlTm_h}{\mlTm_c}                                  \\

    \text{handler}~h                                                 & ::= &
      \mlShorthand                                                   \\ 
  \end{array}
\]

\textbf{Types}
\[
  \begin{array}{r@{~}c@{~}l}
    \text{type}~\mlTyA, \mlTyB                                       & ::= &
      \tyvar                                                         \mid
      \tyUnit                                                        \mid
      \mlTyA \to \mlTyA                                              \mid
      \mlTyA \hto \mlTyB                                             \mid
      \mlCoTy \Rightarrow \mlTyA                                     \mid
      \mkMlCompTy{\mlTyA}                                            \mid
      \forall \tyvar. \mlTyA \\

    \text{coercion type}~\mlCoTy                                     & ::= &
      \mlTyA \le \mlTyB                                              \\
  \end{array}
\]

\textbf{Coercions}
\[
\begin{array}{r@{~}c@{~}l}
    \mlCoercion & ::=  &
       \covar                                   \mid 
       \mlCoUnitRefl                            \mid 
       \refl{\tyvar}                            \mid 
       \mlCoercion_1 \to  \mlCoercion_2         \mid 
       \mlCoercion_1 \hto \mlCoercion_2         \mid 
       \handToFun{\mlCoercion_1}{\mlCoercion_2} \mid 
       \funToHand{\mlCoercion_1}{\mlCoercion_2} \\   
    & \mid &
       \forall \tyvar. \mlCoercion              \mid 
       \mlCoTy \Rightarrow \mlCoercion          \mid 
       \mkMlCompCo{\mlCoercion}                 \mid 
       \mlReturn{\mlCoercion}                   \mid 
       \unsafe{\mlCoercion}                     \\   
\end{array}
\]

\vspace{-5mm}
\caption{\noeff Syntax}
\label{fig:ml-syntax}
\end{myfigure}

Figure~\ref{fig:ml-syntax} presents the syntax of \noeff. Notably \noeff
replaces \target's two syntactic sorts of values and computations by a single
syntactic sort of terms that combines their syntactic forms. The four
absent forms are dirt and skeleton abstraction and application, as \noeff does not feature
either dirt or skeletons. Similarly, \target's syntactic sorts for value types $\vty$ and computation types $\cty$ are merged
into a single sort of types $\mlTyA$. Here \target's computation types of the form
$\withdirt{\vty}{\dirt}$ are replaced by \noeff's computation types $\mkMlCompTy{\mlTyA}$ without dirt.
The absence of dirt can also be seen in \noeff's coercion types $\mlCoTy$, which do not feature a form
for dirt subtyping.

Finally, \noeff features adapted versions of \target's type coercions.  Absent
are those related to dirt and skeletons, and the computation type coercion is abstracted to
the form $(\mkMlCompCo{\mlCoercion})$ which does not feature a dirt coercion.
There are also four new coercion forms
($\handToFun{\mlCoercion_1}{\mlCoercion_2}$, $\funToHand{\mlCoercion_1}{\mlCoercion_2}$,  $\mlReturn{\mlCoercion}$ and
$\unsafe{\mlCoercion}$) which enable the elaboration from \target into \noeff;
we explain their semantics when we discuss typing and their purpose when
explaining the elaboration.

\subsection{Typing of \noeff}\label{sec:ml-typing}

We now turn to typing of \noeff. First, we introduce \noeff typing
environments; they are identical to those for \target, modulo dirt and skeleton
information:
\[
\begin{array}{r@{~}c@{~}l}
  \mlEnv & ::= & \epsilon                             \mid
                 \mlEnv, \tyvar                       \mid
                 \mlEnv, x : \mlTyA                   \mid
                 \mlEnv, \covar : \mlCoTy             \\
\end{array}
\]

\noindent
The remainder of this section gives the typing judgements for terms
(Section~\ref{sec:ml-term-typing}) and coercions
(Section~\ref{sec:ml-coercion-typing}); uninteresting judgements like
well-formedness of types ($\tcNoEffTy{\mlEnv}{\mlTyA}$) and well-formedness of
constraints ($\tcNoEffCoTy{\mlEnv}{\mlCoTy}$) are included in
Appendix~\ref{appendix:ml-additional}.

\subsubsection{Term Typing}\label{sec:ml-term-typing}

\begin{myfigure}[t]
$\ruleform{\tcNoEffTm{\mlEnv}{\mlTm}{\mlTyA}}$ \textbf{Term Typing}
\begin{mathpar}
\inferrule*[right=]
           { (x : \mlTyA) \in \mlEnv }
           { \tcNoEffTm{\mlEnv}{x}{\mlTyA} }

\inferrule*[right=]
           { }
           { \tcNoEffTm{\mlEnv}{\tmUnit}{\tyUnit} }

\inferrule*[right=]
           { \tcNoEffTy{\mlEnv}{\mlTyA} \\
             \tcNoEffTm{\mlEnv, x : \mlTyA}{\mlTm}{\mlTyB}
           } 
           { \tcNoEffTm{\mlEnv}{(\fun{x:\mlTyA}{\mlTm})}{\mlTyA \to \mlTyB} }

\inferrule*[right=]
           { \tcNoEffTm{\mlEnv, \tyvar}{\mlTm}{\mlTyA} }
           { \tcNoEffTm{\mlEnv}{\Lambda \tyvar. \mlTm}{\forall \tyvar. \mlTyA} }

\inferrule*[right=]
           { \tcNoEffTy{\mlEnv}{\mlTyA} \\
             \tcNoEffTm{\mlEnv}{\mlTm}{\forall \tyvar. \mlTyB}
           } 
           { \tcNoEffTm{\mlEnv}{\mlTm~\mlTyA}{\mlTyB[\mlTyA/\tyvar]} }

\inferrule*[right=]
           { \tcNoEffTm{\mlEnv}{\mlTm}{\mlTyA} \\
             \tcNoEffCoercion{\mlEnv}{\mlCoercion}{\mlTyA \le \mlTyB}
           } 
           { \tcNoEffTm{\mlEnv}{\cast{\mlTm}{\mlCoercion}}{\mlTyB} }

\inferrule*[right=]
           { \tcNoEffTm{\mlEnv, x : \mlTyA}{\mlTm_r}{\mkMlCompTy{\mlTyB}} \\
             \left[
               (\op : \mlTyA_1 \to \mlTyA_2) \in \mlSig \qquad
               \tcNoEffTm{\mlEnv, x : \mlTyA_1, k : \mlTyA_2 \to \mkMlCompTy{\mlTyB}}{\mlTm_\op}{\mkMlCompTy{\mlTyB}}
             \right]_{\op \in \ops}
           } 
           { \tcNoEffTm{\mlEnv}{\mlShorthand}{\mlTyA \hto \mlTyB} }

\inferrule*[right=]
           { \tcNoEffCoTy{\mlEnv}{\mlCoTy} \\
             \tcNoEffTm{\mlEnv, \covar : \mlCoTy}{\mlTm}{\mlTyA}
           } 
           { \tcNoEffTm{\mlEnv}{\Lambda (\covar : \mlCoTy). \mlTm}{\mlCoTy \Rightarrow \mlTyA} }

\inferrule*[right=]
           { \tcNoEffTm{\mlEnv}{\mlTm}{\mlCoTy \Rightarrow \mlTyA} \\
             \tcNoEffCoercion{\mlEnv}{\mlCoercion}{\mlCoTy}
           } 
           { \tcNoEffTm{\mlEnv}{\mlTm~\mlCoercion}{\mlTyA} }

\inferrule*[right=]
           { \tcNoEffTm{\mlEnv}{\mlTm_1}{\mlTyA \to \mlTyB} \\
             \tcNoEffTm{\mlEnv}{\mlTm_2}{\mlTyA}
           } 
           { \tcNoEffTm{\mlEnv}{\mlTm_1~\mlTm_2}{\mlTyB} }

\inferrule*[right=]
           { \tcNoEffTm{\mlEnv}{\mlTm_1}{\mlTyA} \\
             \tcNoEffTm{\mlEnv, x : \mlTyA}{\mlTm_2}{\mlTyB}
           } 
           { \tcNoEffTm{\mlEnv}{\letval{x}{\mlTm_1}{\mlTm_2}}{\mlTyB} }

\inferrule*[right=]
           { \tcNoEffTm{\mlEnv}{\mlTm}{\mlTyA} }
           { \tcNoEffTm{\mlEnv}{\return{\mlTm}}{\mkMlCompTy{\mlTyA}} }

\inferrule*[right=]
           { (\op : \mlTyA_1 \to \mlTyA_2) \in \mlSig \\
             \tcNoEffTm{\mlEnv}{\mlTm_1}{\mlTyA_1} \\
             \tcNoEffTm{\mlEnv, y : \mlTyA_2}{\mlTm_2}{\mkMlCompTy{\mlTyB}} %
           } 
           { \tcNoEffTm{\mlEnv}{\operation{\mlTm_1}{y : \mlTyA_2}{\mlTm_2}}{\mkMlCompTy{\mlTyB}} }

\inferrule*[right=]
           { \tcNoEffTm{\mlEnv}{\mlTm_1}{\mkMlCompTy{\mlTyA}} \\
             \tcNoEffTm{\mlEnv, x : \mlTyA}{\mlTm_2}{\mkMlCompTy{\mlTyB}}
           } 
           { \tcNoEffTm{\mlEnv}{\doin{x}{\mlTm_1}{\mlTm_2}}{\mkMlCompTy{\mlTyB}} }

\inferrule*[right=]
           { \tcNoEffTm{\mlEnv}{\mlTm_h}{\mlTyA \hto \mlTyB} \\
             \tcNoEffTm{\mlEnv}{\mlTm_c}{\mkMlCompTy{\mlTyA}}
           } 
           { \tcNoEffTm{\mlEnv}{\withhandle{\mlTm_h}{\mlTm_c}}{\mkMlCompTy{\mlTyB}} }
\end{mathpar}

\vspace{-5mm}
\caption{\noeff Term Typing}
\label{fig:ml-term-typing}
\end{myfigure}

Typing for \noeff terms is given by judgement
$\tcNoEffTm{\mlEnv}{\mlTm}{\mlTyA}$, which is presented in
Figure~\ref{fig:ml-term-typing}. The rules are similar to those of
\source and \target, with the exception of
dirt and skeleton features, which are absent in \noeff.

There is one subtle point: By design, type $\mlTyA \hto \mlTyB$ classifies
handlers that handle terms of type $\mkMlCompTy{\mlTyA}$ and produce results of
type $\mkMlCompTy{\mlTyB}$. This way, we enforce that handlers always take
computations to computations. If the input is not a computation, we can use a
regular function instead of a handler. So this restriction matters little.

More importantly, by forcing the output to be a computation, we avoid a
potential source of unsoundness in \noeff.  Indeed, because the type system
does not track which operations are performed in the input computation, we
cannot tell whether or not they will all be handled. Of course, we do want any
operation that is not handled to be forwarded to the output, just like in
\target. Hence, because we cannot statically tell in \noeff whether any
operations will be forwarded, to remain on the safe side we have to assume that
there may be some. Thus, with forwarded operations, the output must be a
computation. We will see that this causes additional difficulties in the elaboration
from \target to \noeff.

%
\subsubsection{Coercion Typing}\label{sec:ml-coercion-typing}

\begin{myfigure}[t]
$\ruleform{\tcNoEffCoercion{\mlEnv}{\mlCoercion}{\mlCoTy}}$ \textbf{Coercion Typing}
\begin{mathpar}
\inferrule*[right=]
           { (\covar : \mlCoTy) \in \mlEnv }
           { \tcNoEffCoercion{\mlEnv}{\covar}{\mlCoTy} }

\inferrule*[right=]
           { }
           { \tcNoEffCoercion{\mlEnv}{\mlCoUnitRefl}{\tyUnit \le \tyUnit} }

\inferrule*[right=]
           { \tyvar \in \mlEnv }
           { \tcNoEffCoercion{\mlEnv}{\refl{\tyvar}}{\tyvar \le \tyvar} }

\inferrule*[right=]
           { \tcNoEffCoercion{\mlEnv}{\mlCoercion_1}{\mlTyA_2 \le \mlTyA_1} \\
             \tcNoEffCoercion{\mlEnv}{\mlCoercion_2}{\mlTyB_1 \le \mlTyB_2}
           } 
           { \tcNoEffCoercion{\mlEnv}{\mlCoercion_1 \to \mlCoercion_2}{(\mlTyA_1 \to \mlTyB_1) \le (\mlTyA_2 \to \mlTyB_2)} }

\inferrule*[right=]
           { \tcNoEffCoercion{\mlEnv}{\mlCoercion_1}{\mkMlCompTy{\mlTyA_2} \le \mkMlCompTy{\mlTyA_1}} \\
             \tcNoEffCoercion{\mlEnv}{\mlCoercion_2}{\mkMlCompTy{\mlTyB_1} \le \mkMlCompTy{\mlTyB_2}}
           } 
           { \tcNoEffCoercion{\mlEnv}{\mlCoercion_1 \hto \mlCoercion_2}{(\mlTyA_1 \hto \mlTyB_1) \le (\mlTyA_2 \hto \mlTyB_2)} }

\inferrule*[right=]
           { \tcNoEffCoercion{\mlEnv}{\mlCoercion_1}{\mlTyA_2 \le \mlTyA_1} \\
             \tcNoEffCoercion{\mlEnv}{\mlCoercion_2}{\mkMlCompTy{\mlTyB_1} \le \mlTyB_2}
           } 
           { \tcNoEffCoercion{\mlEnv}{\handToFun{\mlCoercion_1}{\mlCoercion_2}}{(\mlTyA_1 \hto \mlTyB_1) \le (\mlTyA_2 \to \mlTyB_2)} }

\inferrule*[right=]
           { \tcNoEffCoercion{\mlEnv}{\mlCoercion_1}{\mlTyA_2 \le \mlTyA_1} \\
             \tcNoEffCoercion{\mlEnv}{\mlCoercion_2}{\mlTyB_1 \le \mkMlCompTy{\mlTyB_2}}
           } 
           { \tcNoEffCoercion{\mlEnv}{\funToHand{\mlCoercion_1}{\mlCoercion_2}} \le (\mlTyA_1 \to \mlTyB_1) \le {(\mlTyA_2 \hto \mlTyB_2)} }

\inferrule*[right=]
           { \tcNoEffCoercion{\mlEnv, \tyvar}{\mlCoercion}{\mlTyA \le \mlTyB} }
           { \tcNoEffCoercion{\mlEnv}{\forall \tyvar. \mlCoercion}{\forall \tyvar. \mlTyA \le \forall \tyvar. \mlTyB} }

\inferrule*[right=]
           { \tcNoEffCoTy{\mlEnv}{\mlCoTy} \\
             \tcNoEffCoercion{\mlEnv}{\mlCoercion}{\mlTyA \le \mlTyB}
           } 
           { \tcNoEffCoercion{\mlEnv}{\mlCoTy \Rightarrow \mlCoercion}{\mlCoTy \Rightarrow \mlTyA \le \mlCoTy \Rightarrow \mlTyB} }

\inferrule*[right=]
           { \tcNoEffCoercion{\mlEnv}{\mlCoercion}{\mlTyA_1 \le \mlTyA_2} }
           { \tcNoEffCoercion{\mlEnv}{\mkMlCompCo{\mlCoercion}}{\mkMlCompTy{\mlTyA_1} \le \mkMlCompTy{\mlTyA_2}} }

\inferrule*[right=]
           { \tcNoEffCoercion{\mlEnv}{\mlCoercion}{\mlTyA_1 \le \mlTyA_2} }
           { \tcNoEffCoercion{\mlEnv}{\mlReturn{\mlCoercion}}{\mlTyA_1 \le \mkMlCompTy{\mlTyA_2}} }

\inferrule*[right=]
           { \tcNoEffCoercion{\mlEnv}{\mlCoercion}{\mlTyA_1 \le \mlTyA_2} }
           { \tcNoEffCoercion{\mlEnv}{\unsafe{\mlCoercion}}{\mkMlCompTy{\mlTyA_1} \le \mlTyA_2} }
\end{mathpar}

\vspace{-5mm}
\caption{\noeff Coercion Typing}
\label{fig:ml-coercion-typing}
\end{myfigure}

Coercion typing is given by judgement
$\tcNoEffCoercion{\mlEnv}{\mlCoercion}{\mlCoTy}$, presented in
Figure~\ref{fig:ml-coercion-typing}. Most of the rules are straightforward so
we only focus on the four new coercion forms.

The first new coercion form ($\handToFun{\mlCoercion_1}{\mlCoercion_2}$) concerns the
issue of handler typing above. It converts a handler, which expects a
computation as input, into a function, which can be applied to a
non-computation. The next coercion form ($\funToHand{\mlCoercion_1}{\mlCoercion_2}$) is its
dual; it turns a function into a handler that only specifies how to handle the $\keyreturn$
case and forwards all operations.

The third new coercion form ($\mlReturn{\mlCoercion}$) promotes a value $\mlTm$ of any type $\mlTyA$
to a computation $\return{\mlTm}$ of type $\mkMlCompTy{\mlTyA}$.
The last new coercion form ($\unsafe{\mlCoercion}$) is the dual of the previous form. It forces
a value of computation type $\mkMlCompTy{\mlTyA}$ to a value of type $\mlTyA$. This only
works when the value is of the form $\return{\mlTm}$ and in that case yields $\mlTm$.
If the computation is of the form $\operation{\mlTm_1}{y : \mlTyB}{\mlTm_2}$, the cast
gets stuck; hence its name. We will see that this is the single source of type unsafety in \noeff, though we claim that programs elaborated from \target into \noeff only use this coercion
in a safe way and never get stuck.

\subsection{Operational Semantics of \noeff}\label{sec:ml-opsem}

\begin{myfigure}[t]
\[
\begin{array}{r@{~}c@{~}l}
  \text{value}~\mlValue & ::= &%
    \tmUnit                                                     \mid
    \mlHandler                                                  \mid
    \fun{x:\mlTyA}{\mlTm}                                       \mid
    \Lambda \tyvar. \mlTm                                       \mid
    \Lambda (\covar : \mlCoTy). \mlTm                           \mid
    \cast{\mlValue}{(\mlCoercion_1 \to  \mlCoercion_2)}         \mid
    \cast{\mlValue}{(\mlCoercion_1 \hto \mlCoercion_2)}         \\
  & \mid &
    \cast{\mlValue}{(\handToFun{\mlCoercion_1}{\mlCoercion_2})} \mid
    \cast{\mlValue}{(\funToHand{\mlCoercion_1}{\mlCoercion_2})} \mid
    \cast{\mlValue}{\forall \tyvar. \mlCoercion}                \mid
    \cast{\mlValue}{(\mlCoTy \Rightarrow \mlCoercion)}          \\ 
    & \mid &
    \mlReturn{\mlValue}                                         \mid
    \operation{\mlValue}{y : \mlTyA}{\mlTm}                     \\
\end{array}
\]

\noindent
$\ruleform{\mlSmallStep{\mlTm}{\mlTm'}}$ \textbf{Operational Semantics}
\begin{mathpar}

\inferrule*[right=]
           {}
           { \mlSmallStep{\cast{(\mlReturn{\mlValue})}{(\mkMlCompCo{\mlCoercion}})}{\mlReturn{(\cast{\mlValue}{\mlCoercion})}} }

\\

\inferrule*[right=]
           {}
           { \mlSmallStep{\doin{x}{\mlReturn{\mlValue}}{\mlTm}}{\mlTm[\mlValue/x]} }

%
%
%

\inferrule*[right=]
           {}
           { \mlSmallStep{\withhandle{\mlHandler}{(\mlReturn{\mlValue})}}{\mlTm_r[\mlValue/x]} }

\inferrule*[right=]
           {}
           { \mlSmallStep{(\cast{\mlValue_1}{(\handToFun{\mlCoercion_1}{\mlCoercion_2})})~\mlValue_2}
                         {\cast{(\withhandle{\mlValue_1}{(\mlReturn{(\cast{\mlValue_2}{\mlCoercion_1})})})}{\mlCoercion_2}}
           } 

\inferrule*[right=]
                      {}
                      { \mlSmallStep{\withhandle{(\cast{\mlValue_2}{(\funToHand{\mlCoercion_1}{\mlCoercion_2})})}{(\return{\mlValue_1})} }
                                    {\cast{(\mlValue_2~(\cast{\mlValue_1}{\mlCoercion_1}))}{\mlCoercion_2}}
                      } 

\inferrule*[right=]           
           {}
           { {\withhandle{(\cast{\mlValue_2}{(\funToHand{\mlCoercion_1}{\mlCoercion_2})})}{(\operation{\mlValue_1}{y : \mlTyB}{\mlTm})} }
             \\ \leadsto
             {\operation{\mlValue_1}{y : \mlTyB}{\withhandle{(\cast{\mlValue_2}{(\funToHand{\mlCoercion_1}{\mlCoercion_2})})}{\mlTm}}}
           }  

\inferrule*[right=]
           {}
           { \mlSmallStep{\cast{\mlValue}{\mlReturn{\mlCoercion}}}{\mlReturn{(\cast{\mlValue}{\mlCoercion})}} }

\inferrule*[right=]
           {}
           { \mlSmallStep{\cast{(\mlReturn{\mlValue})}{(\unsafe{\mlCoercion})}}{\cast{\mlValue}{\mlCoercion}} }


\end{mathpar}

\vspace{-5mm}
\caption{\noeff Operational Semantics (Selected Rules)}
\label{fig:ml-opsem-selected}
\end{myfigure}

Figure~\ref{fig:ml-opsem-selected} presents selected rules of \noeff's
small-step, call-by-value operational semantics. We omit
other rules as they closely follow the rules for \target, except being adjusted for the amalgamation of values and computations. The complete operational semantics can be found in Appendix~\ref{appendix:ml-additional}.

The first rule pushes the cast onto the returned value; in contrast to \target, there is no effect information to lose, making this reduction type-preserving. This allows the second and third rule which are simplified variants of the ones for \noeff: because all the coercions can be pushed into $\mlValue$, there is no need to extract their pure parts before substituting $\mlValue$ for a variable. The remaining five rules capture the semantics of the newly introduced coercion forms, exactly as described in Section~\ref{sec:ml-coercion-typing}.

%

\paragraph{The \noeff Metatheory}

We have proven a weak form of type safety for \noeff in terms
of type preservation and (partial) progress theorems. The latter
characterises the way in which well-typed terms can get stuck.

\begin{theorem}[Preservation]
If $\tcNoEffTm{\mlEnv}{\mlTm}{\mlTyA}$ and
$\mlSmallStep{\mlTm}{\mlTm'}$, then $\tcNoEffTm{\mlEnv}{\mlTm'}{\mlTyA}$.
\label{thm:ml-preservation}
\end{theorem}

\begin{theorem}[Partial Progress]
\label{thm:ml-progress}
If $\tcNoEffTm{\mlEnv}{\mlTm}{\mlTyA}$ then either
\begin{inparaenum}[(a)]
\item
  $\mlTm$ is a value,
\item
  $\mlSmallStep{\mlTm}{\mlTm'}$, or
\item
  $\mlTm$ is {\em``stuck''}.
\end{inparaenum}
\end{theorem}

Stuck terms are defined as follows:
\begin{center}\begin{shaded}\begin{minipage}{\columnwidth}
\vspace{-4mm}
\small
\[
\begin{array}{r@{~}c@{~}l}
  \mlStuck & ::= & \cast{\operation{\mlValue}{y : \mlTyA}{\mlTm}}{\unsafe{\mlCoercion}} \mid
                   \mlStuck~\mlTyA                                                      \mid
                   \cast{\mlStuck}{\mlCoercion}                                         \mid
                   \mlStuck~\mlCoercion                                                 \mid
                   \mlStuck~\mlTm                                                       \mid
                   \mlValue~\mlStuck                                                    \mid
                   \letval{x}{\mlStuck}{\mlTm}                                          \mid
                   \return{\mlStuck}                                                    \\
   & \mid &        \operation{\mlStuck}{y : \mlTyA}{\mlTm}                              \mid
                   \doin{x}{\mlStuck}{\mlTm}                                            \mid
                   \withhandle{\mlStuck}{\mlTm_c}                                       \mid
                   \withhandle{\mlValue}{\mlStuck}                                      \\
\end{array}
\]
\vspace{-4mm}
\end{minipage}\end{shaded}\end{center}
The first case is the essential one, while the remaining ones just provide an
evaluation context around it. Hence, terms only get stuck when an unsafe
coercion is applied to an operation. As we have already indicated, we claim
that elaborated \noeff programs never end up in this situation.

\subsection{Elaboration of \target to \noeff}\label{sec:eff-to-ml}

\subsubsection{Type Elaboration}
\label{sec:eff-to-ml-types}

Figure~\ref{fig:eff-to-ml-types} presents the elaboration of value types
($\valTyToNoEff{\tyEnv}{\vty}{\ety}{\mlTyA}$) and computation types
($\compTyToNoEff{\tyEnv}{\cty}{\ety}{\mlTyA}$). The latter captures the main
idea of the whole elaboration: when the dirt $\dirt$ of a computation type is
empty, the elaboration of the computation type $\vty \bang \dirt$ is just the
elaboration of the value type $\vty$. If it is non-empty, the elaborated
value type $\mlTyA$ is wrapped in a computation type, $\mkMlCompTy{\mlTyA}$.
We cannot always tell whether $\dirt$ is empty or not, namely in case it is
a dirt variable $\dirtvar$. Our conservative solution is to assume that
dirt variables are also non-empty. This works because we can always represent
a term $\mlTm$ of type $\mlTyA$ in terms of a trivial computation $\ret{t}$ of
type $\mkMlCompTy{\mlTyA}$.

Most cases for value types are straightforward, but a few are worth mentioning.
Firstly, to respect the particularities of \noeff handler types explained in
Section~\ref{sec:ml-term-typing}, we distinguish two different cases for elaborating \target.
Recall that if a computation type has an empty dirt, it is elaborated to some pure type~$\mlTyA$, not a computation type~$\mkMlCompTy{\mlTyA}$ that handlers expect. Correspondingly, handler types with empty input dirts are elaborated into
\noeff function types. If the dirt is non-empty, we unavoidably elaborate to a \noeff handler type.
Note that in the latter case, we ignore whether or not the output computation type has
an empty dirt; the \noeff handler type always implicitly assumes an output computation type.

Secondly, since dirts and skeletons are absent from \noeff, the elaboration drops
universal quantification over skeletons and dirts, as well as dirt subtyping qualifiers.



\begin{myfigure}[t]
$\ruleform{\fullDirt{\dirt}}$ \textbf{Conservative Non-Empty Dirt}
\begin{mathpar}
\inferrule*[right=]
           { }
           { \fullDirt{\dirtvar} }

\inferrule*[right=]
           { }
           { \fullDirt{\{ \op \} \cup \dirt} }
\end{mathpar}
$\ruleform{\valTyToNoEff{\tyEnv}{\vty}{\ety}{\mlTyA}}$ \textbf{Value Type Elaboration}
\begin{mathpar}
\inferrule*[right=]
           { (\tyvar : \ety) \in \tyEnv }
           { \valTyToNoEff{\tyEnv}{\tyvar}{\ety}{\tyvar} }

\inferrule*[right=]
           { }
           { \valTyToNoEff{\tyEnv}{\tyUnit}{\tyUnit}{\tyUnit} }

\inferrule*[right=]
           { \valTyToNoEff{\tyEnv}{\vty}{\ety_1}{\mlTyA} \\
             \compTyToNoEff{\tyEnv}{\cty}{\ety_2}{\mlTyB}
           } 
           { \valTyToNoEff{\tyEnv}{\vty \to \cty}{\ety_1 \to \ety_2}{\mlTyA \to \mlTyB} }

\inferrule*[right=]
           { \valTyToNoEff {\tyEnv}{\vty}{\ety_1}{\mlTyA} \\
             \compTyToNoEff{\tyEnv}{\cty}{\ety_2}{\mlTyB}
           } 
           { \valTyToNoEff{\tyEnv}{\vty~!~\emptyset \hto \cty}{\ety_1 \hto \ety_2}{\mlTyA \to \mlTyB} }

\inferrule*[right=]
           { \valTyToNoEff{\tyEnv}{\vty_1}{\ety_1}{\mlTyA} \\
             \valTyToNoEff{\tyEnv}{\vty_2}{\ety_2}{\mlTyB} \\
             \fullDirt{\dirt_1}
           } 
           { \valTyToNoEff{\tyEnv}{(\vty_1~!~\dirt_1 \hto \vty_2~!~\dirt_2)}{\ety_1 \hto \ety_2}{\mlTyA \hto \mlTyB} }

\inferrule*[right=]
           { \valTyToNoEff{\tyEnv, \evar}{\vty}{\ety}{\mlTyA} }
           { \valTyToNoEff{\tyEnv}{\forall \evar. \vty}{\forall \evar. \ety}{\mlTyA} }

\inferrule*[right=]
           { \valTyToNoEff{\tyEnv, \tyvar : \ety_1}{\vty}{\ety_2}{\mlTyA} }
           { \valTyToNoEff{\tyEnv}{\forall (\tyvar : \ety_1). \vty}{\ety_2}{\forall \tyvar. \mlTyA} }

\inferrule*[right=]
           { \valTyToNoEff{\tyEnv, \dirtvar}{\vty}{\ety}{\mlTyA} }
           { \valTyToNoEff{\tyEnv}{\forall \dirtvar. \vty}{\ety}{\mlTyA} }

\inferrule*[right=]
           { \valTyToNoEff{\tyEnv}{\vty}{\ety}{\mlTyA} }
           { \valTyToNoEff{\tyEnv}{(\dirt_1 \le \dirt_2) \Rightarrow \vty}{\ety}{\mlTyA} }

\inferrule*[right=]
           { \valTyToNoEff{\tyEnv}{\vty_1}{\ety_1}{\mlTyB_1} \\
             \valTyToNoEff{\tyEnv}{\vty_2}{\ety_1}{\mlTyB_2} \\
             \valTyToNoEff{\tyEnv}{\vty}{\ety}{\mlTyA}
           } 
           { \valTyToNoEff{\tyEnv}{(\vty_1 \le \vty_2) \Rightarrow \vty}{\ety}{(\mlTyB_1 \le \mlTyB_2) \Rightarrow \mlTyA} }

\inferrule*[right=]
           { \compTyToNoEff{\tyEnv}{\cty_1}{\ety_1}{\mlTyB_1} \\
             \compTyToNoEff{\tyEnv}{\cty_2}{\ety_1}{\mlTyB_2} \\
             \valTyToNoEff{\tyEnv}{\vty}{\ety}{\mlTyA}
           } 
           { \valTyToNoEff{\tyEnv}{(\cty_1 \le \cty_2) \Rightarrow \vty}{\ety}{(\mlTyB_1 \le \mlTyB_2) \Rightarrow \mlTyA} }
\end{mathpar}

\noindent
$\ruleform{\compTyToNoEff{\tyEnv}{\cty}{\ety}{\mlTyA}}$ \textbf{Computation Type Elaboration}
\begin{mathpar}
\inferrule*[right=]
           { \valTyToNoEff{\tyEnv}{\vty}{\ety}{\mlTyA} }
           { \compTyToNoEff{\tyEnv}{\vty~!~\emptyset}{\ety}{\mlTyA} }

\inferrule*[right=]
           { \fullDirt{\dirt} \\
             \valTyToNoEff{\tyEnv}{\vty}{\ety}{\mlTyA}
           } 
           { \compTyToNoEff{\tyEnv}{\vty~!~\dirt}{\ety}{\mkMlCompTy{\mlTyA}} }
\end{mathpar}

\vspace{-5mm}
\caption{Elaboration of \target Types to \noeff Types}
\label{fig:eff-to-ml-types}
\end{myfigure}


\paragraph{Coercion Elaboration}

We now turn to the elaboration of \target coercions to \noeff coercions.
Most cases are straightforward and either copy a \target coercion to its \noeff
counterpart, or drop a dirt- or skeleton-related \target construct that is not
present in \noeff. Hence, we only discuss the interesting cases here; the
complete definition can be found in Appendix~\ref{appendix:elab-additional}.

Two groups of rules do deserve additional explanation. The first group concerns
the elaboration of handler coercions. If we compare the input dirts of the
source and target handler types of the coercion, there are three different
cases: either both are empty, both are non-empty, or the source input dirt is
non-empty and the target input dirt is empty. The fourth
combination is not possible due to the monotonicity
of subtyping and the contravariance in the input argument.

In the first case, both the source and the target \target type elaborate to
\noeff function types, and thus the coercion is elaborated to a function
coercion:
\begin{mathpar}
\small
  \inferrule*[right=]
             { \coToNoEff{\tyEnv}{\coercion_1}{\vty_2~!~\emptyset \le \vty_1~!~\emptyset}{\mlCoercion'_1} \\
               \coToNoEff{\tyEnv}{\coercion_2}{\cty_1 \le \cty_2}{\mlCoercion'_2}
             } 
             { \coToNoEff{\tyEnv}{\coercion_1 \hto \coercion_2}{(\vty_1~!~\emptyset \hto \cty_1) \le (\vty_2~!~\emptyset \hto \cty_2)}{\mlCoercion'_1 \to \mlCoercion'_2} }
\end{mathpar}
In the second case, both types elaborate to \noeff handler types, and thus the
whole coercion is elaborated to a \noeff handler coercion:
\begin{mathpar}
\small
  \inferrule*[right=]
             { \fullDirt{\dirt_1} \\
               \fullDirt{\dirt_2} \\\\
               \coToNoEff{\tyEnv}{\coercion_1}{(\vty_2~!~\dirt_2 \le \vty_1~!~\dirt_1)}{\mlCoercion'_1} \\
               \coToNoEff{\tyEnv}{\coercion_2}{\vty'_1 \le \vty'_2}{\mlCoercion'_2} \\
               \tcTrgCo{\tyEnv}{\coercion_3}{\dirt'_1 \le \dirt'_2}
             } 
             { \coToNoEff{\tyEnv}
                         {(\coercion_1 \hto (\coercion_2~!~\coercion_3))}
                         {((\vty_1~!~\dirt_1) \hto (\vty'_1~!~\dirt'_1)) \le ((\vty_2~!~\dirt_2) \hto (\vty'_2~!~\dirt'_2))}
                         {\mlCoercion'_1 \hto \mkMlCompCo{\mlCoercion'_2}}
             } 
\end{mathpar}
In the third case the elaborated source type is a handler type and the target
type a function type. Here we use the $\keyhandToFun$ coercion to bridge
between the two. There are two subcases to consider though, depending on
whether the source output dirt is empty or not:
\begin{mathpar}
\small
  \inferrule*[right=]
             { \fullDirt{\dirt_1} \\
             	\coToNoEff{\tyEnv}{\coercion_1}{\vty_2 \le \vty_1}{\mlCoercion'_1} \\
               \coToNoEff{\tyEnv}{\coercion_2}{(\vty'_1 \le \vty'_2)}{\mlCoercion'_2} \\
               \tcTrgCo{\tyEnv}{\coercion_3}{\emptyset \le \dirt_1} \\
               \tcTrgCo{\tyEnv}{\coercion_4}{\emptyset \le \dirt'_2} 
             } 
             { \coToNoEff{\tyEnv}
                         {(\coercion_1~!~\coercion_3 \hto \coercion_2~!~\coercion_4)}
                         {((\vty_1~!~\dirt_1 \hto \vty'_1~!~\emptyset) \le (\vty_2~!~\emptyset \hto \vty'_2~!~\dirt'_2))}
                         {\handToFun{\mlCoercion'_1}{(\unsafe{\mlCoercion'_2})}}
             } 

  \inferrule*[right=]
             { \fullDirt{\dirt_1} \\
               \fullDirt{\dirt'_1} \\
               \coToNoEff{\tyEnv}{\coercion_1}{\vty_2 \le \vty_1}{\mlCoercion'_1} \\
               \coToNoEff{\tyEnv}{\coercion_2}{(\vty'_1 \le \vty'_2)}{\mlCoercion'_2} \\
               \tcTrgCo{\tyEnv}{\coercion_3}{\emptyset \le \dirt_1} \\
               \tcTrgCo{\tyEnv}{\coercion_4}{\dirt'_1 \le \dirt'_2}
             } 
             { \coToNoEff{\tyEnv}
                         {(\coercion_1~!~\coercion_3 \hto \coercion_2~!~\coercion_4)}
                         {((\vty_1~!~\dirt_1 \hto \vty'_1~!~\dirt'_1) \le (\vty_2~!~\emptyset \hto \vty'_2~!~\dirt'_2))}
                         {\handToFun{\mlCoercion'_1}{\mlCoercion'_2}}
             } 
\end{mathpar}
In the former case, \noeff does not respect the emptiness in the elaborated
handler type, but does respect it in the elaboration of $\coercion_2$.  To
bridge the discrepancy that arises here, we insert an $\keyunsafe$ coercion. In
the latter case, no discrepancy arises, and no $\keyunsafe$ coercion is needed.

The second group of interest concerns the elaboration of computation type
coercions. Again we distinguish three different cases based on the source and
target dirt. If both are empty, the computation type coercion is elaborated
like the underlying value type coercion $\coercion_1$:
\begin{mathpar}
\small
  \inferrule*[right=]
             { \coToNoEff{\tyEnv}{\coercion_1}{\vty_1 \le \vty_2}{\mlCoercion'_1} \\
               \tcTrgCo{\tyEnv}{\coercion_2}{\emptyset \le \emptyset}
             } 
             { \coToNoEff{\tyEnv}{(\coercion_1~!~\coercion_2)}{(\vty_1~!~\emptyset \le \vty_2~!~\emptyset)}{\mlCoercion'_1} }
\end{mathpar}
If both are non-empty, we elaborate to a \noeff computation type coercion
$\mkMlCompCo{\coercion_1'}$:
\begin{mathpar}
\small
  \inferrule*[right=]
             { \coToNoEff{\tyEnv}{\coercion_1}{\vty_1 \le \vty_2}{\mlCoercion'_1} \\
               \tcTrgCo{\tyEnv}{\coercion_2}{\emptyset \le \dirt_2} \\
               \fullDirt{\dirt_2}
             } 
             { \coToNoEff{\tyEnv}{(\coercion_1~!~\coercion_2)}{(\vty_1~!~\emptyset \le \vty_2~!~\dirt_2)}{\mlReturn{\mlCoercion'_1}} }
\end{mathpar}
In the third case, there is a mismatch because the source is pure and the
target is impure; we bridge this with a $\keyreturn$ coercion:
\begin{mathpar}
\small
  \inferrule*[right=]
             { \coToNoEff{\tyEnv}{\coercion_1}{\vty_1 \le \vty_2}{\mlCoercion'_1} \\
               \tcTrgCo{\tyEnv}{\coercion_2}{\dirt_1 \le \dirt_2} \\
               \fullDirt{\dirt_1} \\
               \fullDirt{\dirt_2}
             } 
             { \coToNoEff{\tyEnv}{(\coercion_1~!~\coercion_2)}{(\vty_1~!~\dirt_1 \le \vty_2~!~\dirt_2)}{\mkMlCompCo{\mlCoercion'_1}} }
\end{mathpar}

\subsubsection{Value Elaboration}\label{sec:eff-to-ml-values}

Again, the elaboration of \target values into \noeff terms is mostly
straightforward, so we only discuss the interesting cases here; the complete
definition can be found in Appendix~\ref{appendix:elab-additional}. There are
two cases of interest: handlers and dirt applications.

\paragraph{\bf Handlers}
We have three rules describing different cases of elaborating handlers of type
$\withdirt{\vty_x}{\ops} \hto \withdirt{\vty}{\dirt}$.  Recall from
Section~\ref{sec:eff-to-ml-types} that if $\ops = \emptyset$, handlers need to
be elaborated into functions, which is described by the first of these three
rules:
\begin{mathpar}
\small
  \inferrule*[right=]
             { \valTyToNoEff{\tyEnv}{\vty}{\ety}{\mlTyA} \\
               \compToNoEff{\tyEnv,x\!:\!\vty}{c_r}{\cty}{\mlTm}
             } 
             { \valToNoEff{\tyEnv}{\{\return{(x : \vty)} \mapsto c_r\}}{\withdirt{\vty}{\emptyset} \hto \cty}{\fun{(x:\mlTyA)}{\mlTm}} }
\end{mathpar}
The second rule describes the case where $\ops$ is non-empty, but $\dirt$ is
empty:
\begin{mathpar}
\small
  \inferrule*[right=]
             { \fullDirt{\ops} \\
               \valTyToNoEff{\tyEnv}{\vty_x}{\ety}{\mlTyA} \\
               \compToNoEff{\tyEnv, x\!:\!\vty_x}{c_r}{\withdirt{\vty}{\emptyset}}{\mlTm_r} \\
               \left[
                 (\op : \vty_1^\op \to \vty_2^\op) \in \sig \quad
                 \valTyToNoEff{}{\vty_i^\op}{\ety_i^\op}{\mlTyA_i^\op} \qquad
                 \compToNoEff{\tyEnv, x : \vty_1^\op, k : \vty_2^\op \to \vty\,!\,\emptyset}{c_\op}{\vty\,!\,\emptyset}{\mlTm_\op}
               \right]_{\op \in \ops}
             } 
             { \tyEnv \vdashNamedD{v} \trgShorthand : \withdirt{\vty_x}{\ops} \hto \withdirt{\vty}{\emptyset} \\
                    \highlight{\rightsquigarrow 
                          {\handler{\return{(x : \mlTyA)} \mapsto \return{\mlTm_r}
                          ,\big[\call{\op}{x}{k} \mapsto \return{\mlTm_\op[\cast{k}{\refl{\mlTyA_1^\op} \to \unsafe{\refl{\mlTyA_2^\op}}}/k]}\big]_{\op \in \ops}}}
                  }
           }
\end{mathpar}
Since $\ops$ is non-empty, we do elaborate a handler into a handler, but there is an
important caveat. Recall from~\ref{sec:ml-term-typing} that to ensure safe forwarding of
unhandled operations, handlers take computations to computations.
But as $\dirt$ is empty, handler clauses of type~$\withdirt{\vty}{\emptyset}$ are
elaborated to terms of type~$A$ (the elaboration of $\vty$), not $\mkMlCompTy{A}$ as expected.
We amend this by wrapping them with a $\keyreturn$. However, the handled continuations
now include an extraneous $\keyreturn$, which we remove with an $\keyunsafe$ coercion
before plugging them into the operation clause that expects $k$ to result in~$A$, not $\mkMlCompTy{A}$.

In the third rule, both $\ops$ and $\dirt$ are non-empty, and the elaboration
is structural:
\begin{mathpar}
\small
  \inferrule*[right=]
             { \fullDirt{\ops} \\
               \fullDirt{\dirt} \\
               \valTyToNoEff{\tyEnv}{\vty_x}{\ety}{\mlTyA} \\
               \compToNoEff{\tyEnv, x\!:\!\vty_x}{c_r}{\withdirt{\vty}{\dirt}}{\mlTm_r} \\
               \left[
                 (\op : \vty_1^\op \to \vty_2^\op) \in \sig \qquad
                 \compToNoEff{\tyEnv, x : \vty_1^\op, k : \vty_2^\op \to \vty\,!\,\dirt}{c_\op}{\vty\,!\,\dirt}{\mlTm_\op}
               \right]_{\op \in \ops}
             } 
             { \tyEnv \vdashNamedD{v} \trgShorthand : \withdirt{\vty_x}{\ops} \hto \withdirt{\vty}{\dirt} \\
                    \highlight{\rightsquigarrow 
                          {\handler{\return{(x : \mlTyA)} \mapsto {\mlTm_r}
                          ,[\call{\op}{x}{k} \mapsto {\mlTm_\op}]_{\op \in \ops}}}
                    }
             }
\end{mathpar}

\begin{myfigure}[t!]
\begin{center}\begin{shaded}\begin{minipage}{\columnwidth}
\vspace{-3mm}
$\ruleform{\fromImpureVal {\vty}{\dirt}{\mlCoercion}}$~\textbf{Value Type Coercion from Impure Dirt Instantiation}
\begin{mathpar}
\inferrule*[right=FiUnit]
           { }
           { \fromImpureVal{\tyUnit}{\dirt}{\mlCoUnitRefl} }

\inferrule*[right=FiArr]
           { \toImpureVal{\vty}{\dirt}{\mlCoercion_1} \\
             \fromImpureComp{\cty}{\dirt}{\mlCoercion_2}
           } 
           { \fromImpureVal{\vty \to \cty}{\dirt}{\mlCoercion_1 \to \mlCoercion_2}  }

\inferrule*[right=FiHand1]
           { \toImpureVal{\vty}{\dirt}{\mlCoercion_1} \\
             \fromImpureComp{\cty}{\dirt}{\mlCoercion_2}
           } 
           { \fromImpureVal{\withdirt{\vty}{\emptyset} \hto \cty}{\dirt}{\mlCoercion_1 \to \mlCoercion_2}  }

\inferrule*[right=FiHand2]
           { \dirt_2[\emptyset/\dirtvar] = \emptyset \\
             \toImpureVal{\vty_1}{\emptyset}{\mlCoercion_1} \\
             \fromImpureVal{\vty_2}{\emptyset}{\mlCoercion_2}
           } 
           { \fromImpureVal{\withdirt{\vty_1}{\dirtvar} \hto \withdirt{\vty_2}{\dirt_2}}{\emptyset}{\handToFun{\mlCoercion_1}{(\unsafe{\mlCoercion_2})}}  }

\inferrule*[right=FiHand3]
           { \fullDirt{\dirt_2[\emptyset/\dirtvar]} \\
             \toImpureVal{\vty_1}{\emptyset}{\mlCoercion_1} \\
             \fromImpureVal{\vty_2}{\emptyset}{\mlCoercion_2}
           } 
           { \fromImpureVal{\withdirt{\vty_1}{\dirtvar} \hto \withdirt{\vty_2}{\dirt_2}}{\emptyset}{\handToFun{\mlCoercion_1}{(\mkMlCompCo{\mlCoercion_2})}}  }

\inferrule*[right=FiHand4]
           { \fullDirt{\dirt_1[\dirt/\dirtvar]} \\
             \toImpureComp{\withdirt{\vty_1}{\dirt_1}}{\dirt}{\mlCoercion_1} \\
             \fromImpureComp[\ctx, \dirtvar']{\withdirt{\vty_2}{\dirtvar'}}{\dirt}{\mlCoercion_2} \quad \text{fresh}\,\dirtvar'
           } 
           { \fromImpureVal{\withdirt{\vty_1}{\dirt_1} \hto \withdirt{\vty_2}{\dirt_2}}{\dirt}{\mlCoercion_1 \hto \mlCoercion_2}  }

\inferrule*[right=FiSkelAbs]
           { \fromImpureVal[\ctx, \evar]{\vty}{\dirt}{\mlCoercion}
           } 
           { \fromImpureVal{\forall \evar. \vty}{\dirt}{\mlCoercion}  }

\inferrule*[right=FiTyAbs]
           { \fromImpureVal[\ctx, \tyvar\!:\!\ety]{\vty}{\dirt}{\mlCoercion}
           } 
           { \fromImpureVal{\forall \tyvar\!:\!\ety. \vty}{\dirt}{\forall \tyvar. \mlCoercion}  }

\inferrule*[right=FiDirtAbs]
           { \fromImpureVal[\ctx, \dirtvar']{\vty}{\dirt}{\mlCoercion}
           } 
           { \fromImpureVal{\forall \dirtvar'.\vty}{\dirt}{\mlCoercion}  }

\inferrule*[right=FiCoAbsTy]
           { \fromImpureVal{\vty}{\dirt}{\mlCoercion}
           } 
           { \fromImpureVal{\tyvar_1 \le \tyvar_2 \Rightarrow \vty}{\dirt}{\tyvar_1 \le \tyvar_2 \Rightarrow \mlCoercion}  }


\inferrule*[right=FiCoAbsDirt]
           { \fromImpureVal{\vty}{\dirt}{\mlCoercion}
           } 
           { \fromImpureVal{\dirt_1 \le \dirt_2 \Rightarrow \vty}{\dirt}{\mlCoercion}  }


\end{mathpar}
$\ruleform{\fromImpureComp {\cty}{\dirt}{\mlCoercion}}$~\textbf{Computation Type Coercion from Impure Dirt Instantiation}
\begin{mathpar}
\inferrule*[right=FiCmp1]
           { \fromImpureVal{\vty}{\dirt}{\mlCoercion} }
           { \fromImpureComp{\vty~!~\emptyset}{\dirt}{\mlCoercion} }

\inferrule*[right=FiCmp2]
           { \fromImpureVal{\vty}{\emptyset}{\mlCoercion} }
           { \fromImpureComp{\vty~!~\dirtvar}{\emptyset}{\unsafe{\mlCoercion}} }

\inferrule*[right=FiCmp3]
           { \fullDirt{\dirt'[\dirt/\dirtvar]} \\
             \fromImpureVal{\vty}{\dirt}{\mlCoercion}
           } 
           { \fromImpureComp{\vty~!~\dirt'}{\dirt}{\mkMlCompCo{\mlCoercion}} }


\end{mathpar}
$\ruleform{\toImpureVal {\vty}{\dirt}{\mlCoercion}}$~\textbf{Value Type Coercion to Impure Dirt Instantiation}
\[
  \text{defined dually to $\fromImpureVal {\vty}{\dirt}{\mlCoercion}$}
\]
$\ruleform{\toImpureComp {\cty}{\dirt}{\mlCoercion}}$~\textbf{Computation Type Coercion to Impure Dirt Instantiation}
\[
  \text{defined dually to $\fromImpureComp{\vty}{\dirt}{\mlCoercion}$}
\]
\vspace{-5mm}
\end{minipage}\end{shaded}\end{center}
\vspace{-5mm}
\caption{Type Coercions from and to an Impure Dirt Instantiation}
\label{fig:eff-to-ml-impure}
\end{myfigure}

\paragraph{\bf Dirt applications}

The elaboration of dirt applications possibly needs to bridge between an impure
and a pure type. Consider for instance a \target value $v$ of type $\forall
\dirtvar.  \tyUnit \to \withdirt{\tyUnit}{\dirtvar}$ which is applied to the
empty dirt; thus the type of the dirt application is $\tyUnit \to
\withdirt{\tyUnit}{\emptyset}$. The elaboration of the former type is $\tyUnit
\to \mkMlCompTy{\tyUnit}$, while the latter is $\tyUnit \to \tyUnit$.

Such elaborations are handled by the following rule:
\begin{mathpar}
\small
  \inferrule*[right=]
             { \valToNoEff{\tyEnv}{v}{\forall \dirtvar. \vty}{\mlTm} \\
               \fromImpureVal{\vty}{\dirt}{\mlCoercion}
             } 
             { \valToNoEff{\tyEnv}{v~\dirt}{\vty[\dirt/\dirtvar]}{\cast{\mlTm}{\mlCoercion}} }
\end{mathpar}
where for a given $v$ of type $\forall \dirtvar. \vty$, we need a
coercion~$\mlCoercion$ from the elaboration of $\vty$ (recall this is done
under the assumption $\fullDirt{\dirtvar}$) to the elaboration of $\vty[\dirt /
\dirtvar]$.  Such coercion is produced by a judgement
$\fromImpureVal{\vty}{\dirt}{\mlCoercion}$, driven by the structure of $\vty$.
This judgement is defined in Figure~\ref{fig:eff-to-ml-impure} alongside with
the judgment $\fromImpureComp{\cty}{\dirt}{\mlCoercion}$ for computation types.
In addition, there are two dual judgements
$\toImpureVal{\vty}{\dirt}{\mlCoercion}$ and
$\toImpureComp{\cty}{\dirt}{\mlCoercion}$ for the opposite coercions, which are
used on types in contravariant positions. We have omitted their definitions
because they are obtained by flipping the sides of all
$\mapsto$ arrows, and replacing $\keyunsafe$ with $\keyreturn$ and $\keyhandToFun$
with $\keyfunToHand$.
Most rules of these judgements are straightforward congruences.

The main rule of interest is the one that produces an $\keyunsafe$ coercion
where the dirt variable $\delta$ in a computation type
$\withdirt{\vty}{\dirtvar}$ is instantiated to the empty dirt $\emptyset$
(Rule~\textsc{FiCmp2}).  In that case, the elaboration of the polymorphic
abstraction conservatively assumes the computation is impure, while the
elaboration of its instantiation accurately knows it is pure.

A further case that deserves attention is that of the handler type, where four
different rules (Rules~\textsc{FiHand1},~\textsc{FiHand2},~\textsc{FiHand3},
and~\textsc{FiHand4}) cover the possible scenarios related to elaboration into
handler and function types.

Note that in Rule~\textsc{FiCoAbsTy} we have restricted the case of $\vty_1
\le \vty_2 \Rightarrow \vty$ to situations where $\vty_1$ and $\vty_2$ are both
types variables. This is not a severe restriction because subtyping constraints
can be simplified to this form; this simplification is precisely what our type
inference algorithm does. Moreover, there is a good reason to impose the
syntactic restriction. Consider the trivial reflexive subtyping constraint
$(\tyUnit \to \withdirt{\tyUnit}{\dirtvar}) \le (\tyUnit \to
\withdirt{\tyUnit}{\dirtvar})$. If we conservatively assume that $\dirtvar$ is
non-empty, the constraint is elaborated to $(\tyUnit \to \mkMlCompTy{\tyUnit})
\le (\tyUnit \to \mkMlCompTy{\tyUnit})$, whereas, if $\dirtvar$ is instantiated
to $\emptyset$, the constraint is elaborated to $(\tyUnit \to \tyUnit) \le
(\tyUnit \to \tyUnit)$. Hence, we would need to be able to coerce a coercion
for the former constraint to a coercion for the latter, and vice versa.  This
would require a complication of the \noeff language with additional coercion
forms to accomplish this coercion of coercions, which, happily, the above
syntactic restriction allows us to avoid.

\subsubsection{Computation Elaboration}\label{sec:eff-to-ml-computations}

\begin{myfigure}[t]
$\ruleform{\compToNoEff{\tyEnv}{c}{\cty}{\mlTm}}$ \textbf{Computation Elaboration}
\begin{mathpar}
\inferrule*[right=CApp]
           { \valToNoEff{\tyEnv}{v_1}{\vty \to \cty}{\mlTm_1} \\
             \valToNoEff{\tyEnv}{v_2}{\vty}{\mlTm_2}
           } 
           { \compToNoEff{\tyEnv}{v_1~v_2}{\cty}{\mlTm_1~\mlTm_2} }

\inferrule*[right=CLet]
           { \valToNoEff{\tyEnv}{v}{\vty}{\mlTm_1} \\
             \compToNoEff{\tyEnv, x : \vty}{c}{\cty}{\mlTm_2}
           } 
           { \compToNoEff{\tyEnv}{\letval{x}{v}{c}}{\cty}{\letval{x}{\mlTm_1}{\mlTm_2}} }

\inferrule*[right=CRet]
           { \valToNoEff{\tyEnv}{v}{\vty}{\mlTm} }
           { \compToNoEff{\tyEnv}{\return{v}}{\vty~!~\emptyset}{\mlTm} }

\inferrule*[right=COp]
           { (\op : \vty_1 \to \vty_2) \in \mlSig \\
             \valTyToNoEff{\tyEnv}{\vty_1}{\ety_1}{\mlTyA} \\
             \valTyToNoEff{\tyEnv}{\vty_2}{\ety_2}{\mlTyB} \\
             \valToNoEff{\tyEnv}{v}{\vty_1}{\mlTm_v} \\
             \compToNoEff{\tyEnv, x : \vty_2}{c}{\vty~!~\dirt}{\mlTm_c} \\
             \op \in \dirt
           } 
           { \compToNoEff{\tyEnv}{\operation{v}{y : \vty_2}{c}}{\vty~!~\dirt}{\operation{\mlTm_v}{y : \mlTyB}{\mlTm_c}} }

\inferrule*[right=CDo1]
           { \compToNoEff{\tyEnv}{c_1}{\vty_1~!~\emptyset}{\mlTm_1} \\
             \compToNoEff{\tyEnv, x : \vty_1}{c_2}{\vty_2~!~\emptyset}{\mlTm_2}
           } 
           { \compToNoEff{\tyEnv}{(\doin{x}{c_1}{c_2})}{\vty_2~!~\emptyset}{\letval{x}{\mlTm_1}{\mlTm_2}} }

\inferrule*[right=CDo2]
           { \fullDirt{\dirt} \\
             \compToNoEff{\tyEnv}{c_1}{(\vty_1~!~\dirt)}{\mlTm_1} \\
             \compToNoEff{\tyEnv, x : \vty_1}{c_2}{(\vty_2~!~\dirt)}{\mlTm_2}
           } 
           { \compToNoEff{\tyEnv}{\doin{x}{c_1}{c_2}}{(\vty_2~!~\dirt)}{\doin{x}{\mlTm_1}{\mlTm_2}} }

\inferrule*[right=CHandle1]
           { \compToNoEff{\tyEnv}{c}{\vty~!~\emptyset}{\mlTm_2} \\
             \valToNoEff{\tyEnv}{v}{(\vty~!~\emptyset \hto \cty)}{\mlTm_1}
           } 
           { \compToNoEff{\tyEnv}{(\withhandle{v}{c})}{\cty}{\mlTm_1~\mlTm_2} }

\inferrule*[right=CHandle2]
           { \compToNoEff{\tyEnv}{c}{\vty_1~!~\dirt_1}{\mlTm_c} \\
             \fullDirt{\dirt_1} \\
             \valToNoEff{\tyEnv}{v}{(\vty_1~!~\dirt_1 \hto \vty_2~!~\emptyset)}{\mlTm_v} \\
             \valTyToNoEff{\tyEnv}{\vty_2}{\ety}{\mlTyA}
           } 
           { \compToNoEff{\tyEnv}{(\withhandle{v}{c})}{\vty_2~!~\emptyset}{\cast{(\withhandle{\mlTm_v}{\mlTm_c})}{\unsafe{\refl{\mlTyA}}}} }

\inferrule*[right=CHandle3]
           { \fullDirt{\dirt_2} \\
             \compToNoEff{\tyEnv}{c}{\vty_1~!~\dirt_1}{\mlTm_c} \\
             \valToNoEff{\tyEnv}{v}{(\vty_1~!~\dirt_1 \hto \vty_2~!~\dirt_2)}{\mlTm_v} \\
             \fullDirt{\dirt_1}
           } 
           { \compToNoEff{\tyEnv}{(\withhandle{v}{c})}{\vty_2~!~\dirt_2}{\withhandle{\mlTm_v}{\mlTm_c}} }

\inferrule*[right=CCast]
           { \compToNoEff{\tyEnv}{c}{\cty_1}{\mlTm} \\
             \coToNoEff{\tyEnv}{\coercion}{\cty_1 \le \cty_2}{\mlCoercion'}
           } 
           { \compToNoEff{\tyEnv}{\cast{c}{\coercion}}{\cty_2}{\cast{\mlTm}{\mlCoercion'}} }
\end{mathpar}

\vspace{-5mm}
\caption{Elaboration of \target Computations to \noeff Terms}
\label{fig:eff-to-ml-computations}
\end{myfigure}

Finally, Figure~\ref{fig:eff-to-ml-computations} defines how \target computations
are elaborated into \noeff terms. There are a number of interesting cases.

Firstly, because $(\return{v})$ has an empty dirt, its elaborated form drops
the $\keyreturn$ (Rule~\textsc{CRet}).
Secondly, $\keydo$- computations are translated to either $\keylet$- or
$\keydo$- terms, depending on whether the dirt is empty or not
(Rules~\textsc{CDo1} and~\textsc{CDo2}, respectively).
Thirdly, handler applications are elaborated in three possible ways. If the
input dirt of the handler is empty, it is elaborated as a function and thus the
handler application too should be elaborated as function application
(Rule~\textsc{CHandle1}).
Otherwise, a handler application is indeed elaborated as a handler application.
If the output dirt is empty, the translation is straightforward
(Rule~\textsc{CHandle3}).
However, if the output dirt is empty, then the elaboration of the handler still
produces a computation where none is expected.  Hence, we insert an
$\keyunsafe$ coercion to bridge the gap (Rule~\textsc{CHandle2}).

\begin{example}
  \label{exa:running-noeff}
  Elaboration of terms to \noeff again depends on the type of a \target term. A monomorphic function
  \[
  \begin{array}{l@{\hspace{1mm}}c@{\hspace{1mm}}l}
    \keylet~~\mathit{f} & : & (\tyUnit \to \tyUnit~!~\emptyset) \to \tyUnit~!~\emptyset \\
                        & = & \fun{(g : \tyUnit \to \tyUnit~!~\emptyset)}{g~\tmUnit} \\
    \keyin~\ldots \\
  \end{array}
  \]
  is erased to
    \[
    \begin{array}{l@{\hspace{1mm}}c@{\hspace{1mm}}l}
      \keylet~~\mathit{f} & : & (\tyUnit \to \tyUnit) \to \tyUnit \\
                          & = & \fun{(g : \tyUnit \to \tyUnit)}{g~\tmUnit} \\
      \keyin~\ldots \\
    \end{array}
    \]
  as before, while its polymorphic variant
  \[
    \begin{array}{l@{\hspace{1mm}}c@{\hspace{1mm}}l}
      \keylet~~\mathit{f} & : & \forall \evar. \forall \tyvar : \evar. \forall \tyvar' : \evar. \forall \dirtvar. \forall \dirtvar'.
                                  (\tyvar \le \tyvar') \Rightarrow (\dirtvar \le \dirtvar') \Rightarrow (\tyUnit \to \tyvar~!~\dirtvar) \to \tyvar' ~!~ \dirtvar' \\
                          & = & \Lambda \evar. \Lambda (\tyvar : \evar). \Lambda (\tyvar' : \evar). \Lambda \dirtvar. \Lambda \dirtvar'. \Lambda (\covar : \tyvar \le \tyvar'). \Lambda (\covar' : \dirtvar \le \dirtvar'). \\
                          &   & \quad \fun{(g : \tyUnit \to \tyvar \,!\, \dirtvar)}{(\cast{(g~\tmUnit)}{(\covar \,!\, \covar')})} \\
      \keyin~\ldots \\
    \end{array}
  \]
  is conservatively elaborated to an impure
  \[
  \begin{array}{l@{\hspace{1mm}}c@{\hspace{1mm}}l}
    \keylet~~\mathit{f} & : & \forall \tyvar. \forall \tyvar'.
                                (\tyvar \le \tyvar') \Rightarrow (\tyUnit \to \mkMlCompTy{\tyvar}) \to \mkMlCompTy{\tyvar'} \\
                        & = & \Lambda \tyvar. \Lambda \tyvar'. \Lambda (\covar : \tyvar \le \tyvar'). \\
                        &   & \quad \fun{(g : \tyUnit \to \mkMlCompTy{\tyvar})}{(\cast{(g~\tmUnit)}{\covar})} \\
    \keyin~\ldots \\
  \end{array}
  \]
  Note that in contrast to the erasure to \erased, we keep type variables $\tyvar$ and $\tyvar'$, while removing any mention of their skeleton $\evar$.
  As before, we remove any effect annotations, conservatively assuming that dirt variables are impure,
  but keep an explicit coercion $\covar$ between types.
  
  Recall that in \target, the application $f~\mathit{id}$ was pure, and so must be its elaboration. However, since $f$ itself was conservatively assumed to be impure,
  the application must be suitably coerced. In particular, the elaboration of
  \[
    f~\tyUnit~\tyUnit~\tyUnit~\emptyset~\emptyset~\trgUnitRefl~\emptyset_\emptyset~(\fun{(x:\tyUnit)}{\return{x}})
  \]
  is
  \[
    (\cast{(\cast{(f~\tyUnit~\tyUnit)}{\mlCoercion_1})}{\mlCoercion_2})~\mlCoUnitRefl~(\fun{(x:\tyUnit)}{x})
  \]
  where for $\mlCoTy = \tyUnit \le \tyUnit$, the coercion $\mlCoercion_1$, which lifts a pure function into one that
  returns a computation, is given by
  \begin{align*}
    &\mlCoTy \Rightarrow (\mlCoUnitRefl \to \mlReturn{\mlCoUnitRefl}) \to \mkMlCompCo{\mlCoUnitRefl} \\
    &: (\mlCoTy \Rightarrow (\tyUnit \to \mkMlCompTy{\tyUnit}) \to \mkMlCompTy{\tyUnit})
    \le (\mlCoTy \Rightarrow (\tyUnit \to \tyUnit) \to \mkMlCompTy{\tyUnit})
  \end{align*}
  while $\mlCoercion_2$, which extracts back the value from a pure computation is:
  \begin{align*}
    &\mlCoTy \Rightarrow (\mlCoUnitRefl \to \mlCoUnitRefl) \to \unsafe{\mlCoUnitRefl} \\
    &: (\mlCoTy \Rightarrow (\tyUnit \to \tyUnit) \to \mkMlCompTy{\tyUnit})
    \le (\mlCoTy \Rightarrow (\tyUnit \to \tyUnit) \to \tyUnit)
  \end{align*}
  On a side note, observe the removal of $\keyreturn$ in the identity as its elaboration is a pure function.
  For an impure function
  \begin{align*}
    &f~\tyUnit~\tyUnit~\tyUnit~\{\texttt{Tick}\}~\{\texttt{Tick}, \texttt{Tock}\}~\trgUnitRefl~(\{\texttt{Tick}\} \cup \emptyset_{\{\texttt{Tock}\}}) \\
    &\quad (\fun{x : \tyUnit}{\operation[\mathtt{Tick}]{x}{y : \tyUnit}{(\cast{(\return{y})}{\trgUnitRefl~!~\emptyset_{\{\texttt{Tick}\}}})}})
  \end{align*}
  the elaboration
  \begin{align*}
    &(\cast{(\cast{(f~\tyUnit~\tyUnit)}{\mlCoercion_1'})}{\mlCoercion_2'})~\mlCoUnitRefl \\
    &\quad (\fun{x : \tyUnit}{\operation[\mathtt{Tick}]{x}{y : \tyUnit}{(\cast{y}{\mlReturn{\trgUnitRefl}})}})
  \end{align*}
  is similar, except that the coercions $\mlCoercion_1' = \mlCoercion_2'$ are both trivial:
  \begin{align*}
    &\mlCoTy \Rightarrow (\mlCoUnitRefl \to \mkMlCompCo{\mlCoUnitRefl}) \to \mkMlCompCo{\mlCoUnitRefl} \\
    &: (\mlCoTy \Rightarrow (\tyUnit \to \mkMlCompTy{\tyUnit}) \to \mkMlCompTy{\tyUnit})
    \le (\mlCoTy \Rightarrow (\tyUnit \to \mkMlCompTy{\tyUnit}) \to \mkMlCompTy{\tyUnit})
  \end{align*}
  and may be removed by an optimizer.
  Also note that just as in $\mathit{id}$, the $\keyreturn$ vanishes in the elaboration, though in this case it is reintroduced by the elaboration of the coercion~$\emptyset_{\{\texttt{Tick}\}}$.
\end{example}

\subsubsection{Metatheory of Elaboration}

We have proven in Abella that the elaboration of \target values and
computations into \noeff terms preserves typing.
\begin{theorem}[Type Preservation]
\label{thm:eff-to-ml-type-preservation}
\begin{itemize}
\item
  If  $\valToNoEff{\tyEnv}{v}{\vty}{\mlTm}$
  and $\tyEnvToNoEff{\tyEnv}{\mlEnv'}$
  then $\valTyToNoEff{\tyEnv}{\vty}{\ety}{\mlTyA}$
  and  $\tcNoEffTm{\mlEnv'}{\mlTm}{\mlTyA}$.
\item
  If  $\compToNoEff{\tyEnv}{c}{\cty}{\mlTm}$
  and $\tyEnvToNoEff{\tyEnv}{\mlEnv'}$
  then $\compTyToNoEff{\tyEnv}{\cty}{\ety}{\mlTyB}$
  and  $\tcNoEffTm{\mlEnv'}{\mlTm}{\mlTyB}$.
\end{itemize}
\end{theorem}

A key lemma in the theorem's proof establishes the appropriate typing of the coercion produced by the
$\fromImpureVal {\vty}{\dirt}{\mlCoercion}$ judgement.
\begin{lemma}[From Impure Coercion Typing]
If
      $\fromImpureVal {\vty}{\dirt}{\mlCoercion}$    
and
      $\valTyToNoEff{\ctx,\dirtvar}{\vty}{\ety}{\mlTyA}$ 
then there exists a $\mlTyB$ such that
      $\valTyToNoEff{\ctx}{\vty[\dirt/\dirtvar]}{\ety}{\mlTyB}$ 
and
      $\tcNoEffCoercion{\mlEnv}{\mlCoercion}{\mlTyA \le \mlTyB}$. 
\end{lemma}

Semantic preservation for the elaboration from \target to \noeff turns out to
be much more complicated than for the elaboration to \erased. Indeed, the
congruence closure of the step relation is not sufficient in
the case of \noeff.

For instance, consider the following \target evaluation step:
\[\smallStepVal{(\Lambda \dirtvar. \fun{x:\tyUnit}{v})\,\emptyset}{\fun{x:\tyUnit}{v[\emptyset/\dirtvar]}}\]
where the dirt abstraction $(\Lambda \dirtvar. \fun{x:\tyUnit}{v})$ has type $\forall \dirtvar.  \tyUnit
\to \withdirt{\tyUnit}{\set{\op} \cup \dirtvar}$ and its application to $\emptyset$
has type $\tyUnit \to \withdirt{\tyUnit}{\set{\op} \cup \emptyset}$.  Suppose that
the right-hand side elaborates to the \noeff term $\fun{x:\tyUnit}{v'}$.  Then the left-hand
side elaborates to $\cast{\fun{(x:\tyUnit}{v'})}{(\mlCoUnitRefl \to \mkMlCompCo{\mlCoUnitRefl})}$;
observe that the function coercion is nothing more than a reflexivity coercion.
Neither of these two elaborated \noeff terms is reducible.
In particular, we cannot eliminate the reflexivity coercion by reduction and
thus the two terms are not related by a congruence closure of the step
relation.

Instead, we believe that a semantic notion of equivalence is needed: contextual
equivalence~\cite{morris1969lambda}. Informally, this notion expresses that two terms are
equivalent iff, when placed in any ``appropriate'' program context, the
resulting programs reduce to normal forms that are equivalent under some other,
simpler notion of equivalence such as syntactic equality. 

The precise formal definition depends on the particular setting it is used in.
In our setting there are a number of complicating factors that need to be taken
into account.
\begin{itemize}
\item
Firstly, we are dealing with two mutually recursive syntactic
sorts for terms, values and computations. This calls for four different
mutually recursive sorts of program contexts: ones that take a
value/computation and yield a value/computation. 
\item
Secondly, we need to consider
what simpler notion of equivalence to use and how to restrict program contexts
so that we can use it. A common approach is to consider only contexts that have
some atomic type as a result, such as naturals or integers, where syntactic
equality is appropriate. We believe that approach works here too. Indeed, we can expect that
an appropriate computation context handles all
operations and yields a pure program. 
\item
Thirdly, we do not want to admit all possible \noeff contexts. In particular,
we do not want to admit those that get stuck because of an inappropriate use of
an $\keyunsafe$ coercion. Hence, we want to restrict ourselves to those that are
the image of a \target program context. 
\end{itemize}
We leave working out the precise formal definition of contextual equivalence
and proving semantic preservation on top of it a substantial open challenge.
Yet, we point to the work of Bi et al.~\shortcite{ecoop2018:it} as an important source of
inspiration. They also deal with an elaboration-based setting, for disjoint intersection
types, and use logical relations as the basis of their proofs.

%
%

\section{Related Work \& Conclusion}\label{sec:conclusion}
\paragraph{\bf Eff's Implicit Type System}

The most closely related work is that of \citet{Pretnar}{DBLP:journals/corr/Pretnar13}
on inferring algebraic effects for Eff, which is the basis for our
implicitly-typed \source calculus, its type system and the type inference
algorithm. There are three major differences with Pretnar's inference algorithm.

Firstly, our work introduces an explicitly-typed calculus. For this reason we
have extended the constraint generation phase with the elaboration into \target
and the constraint solving phase with the construction of coercions.

Secondly, we add skeletons to guarantee erasure. Skeletons also allow us to use
the standard occurs-check during
unification. In contrast, unification in Pretnar's algorithm performs the occurs-check up to the
equivalence closure of the subtyping relation~\cite{DBLP:journals/tcs/FuhM90,simonet2003type},
and needs to take care of appropriately instantiating all variables in an equivalence class
(also called a skeleton). As these classes turn out to be surrogates for the underlying skeleton
types, we have decided to keep the name. \citet{Traytel \emph{et al.}}{DBLP:conf/aplas/TraytelBN11}
propose an alternative approach and first perform a weak unification algorithm, which is unification
with the standard occurs check on what are essentially skeletons, although this is not an explicit
concept in their work.

Finally, Pretnar incorporates garbage collection of constraints~\cite{pottier2001simplifying}. The aim
of this approach is to obtain unique and simple type schemes by eliminating redundant
constraints. Garbage collection is not suitable for our use
as type variables and coercions witnessing subtyping constraints
cannot simply be dropped, but must be instantiated in a
suitable manner, which cannot be done in general.

Consider for instance a situation with type variables
$\tyvar_1$,
$\tyvar_2$,
$\tyvar_3$,
$\tyvar_4$, and
$\tyvar_5$
where
$\tyvar_1 \le \tyvar_3$,
$\tyvar_2 \le \tyvar_3$,
$\tyvar_3 \le \tyvar_4$, and
$\tyvar_3 \le \tyvar_5$.
Suppose that $\tyvar_3$ does not appear in the type. Then garbage collection would eliminate
it and replace the constraints by
$\tyvar_1 \le \tyvar_4$,
$\tyvar_2 \le \tyvar_4$,
$\tyvar_1 \le \tyvar_5$, and
$\tyvar_2 \le \tyvar_5$. While garbage collection guarantees that for any ground
instantiation of the remaining type variables, there exists a valid ground
instantiation for $\tyvar_3$, \target would need to be extended with joins (or meets) to express a
generically valid instantiation like $\tyvar_1 \sqcup \tyvar_2$. Moreover, we would need additional
coercion formers to establish $\tyvar_1 \le (\tyvar_1 \sqcup \tyvar_2)$ or $(\tyvar_1 \sqcup \tyvar_2) \le \tyvar_4$.

As these additional constructs considerably complicate the calculus,
we propose a simpler solution, especially as we have experienced no blow-up
in inference times during our initial experiments. We use \target as it is for
internal purposes, but display types to programmers in their garbage-collected
form.

\paragraph{\bf Calculi with Explicit Coercions}

The notion of explicit coercions is not new;
\citet{Mitchell}{mitchell} introduced the idea of inserting coercions during type
inference for ML-based languages, as a means for explicit casting between
different numeric types.

\citet{Breazu-Tannen et al.}{coercions} also present a translation of languages
with inheritance polymorphism into \SystemF, extended with coercions. Although
their coercion combinators are very similar to our coercion forms,
Breazu-Tannen et al.'s coercions are terms, and thus cannot be erased.

Much closer to \target is Crary's coercion calculus for inclusive
subtyping~\citep{crary}, from which we borrowed the stratification of value
results. Though the coercion calculus does not support coercion abstraction and
other coercion forms that we need for supporting effects, coercions in Crary's
system are also erasable so they have no runtime effect.

\SystemFC~\citep{systemfc} uses explicit type-equality coercions to encode
complex language features (e.g. GADTs~\citep{unigadts} or type
families~\citep{typecheckingwithopentf}). Though \target's coercions are proofs
of subtyping rather than type equality, our system has a lot in common with it,
and in particular the ``push'' rules. A difference between the two lies in the
presence of inversion coercions (that is, coercions that allow for
decomposition of type inequalities), which \SystemFC (and our own earlier
work~\cite{esop2018}) includes.

The \noeff $\keyunsafe$ coercion shows similarities with downcasts in
object-oriented languages and calculi like Featherweight Java~\cite{fj}, which
get stuck when the object is from the wrong class. A difference to
Featherweight Java is that, when successful, $\keyunsafe$ also destructures a
value. This shares similarities with the explicit coercions in the recent backend calculi
for (disjoint) intersection types \cite{icfp2016:it,ecoop2018:it}, which also extract relevant
components from composite values.

\paragraph{\bf Future Work}

Our plans focus on resuming the postponed work on efficient compilation of handlers.
First, we intend to adjust program transformations to the explicit type information. We
hope that this will not only make the optimizer more robust, but also expose new
optimization opportunities. Next, we plan to write compilers to both Multicore
OCaml and standard OCaml.
Finally, once the compiler shows promising preliminary results,
we plan to extend it to other Eff features such as user-defined types
or recursion, allowing us to benchmark it on more realistic programs.

\paragraph{\bf Acknowledgements}

We would like thank the anonymous reviewers, members of the IFIP WG 2.1 group, participants of Dagstuhl seminars 16112~\cite{DBLP:journals/dagstuhl-reports/Bauer0PY16} and 18172~\cite{DBLP:journals/dagstuhl-reports/ChandrasekaranL18}, Gert-Jan Bottu, Mauro Jaskelioff, Filip Koprivec, Žiga Lukšič, Leonidas Lampropoulos, Klara Mardirosian, Ruben Pieters, Alexander Vandenbroucke, Nicolas Wu, and Žiga Zupančič for all their helpful comments and suggestions. Part of this work was funded by the KU Leuven Special Research Fund (BOF), project 3E160354, and by the Fund for Scientific Research - Flanders, project G0D1419N. This material is based upon work supported by the Air Force Office of Scientific Research under award number FA9550-17-1-0326.

\bibliography{bibliography}

\begin{thebibliography}{10}

\bibitem{barendregt}
Henk Barendregt.
\newblock {\em The Lambda Calculus: its Syntax and Semantics, volume 103 of
  Studies in Logic and the Foundations of Mathematics}.
\newblock North-Holland, 1981.

\bibitem{DBLP:journals/dagstuhl-reports/Bauer0PY16}
Andrej Bauer, Martin Hofmann, Matija Pretnar, and Jeremy Yallop.
\newblock From theory to practice of algebraic effects and handlers (dagstuhl
  seminar 16112).
\newblock {\em Dagstuhl Reports}, 6(3):44--58, 2016.

\bibitem{bauer15}
Andrej Bauer and Matija Pretnar.
\newblock Programming with algebraic effects and handlers.
\newblock {\em Journal of Logic and Algebraic Programming}, 84(1):108--123,
  2015.

\bibitem{ecoop2018:it}
Xuan Bi, Bruno~C. d.~S.~Oliveira, and Tom Schrijvers.
\newblock The essence of nested composition.
\newblock In Todd~D. Millstein, editor, {\em 32nd European Conference on
  Object-Oriented Programming, {ECOOP} 2018, July 16-21, 2018, Amsterdam, The
  Netherlands}, volume 109 of {\em LIPIcs}, pages 22:1--22:33. Schloss Dagstuhl
  - Leibniz-Zentrum fuer Informatik, 2018.

\bibitem{coercions}
Val Breazu-Tannen, Thierry Coquand, Carl~A. Gunter, and Andre Scedrov.
\newblock Inheritance as implicit coercion.
\newblock {\em Information and Computation vol}, 93:172--221, 1991.

\bibitem{DBLP:journals/dagstuhl-reports/ChandrasekaranL18}
Sivaramakrishnan~Krishnamoorthy Chandrasekaran, Daan Leijen, Matija Pretnar,
  and Tom Schrijvers.
\newblock Algebraic effect handlers go mainstream (dagstuhl seminar 18172).
\newblock {\em Dagstuhl Reports}, 8(4):104--125, 2018.

\bibitem{crary}
Karl Crary.
\newblock Typed compilation of inclusive subtyping.
\newblock In {\em Proceedings of the Fifth ACM SIGPLAN International Conference
  on Functional Programming}, ICFP '00, pages 68--81, NY, USA, 2000. ACM.

\bibitem{icfp2016:it}
Bruno~C. d.~S.~Oliveira, Zhiyuan Shi, and Jo{\~{a}}o Alpuim.
\newblock Disjoint intersection types.
\newblock In Jacques Garrigue, Gabriele Keller, and Eijiro Sumii, editors, {\em
  Proceedings of the 21st {ACM} {SIGPLAN} International Conference on
  Functional Programming, {ICFP} 2016, Nara, Japan, September 18-22, 2016},
  pages 364--377. {ACM}, 2016.

\bibitem{DamasMilner}
Luis Damas and Robin Milner.
\newblock Principal type-schemes for functional programs.
\newblock In {\em Proceedings of the 9th ACM SIGPLAN-SIGACT Symposium on
  Principles of Programming Languages}, POPL '82, pages 207--212, NY, USA,
  1982. ACM.

\bibitem{ocaml}
Stephen Dolan, Leo White, KC~Sivaramakrishnan, Jeremy Yallop, and Anil
  Madhavapeddy.
\newblock Effective concurrency through algebraic effects.
\newblock In {\em OCaml Workshop}, 2015.

\bibitem{DBLP:journals/tcs/FuhM90}
You{-}Chin Fuh and Prateek Mishra.
\newblock Type inference with subtypes.
\newblock {\em Theor. Comput. Sci.}, 73(2):155--175, 1990.

\bibitem{girardthesis}
Jean-Yves Girard.
\newblock {\em Interpr{\'e}tation fonctionelle et {\'e}limination des coupures
  de l’arithm{\'e}tique d’ordre sup{\'e}rieur}.
\newblock PhD thesis, PhD thesis, Universit{\'e} Paris VII, 1972.

\bibitem{systemf}
Jean-Yves Girard, Paul Taylor, and Yves Lafont.
\newblock {\em Proofs and Types}.
\newblock Cambridge University Press, 1989.

\bibitem{CoercionCalculus}
Fritz Henglein.
\newblock Dynamic typing: Syntax and proof theory.
\newblock In {\em Selected Papers of the Symposium on Fourth European Symposium
  on Programming}, ESOP'92, pages 197--230, Amsterdam, The Netherlands, The
  Netherlands, 1994. Elsevier Science Publishers B. V.

\bibitem{links_rows}
Daniel Hillerstr{\"{o}}m and Sam Lindley.
\newblock Liberating effects with rows and handlers.
\newblock In James Chapman and Wouter Swierstra, editors, {\em Proceedings of
  the 1st International Workshop on Type-Driven Development, TyDe@ICFP 2016,
  Nara, Japan, September 18, 2016}, pages 15--27. {ACM}, 2016.

\bibitem{hindley}
R.~Hindley.
\newblock {The Principal Type-Scheme of an Object in Combinatory Logic}.
\newblock {\em Transactions of the American Mathematical Society}, 146:29--60,
  1969.

\bibitem{fj}
Atsushi Igarashi, Benjamin~C. Pierce, and Philip Wadler.
\newblock Featherweight java: A minimal core calculus for java and gj.
\newblock {\em ACM Trans. Program. Lang. Syst.}, 23(3):396--450, May 2001.

\bibitem{markjones}
Mark~P. Jones.
\newblock A theory of qualified types.
\newblock In Bernd Krieg{-}Br{\"{u}}ckner, editor, {\em {ESOP} '92, 4th
  European Symposium on Programming, Rennes, France, February 26-28, 1992,
  Proceedings}, volume 582 of {\em Lecture Notes in Computer Science}, pages
  287--306. Springer, 1992.

\bibitem{kammar}
Ohad Kammar, Sam Lindley, and Nicolas Oury.
\newblock Handlers in action.
\newblock In {\em Proceedings of the 18th ACM SIGPLAN International Conference
  on Functional programming}, ICFP '14, pages 145--158. ACM, 2013.

\bibitem{eff2ocaml}
Oleg Kiselyov and KC~Sivaramakrishnan.
\newblock Eff directly in ocaml.
\newblock In {\em OCaml Workshop}, 2016.

\bibitem{DBLP:journals/corr/Leijen14}
Daan Leijen.
\newblock Koka: Programming with row polymorphic effect types.
\newblock In Paul Levy and Neel Krishnaswami, editors, {\em Proceedings 5th
  Workshop on Mathematically Structured Functional Programming, MSFP@ETAPS
  2014, Grenoble, France, 12 April 2014.}, volume 153 of {\em {EPTCS}}, pages
  100--126, 2014.

\bibitem{koka2017}
Daan Leijen.
\newblock Type directed compilation of row-typed algebraic effects.
\newblock In Giuseppe Castagna and Andrew~D. Gordon, editors, {\em Proceedings
  of the 44th {ACM} {SIGPLAN} Symposium on Principles of Programming Languages,
  {POPL} 2017, Paris, France, January 18-20, 2017}, pages 486--499. {ACM},
  2017.

\bibitem{DBLP:conf/popl/LindleyMM17}
Sam Lindley, Conor McBride, and Craig McLaughlin.
\newblock Do be do be do.
\newblock In Giuseppe Castagna and Andrew~D. Gordon, editors, {\em Proceedings
  of the 44th {ACM} {SIGPLAN} Symposium on Principles of Programming Languages,
  {POPL} 2017, Paris, France, January 18-20, 2017}, pages 500--514. {ACM},
  2017.

\bibitem{milner}
Robin Milner.
\newblock A theory of type polymorphism in programming.
\newblock {\em Journal of Computer and System Sciences}, 17:348--375, 1978.

\bibitem{mitchell}
John~C. Mitchell.
\newblock Coercion and type inference.
\newblock In {\em Proceedings of the 11th ACM SIGACT-SIGPLAN Symposium on
  Principles of Programming Languages}, POPL '84, pages 175--185, New York, NY,
  USA, 1984. ACM.

\bibitem{morris1969lambda}
James~Hiram Morris~Jr.
\newblock {\em Lambda-calculus models of programming languages}.
\newblock PhD thesis, Massachusetts Institute of Technology, 1969.

\bibitem{unigadts}
Simon Peyton~Jones, Dimitrios Vytiniotis, Stephanie Weirich, and Geoffrey
  Washburn.
\newblock Simple unification-based type inference for gadts.
\newblock In {\em ICFP '06}, 2006.

\bibitem{DBLP:journals/acs/PlotkinP03}
Gordon~D. Plotkin and John Power.
\newblock Algebraic operations and generic effects.
\newblock {\em Applied Categorical Structures}, 11(1):69--94, 2003.

\bibitem{DBLP:journals/corr/PlotkinP13}
Gordon~D. Plotkin and Matija Pretnar.
\newblock Handling algebraic effects.
\newblock {\em Logical Methods in Computer Science}, 9(4), 2013.

\bibitem{pottier2001simplifying}
Fran{\c{c}}ois Pottier.
\newblock Simplifying subtyping constraints: {A} theory.
\newblock {\em Information and Computation}, 170(2):153--183, 2001.

\bibitem{DBLP:journals/corr/Pretnar13}
Matija Pretnar.
\newblock Inferring algebraic effects.
\newblock {\em Logical Methods in Computer Science}, 10(3), 2014.

\bibitem{pretnar:tutorial}
Matija Pretnar.
\newblock An introduction to algebraic effects and handlers, invited tutorial.
\newblock {\em Electronic Notes in Theoretical Computer Science}, 319:19--35,
  2015.

\bibitem{optimization}
Matija Pretnar, Amr~Hany Saleh, Axel Faes, and Tom Schrijvers.
\newblock Efficient compilation of algebraic effects and handlers.
\newblock Technical Report CW 708, KU Leuven Department of Computer Science,
  2017.

\bibitem{reynolds-systemf-1}
John~C. Reynolds.
\newblock Towards a theory of type structure.
\newblock In {\em Programming Symposium, Proceedings Colloque Sur La
  Programmation}, pages 408--423, London, UK, UK, 1974. Springer-Verlag.

\bibitem{reynolds-systemf-2}
John~C. Reynolds.
\newblock Types, abstraction, and parametric polymorphism.
\newblock In R.E.A. Mason, editor, {\em Information Processing 83}, pages
  513--523, North Holland, Amsterdam, 1983.

\bibitem{esop2018}
Amr~Hany Saleh, Georgios Karachalias, Matija Pretnar, and Tom Schrijvers.
\newblock Explicit effect subtyping.
\newblock In {\em {ESOP}}, volume 10801 of {\em Lecture Notes in Computer
  Science}, pages 327--354. Springer, 2018.

\bibitem{typecheckingwithopentf}
Tom Schrijvers, Simon Peyton~Jones, Manuel Chakravarty, and Martin Sulzmann.
\newblock Type checking with open type functions.
\newblock In {\em ICFP '08}, pages 51--62. ACM, 2008.

\bibitem{simonet2003type}
Vincent Simonet.
\newblock Type inference with structural subtyping: {A} faithful formalization
  of an efficient constraint solver.
\newblock In Atsushi Ohori, editor, {\em Programming Languages and Systems,
  First Asian Symposium, {APLAS} 2003, Beijing, China, November 27--29, 2003,
  Proceedings}, pages 283--302. Springer, 2003.

\bibitem{systemfc}
Martin Sulzmann, Manuel M.~T. Chakravarty, Simon Peyton~Jones, and Kevin
  Donnelly.
\newblock System f with type equality coercions.
\newblock In {\em Proceedings of the 2007 ACM SIGPLAN International Workshop on
  Types in Languages Design and Implementation}, TLDI '07, pages 53--66, New
  York, NY, USA, 2007. ACM.

\bibitem{DBLP:conf/aplas/TraytelBN11}
Dmitriy Traytel, Stefan Berghofer, and Tobias Nipkow.
\newblock Extending hindley-milner type inference with coercive structural
  subtyping.
\newblock In Hongseok Yang, editor, {\em Programming Languages and Systems -
  9th Asian Symposium, {APLAS} 2011, Kenting, Taiwan, December 5-7, 2011.
  Proceedings}, volume 7078 of {\em Lecture Notes in Computer Science}, pages
  89--104. Springer, 2011.

\bibitem{DBLP:conf/popl/WansbroughJ99}
Keith Wansbrough and Simon~L. Peyton~Jones.
\newblock Once upon a polymorphic type.
\newblock In {\em {POPL}}, pages 15--28. {ACM}, 1999.

\end{thebibliography}


\FloatBarrier

\appendix

\section{\source Additional Judgements}\label{appendix:source-additional}
\paragraph{Type Well-formedness and Elaboration}

Since our system discriminates between value types and computation types,
well-formedness of types is checked using two mutually recursive relations:
$\tcElabPolyTy{\tyEnv}{\aVty}{\ety}{\vty}$ (values), and
$\tcElabCompTy{\tyEnv}{\cCty}{\ety}{\cty}$ (computations). We discuss each one
separately below.

Well-formedness for value types is given by the following rules:
\begin{center}\begin{shaded}\begin{minipage}{\columnwidth}
\vspace{-4mm}
\small
\begin{mathpar}
\inferrule*[right=]
           { \tyvar\!:\!\ety \in \tyEnv }
           { \tcElabPolyTy{\tyEnv}{\tyvar}{\ety}{\tyvar} }

\inferrule*[right=]
           { \tcElabPolyTy{\tyEnv}{\aVty}{\ety_1}{\vty} \\
             \tcElabCompTy{\tyEnv}{\cCty}{\ety_2}{\cty}
           } 
           { \tcElabPolyTy{\tyEnv}{\aVty \to \cCty}{\ety_1 \to \ety_2}{\vty \to \cty} }

\inferrule*[right=]
           { \tcElabCompTy{\tyEnv}{\cCty}{\ety_1}{\cty_1} \\
             \tcElabCompTy{\tyEnv}{\dCty}{\ety_2}{\cty_2}
           } 
           { \tcElabPolyTy{\tyEnv}{\cCty \hto \dCty}{\ety_1 \hto \ety_2}{\cty_1 \hto \cty_2} }

\inferrule*[right=]
           { }
           { \tcElabPolyTy{\tyEnv}{\tyUnit}{\tyUnit}{\tyUnit} }

\inferrule*[right=]
           { \tcElabConstraint{\tyEnv}{\srcPartialCt}{\trgPartialCt} \\
             \tcElabPolyTy{\tyEnv}{\qualTy}{\ety}{\vty}
           } 
           { \tcElabPolyTy{\tyEnv}{\srcPartialCt \Rightarrow \qualTy}{\ety}{\trgPartialCt \Rightarrow \vty} }

\inferrule*[right=]
           { \tcElabPolyTy{\tyEnv, \tyvar:\ety_1}{\polyTy}{\ety_2}{\vty} }
           { \tcElabPolyTy{\tyEnv}{\forall \tyvar\!:\!\ety_1. \polyTy}{\ety_2}{\forall \tyvar\!:\!\ety_1. \vty} }

\inferrule*[right=]
           { \tcElabPolyTy{\tyEnv, \dirtvar}{\polyTy}{\ety}{\vty} }
           { \tcElabPolyTy{\tyEnv}{\forall \dirtvar. \polyTy}{\ety}{\forall \dirtvar. \vty} }

\inferrule*[right=]
           { \tcElabPolyTy{\tyEnv, \evar}{\polyTy}{\ety}{\vty} }
           { \tcElabPolyTy{\tyEnv}{\forall \evar. \polyTy}{\forall \evar.\ety}{\forall \evar. \vty} }
\end{mathpar}
\end{minipage}\end{shaded}\end{center}
The judgement is syntax-directed on the structure of types; each rule
corresponds to a value type syntactic form. Since \target types are a superset
of \source types, the elaboration-part (highlighted in gray) is the
identity transformation. Hence, the essence of the judgement is to check the
well-scopedness of source types.

Well-formedness for computation types is given by the following rule:
\begin{center}\begin{shaded}\begin{minipage}{\columnwidth}
\vspace{-4mm}
\small
\begin{mathpar}
\inferrule*[right=]
           { \tcElabPolyTy{\tyEnv}{\aVty}{\ety}{\vty} \\
             \wfDirt{\tyEnv}{\dirt}
           } 
           { \tcElabCompTy{\tyEnv}{\aVty~!~\dirt}{\ety}{\vty~!~\dirt} }
\end{mathpar}
\end{minipage}\end{shaded}\end{center}
We ensure that both parts of a computation type (the value type and the dirt)
are well-scoped under $\tyEnv$, while elaborating the value-type into a proper
\target representation.

\paragraph{Constraint Well-formedness and Elaboration}

Well-formedness for constraints is given by judgement
$\tcElabConstraint{\tyEnv}{\srcFullCt}{\trgFullCt}$, given by the following
rules:
\begin{center}\begin{shaded}\begin{minipage}{\columnwidth}
\vspace{-4mm}
\small
\begin{mathpar}
\inferrule*[right=]
           { \tcElabPolyTy{\tyEnv}{\aVty}{\ety}{\vty_1} \\
             \tcElabPolyTy{\tyEnv}{\bVty}{\ety}{\vty_2}
           } 
           { \tcElabConstraint{\tyEnv}{\aVty \le \bVty}{\vty_1 \le \vty_2} }

\inferrule*[right=]
           { \tcElabCompTy{\tyEnv}{\cCty}{\ety}{\cty_1} \\
             \tcElabCompTy{\tyEnv}{\dCty}{\ety}{\cty_2}
           } 
           { \tcElabConstraint{\tyEnv}{\cCty \le \dCty}{\cty_1 \le \cty_2} }

\inferrule*[right=]
           { \wfDirt{\tyEnv}{\dirt_1} \\
             \wfDirt{\tyEnv}{\dirt_2}
           } 
           { \tcElabConstraint{\tyEnv}{\dirt_1 \le \dirt_2}{\dirt_1 \le \dirt_2} }
\end{mathpar}
\end{minipage}\end{shaded}\end{center}
Since the dirt syntax is shared between \source and \target, all three rules
check the constraint components for well-scopedness, but only the type-related
constraints are elaborated: the elaboration of a dirt constraint is the
identity.

\paragraph{Dirt Well-formedness}

Judgement $\wfDirt{\tyEnv}{\dirt}$ checks dirt well-formedness and is given by the following rules:
\begin{center}\begin{shaded}\begin{minipage}{\columnwidth}
\vspace{-4mm}
\small
\begin{mathpar}
\inferrule*[right=]
           { }
           { \wfDirt{\tyEnv}{\emptyset} }

\inferrule*[right=]
           { \dirtvar \in \tyEnv }
           { \wfDirt{\tyEnv}{\dirtvar} }

\inferrule*[right=]
           { (\op : \aVty_\op \to \bVty_\op) \in \sig \\
             \wfDirt{\tyEnv}{\dirt}
           } 
           { \wfDirt{\tyEnv}{\{ \op \} \cup \dirt} }
\end{mathpar}
\end{minipage}\end{shaded}\end{center}
In addition to checking that the dirt is well-scoped (illustrated by the second
rule), we also make sure that all operations in a dirt set are already defined,
by looking them up in the globally visible signature $\sig$ (third rule).

\paragraph{Skeleton Well-formedness}

Finally, skeleton well-formedness is performed by judgement
$\tcSkeleton{\tyEnv}{\ety}$, as given by the following rules:
\begin{center}\begin{shaded}\begin{minipage}{\columnwidth}
\vspace{-4mm}
\small
\begin{mathpar}
\inferrule*[right=]
           { \evar \in \tyEnv }
           { \tcSkeleton{\tyEnv}{\evar} }

\inferrule*[right=]
           { }
           { \tcSkeleton{\tyEnv}{\tyUnit} }

\inferrule*[right=]
           { \tcSkeleton{\tyEnv}{\ety_1} \\
             \tcSkeleton{\tyEnv}{\ety_2}
           } 
           { \tcSkeleton{\tyEnv}{\ety_1 \to \ety_2} }

\inferrule*[right=]
           { \tcSkeleton{\tyEnv}{\ety_1} \\
             \tcSkeleton{\tyEnv}{\ety_2}
           } 
           { \tcSkeleton{\tyEnv}{\ety_1 \hto \ety_2} }
\end{mathpar}
\end{minipage}\end{shaded}\end{center}
Since skeletons are uni-kinded, this relation is entirely straightforward and
is in fact identical to the well-formedness of \SystemF simple types.

\section{\target Additional Judgements}\label{appendix:target-additional}
\paragraph{Type Well-formedness}

Again, preserving the separation between value and computation types, \target
comes with two mutually recursive relation for checking the well-formedness of
types: $\tcTrgVty{\tyEnv}{\vty}{\ety}$ (values), and
$\tcTrgCty{\tyEnv}{\cty}{\ety}$ (computations). We discuss each one separately.

Well-formedness for value types is given by the following rules:
\begin{center}\begin{shaded}\begin{minipage}{\columnwidth}
\vspace{-4mm}
\small
\begin{mathpar}
\inferrule*[right=]
           { (\tyvar : \ety)\in \tyEnv }
           { \tcTrgVty{\tyEnv}{\tyvar}{\ety} }

\inferrule*[right=]
           { \tcTrgVty{\tyEnv}{\vty}{\ety_1} \\
             \tcTrgCty{\tyEnv}{\cty}{\ety_2}
           } 
           { \tcTrgVty{\tyEnv}{\vty \to \cty}{\ety_1 \to \ety_2} }

\inferrule*[right=]
           { \tcTrgCty{\tyEnv}{\cty_1}{\ety_1} \\
             \tcTrgCty{\tyEnv}{\cty_2}{\ety_2}
           } 
           { \tcTrgVty{\tyEnv}{\cty_1 \hto \cty_2}{\ety_1 \hto \ety_2} }

\inferrule*[right=]
           { }
           { \tcTrgVty{\tyEnv}{\tyUnit}{\tyUnit} }

\inferrule*[right=]
           { \wfTrgCt{\tyEnv}{\trgPartialCt} \\
             \tcTrgVty{\tyEnv}{\vty}{\ety}
           } 
           { \tcTrgVty{\tyEnv}{\trgPartialCt \Rightarrow \vty}{\ety} }

\inferrule*[right=]
           { \tcTrgVty{\tyEnv, \evar}{\vty}{\ety} }
           { \tcTrgVty{\tyEnv}{\forall \evar. \vty}{\forall \evar. \ety} }

\inferrule*[right=]
           { \tcTrgVty{\tyEnv, \tyvar : \ety_1}{\vty}{\ety_2} }
           { \tcTrgVty{\tyEnv}{\forall \tyvar : \ety_1. \vty}{\ety_2} }

\inferrule*[right=]
           { \tcTrgVty{\tyEnv, \dirtvar}{\vty}{\ety} }
           { \tcTrgVty{\tyEnv}{\forall \dirtvar. \vty}{\ety} }
\end{mathpar}
\end{minipage}\end{shaded}\end{center}
The relation is almost identical to the corresponding one for \source value
types. The only difference between the two lies in the \target's impredicative
polymorphism and higher-rank types.

Similarly, well-formedness of computation types is checked via relation
$\tcTrgCty{\tyEnv}{\cty}{\ety}$, given by a single rule, which is identical to
the corresponding one of \source:
\begin{center}\begin{shaded}\begin{minipage}{\columnwidth}
\vspace{-4mm}
\small
\begin{mathpar}
\inferrule*[right=]
           { \tcTrgVty{\tyEnv}{\vty}{\ety} \\
             \wfDirt{\tyEnv}{\dirt}
           } 
           { \tcTrgCty{\tyEnv}{\vty~!~\dirt}{\ety} }
\end{mathpar}
\end{minipage}\end{shaded}\end{center}
The only difference, again, is that instead of a monotype $\aVty$, computation
types are allowed to refer to arbitrary \SystemF types $\vty$.

\paragraph{Constraint Well-formedness}

Well-formedness for constraints is given by judgement
$\wfTrgCt{\tyEnv}{\trgFullCt}$:
\begin{center}\begin{shaded}\begin{minipage}{\columnwidth}
\vspace{-4mm}
\small
\begin{mathpar}
\inferrule*[right=]
           { \tcTrgVty{\tyEnv}{\vty_1}{\ety} \\
             \tcTrgVty{\tyEnv}{\vty_2}{\ety}
           } 
           { \wfTrgCt{\tyEnv}{\vty_1 \le \vty_2} }

\inferrule*[right=]
           { \tcTrgCty{\tyEnv}{\cty_1}{\ety} \\
             \tcTrgCty{\tyEnv}{\cty_2}{\ety}
           } 
           { \wfTrgCt{\tyEnv}{\cty_1 \le \cty_2} }

\inferrule*[right=]
           { \wfDirt{\tyEnv}{\dirt_1} \\
             \wfDirt{\tyEnv}{\dirt_2}
           } 
           { \wfTrgCt{\tyEnv}{\dirt_1 \le \dirt_2} }
\end{mathpar}
\end{minipage}\end{shaded}\end{center}

\paragraph{Dirt Well-formedness}

Dirt well-formedness takes the form $\wfDirt{\tyEnv}{\dirt}$ and is given by
the following rules:
\begin{center}\begin{shaded}\begin{minipage}{\columnwidth}
\vspace{-4mm}
\small
\begin{mathpar}
\inferrule*[right=]
           { }
           { \wfDirt{\tyEnv}{\emptyset} }

\inferrule*[right=]
           { \dirtvar \in \tyEnv }
           { \wfDirt{\tyEnv}{\dirtvar} }

\inferrule*[right=]
           { (\op : \vty_1 \to \vty_2) \in \sig \\
             \wfDirt{\tyEnv}{\dirt}
           } 
           { \wfDirt{\tyEnv}{\{ \op \} \cup \dirt} }
\end{mathpar}
\end{minipage}\end{shaded}\end{center}
The only difference with the corresponding relation for \source is that instead
of operations $\op$ having \source types, they now have \target types. We abuse
notation and use $\sig$ for both the \source and the \target top-level
signature set.

\paragraph{Skeleton Well-formedness}

Skeleton well-formedness is checked via relation $\tcSkeleton{\tyEnv}{\ety}$, given by the following rules:
\begin{center}\begin{shaded}\begin{minipage}{\columnwidth}
\vspace{-4mm}
\small
\begin{mathpar}
\inferrule*[right=]
           { \evar \in \tyEnv }
           { \tcSkeleton{\tyEnv}{\evar} }

\inferrule*[right=]
           { }
           { \tcSkeleton{\tyEnv}{\tyUnit} }

\inferrule*[right=]
           { \tcSkeleton{\tyEnv}{\ety_1} \\
             \tcSkeleton{\tyEnv}{\ety_2}
           } 
           { \tcSkeleton{\tyEnv}{\ety_1 \to \ety_2} }

\inferrule*[right=]
           { \tcSkeleton{\tyEnv}{\ety_1} \\
             \tcSkeleton{\tyEnv}{\ety_2}
           } 
           { \tcSkeleton{\tyEnv}{\ety_1 \hto \ety_2} }

\inferrule*[right=]
           { \tcSkeleton{\tyEnv, \evar}{\ety} }
           { \tcSkeleton{\tyEnv}{\forall \evar. \ety} }
\end{mathpar}
\end{minipage}\end{shaded}\end{center}
The only noticeable difference between this judgement and the corresponding for
\source skeletons, is captured in the last rule. We have opted for a
\SystemF-based skeleton structure, thus this relation is identical to the
well-formedness of \SystemF types.

\paragraph{Coercion Typing}

\begin{myfigure}[t]
$\ruleform{\tcTrgCo{\tyEnv}{\coercion}{\trgFullCt}}$ \textbf{Coercion Typing}
\begin{mathpar}
\inferrule*[right=]
           { (\covar : \trgPartialCt) \in \tyEnv }
           { \tcTrgCo{\tyEnv}{\covar}{\trgPartialCt} }

\inferrule*[right=]
           { \tcTrgVty{\tyEnv}{\tyvar}{\ety} }
           { \tcTrgCo{\tyEnv}{\refl{\tyvar}}{\tyvar \le \tyvar} }

\inferrule*[right=]
           { \wfDirt{\tyEnv}{\dirt} }
           { \tcTrgCo{\tyEnv}{\dirtRefl{\dirt}}{\dirt \le \dirt} }

\inferrule*[right=]
           { }
           { \tcTrgCo{\tyEnv}{\trgUnitRefl}{\tyUnit \le \tyUnit} }
%
%
%

\inferrule*[right=]
           { \tcTrgCo{\tyEnv}{\coercion_1}{\vty_2 \le \vty_1} \\
             \tcTrgCo{\tyEnv}{\coercion_2}{\cty_1 \le \cty_2}
           } 
           { \tcTrgCo{\tyEnv}{\coercion_1 \to \coercion_2}{\vty_1 \to \cty_1 \le \vty_2 \to \cty_2} }
%
%

\inferrule*[right=]
           { \tcTrgCo{\tyEnv}{\coercion_1}{\cty_3 \le \cty_1} \\
             \tcTrgCo{\tyEnv}{\coercion_2}{\cty_2 \le \cty_4}
           } 
           { \tcTrgCo{\tyEnv}{\coercion_1 \hto \coercion_2}{\cty_1 \hto \cty_2 \le \cty_3 \hto \cty_4} }
%
%

\inferrule*[right=]
           { \tcTrgCo{\tyEnv, \evar}{\coercion}{\vty_1 \le \vty_2} }
           { \tcTrgCo{\tyEnv}{\forall \evar. \coercion}{\forall \evar. \vty_1 \le \forall \evar. \vty_2} }
%

\inferrule*[right=]
           { \tcTrgCo{\tyEnv, \tyvar : \ety}{\coercion}{\vty_1 \le \vty_2} }
           { \tcTrgCo{\tyEnv}{\forall \tyvar : \ety. \coercion}{\forall \tyvar : \ety. \vty_1 \le \forall \tyvar : \ety. \vty_2} }
%

\inferrule*[right=]
           { \tcTrgCo{\tyEnv, \dirtvar}{\coercion}{\vty_1 \le \vty_2} }
           { \tcTrgCo{\tyEnv}{\forall \dirtvar. \coercion}{\forall \dirtvar. \vty_1 \le \forall \dirtvar. \vty_2} }
%

\inferrule*[right=]
           { \tcTrgCo{\tyEnv}{\coercion}{\vty_1 \le \vty_2} \\
             \wfTrgCt{\tyEnv}{\trgPartialCt}
           } 
           { \tcTrgCo{\tyEnv}{\trgPartialCt \Rightarrow \coercion}{\trgPartialCt \Rightarrow \vty_1 \le \trgPartialCt \Rightarrow \vty_2} }
%

\inferrule*[right=]
           { \wfDirt{\tyEnv}{\dirt} }
           { \tcTrgCo{\tyEnv}{\emptyset_\dirt}{\emptyset \le \dirt} }

\inferrule*[right=]
           { \tcTrgCo{\tyEnv}{\coercion_1}{\vty_1 \le \vty_2} \\
             \tcTrgCo{\tyEnv}{\coercion_2}{\dirt_1 \le \dirt_2}
           } 
           { \tcTrgCo{\tyEnv}{\coercion_1~!~\coercion_2}{\vty_1~!~\dirt_1 \le \vty_2~!~\dirt_2} }
%
%

\inferrule*[right=]
           { \tcTrgCo{\tyEnv}{\coercion}{\dirt_1 \le \dirt_2} \\
             (\op : \vty_1 \to \vty_2) \in \sig
           } 
           { \tcTrgCo{\tyEnv}{\{\op\} \cup \coercion}{\{\op\} \cup \dirt_1 \le \{\op\} \cup \dirt_2} }
\end{mathpar}

\vspace{-5mm}
\caption{\target Coercion Typing}
\label{fig:target-coercion-typing}
\end{myfigure}

Coercion typing is presented in Figure~\ref{fig:target-coercion-typing} and
formalizes the intuitive interpretation of coercions we gave in
Section~\ref{subsec:target-syntax}.


\begin{myfigure}[t]
$\ruleform{\smallStepVal{v}{v'}}$ \textbf{Values}
\begin{mathpar}
\inferrule*[right=]
           { \smallStepVal{v}{v'} }
           { \smallStepVal{\cast{v}{\coercion}}{\cast{v'}{\coercion}} }

\inferrule*[right=]
           {}
           { \smallStepVal{\cast{\valRes}{\trgUnitRefl}}{\valRes} }

\inferrule*[right=]
           { \smallStepVal{v}{v'} }
           { \smallStepVal{v~\ety}{v'~\ety} }

\inferrule*[right=]
           { \smallStepVal{v}{v'} }
           { \smallStepVal{v~\vty}{v'~\vty} }

\inferrule*[right=]
           { \smallStepVal{v}{v'} }
           { \smallStepVal{v~\dirt}{v'~\dirt} }

\inferrule*[right=]
           { \smallStepVal{v}{v'} }
           { \smallStepVal{v~\coercion}{v'~\coercion} }

\inferrule*[right=]
           {}
           { \smallStepVal{(\cast{\valRes}{(\forall \evar. \coercion)})~\ety}{\cast{\valRes~\ety}{\coercion[\ety/\evar]}} }

\inferrule*[right=]
           {}
           { \smallStepVal{(\cast{\valRes}{(\forall (\tyvar : \ety). \coercion)})~\vty}{\cast{\valRes~\vty}{\coercion[\vty/\tyvar]}} }

\inferrule*[right=]
           {}
           { \smallStepVal{(\cast{\valRes}{(\forall \dirtvar. \coercion)})~\dirt}{\cast{\valRes~\dirt}{\coercion[\dirt/\dirtvar]}} }

\inferrule*[right=]
           {}
           { \smallStepVal{(\cast{\valRes}{(\trgPartialCt \Rightarrow \coercion_1)})~\coercion_2}{\cast{\valRes~\coercion_2}{\coercion_1}} }

\inferrule*[right=]
           {}
           { \smallStepVal{(\Lambda \evar. v)~\ety}{v[\ety/\evar]} }

\inferrule*[right=]
           {}
           {\smallStepVal{(\Lambda \tyvar : \ety. v)~\vty}{v[\vty/\tyvar]}}

\inferrule*[right=]
           {}
           {\smallStepVal{(\Lambda \dirtvar. v)~\dirt}{v[\dirt/\dirtvar]}}

\inferrule*[right=]
           {}
           {\smallStepVal{(\Lambda (\covar : \trgPartialCt). v)~\coercion}{v[\coercion/\covar]}}
\end{mathpar}

\vspace{-5mm}
\caption{\target Operational Semantics (Values)}
\label{fig:target-opsem-values}
\end{myfigure}

\begin{myfigure}[t]
$\ruleform{\smallStepComp{c}{c'}}$ \textbf{Computations}
\begin{mathpar}
\inferrule*[right=]
           { \smallStepComp{c}{c'} }
           { \smallStepComp{\cast{c}{\coercion}}{\cast{c'}{\coercion}} }

\inferrule*[right=]
           { \smallStepVal{v_1}{v_1'} }
           { \smallStepComp{v_1~v_2}{v_1'~v_2} }

\inferrule*[right=]
           {}
           { \smallStepComp{(\cast{\valRes}{(\coercion_1 \to \coercion_2)})~v}
                           {\cast{(\valRes~(\cast{v}{\coercion_1}))}{\coercion_2}}
           } 

\inferrule*[right=]
           { \smallStepVal{v_2}{v_2'} }
           { \smallStepComp{\termVal~v_2}{\termVal~v_2'} }

\inferrule*[right=]
           {}
           { \smallStepComp{(\fun{x : \vty}{c})~\valRes}{c[\valRes/x]} }

\inferrule*[right=]
  { \smallStepVal{v}{v'} }
  { \smallStepComp{\letval{x}{v}{c}}{\letval{x}{v'}{c}} }

\inferrule*[right=]
           {}
           { \smallStepComp{\letval{x}{\valRes}{c}}{c[\valRes/x]} }

\inferrule*[right=]
  { \smallStepVal{v}{v'} }
  { \smallStepComp{\return{v}}{\return{v'}} }

\inferrule*[right=]
  { \smallStepVal{v}{v'} }
  { \smallStepComp{\operation{v}{y : \vty}{c}}{\operation{v'}{y : \vty}{c}} }

\inferrule*[right=]
           {}
           { \smallStepComp{\cast{(\operation{\valRes}{x : \vty}{c})}{\coercion}}
                           {\operation{\valRes}{x : \vty}{(\cast{c}{\coercion})}}
           } 

\inferrule*[right=]
  { \smallStepComp{c_1}{c_1'} }
  { \smallStepComp{\doin{x}{c_1}{c_2}}{\doin{x}{c_1'}{c_2}} }

\inferrule*[right=]
           {} 
           { \smallStepComp{\doin{x}
                                 {(\cast{\cast{\cast{(\return{\valRes})}{(\coercion_1~!~\coercion_1')}}{\ldots}}{(\coercion_n~!~\coercion_n')})}
                                 {c_2}}
                           {c_2[(\cast{\cast{\cast{\valRes}{\coercion_1}}{\ldots}}{\coercion_n})/x]} }

\inferrule*[right=]
  {}
  { \smallStepComp
       {\doin{x}{\operation{\valRes}{y : \vty}{c_1}}{c_2}}
       {\operation{\valRes}{y : \vty}{\doin{x}{c_1}{c_2}}}
  } 

\inferrule*[right=]
  { \smallStepVal{v}{v'} }
  { \smallStepComp{\withhandle{v}{c}}{\withhandle{v'}{c}} }

\inferrule*[right=]
           {}
           { \smallStepComp{\withhandle{(\cast{\valRes}{(\coercion_1 \hto \coercion_2)})}{c}}
                           {\cast{(\withhandle{\valRes}{(\cast{c}{\coercion_1})})}{\coercion_2}}
           } 

\inferrule*[right=]
  { \smallStepComp{c}{c'} }
  { \smallStepComp{\withhandle{\termVal}{c}}{\withhandle{\termVal}{c'}} }

\inferrule*[right=]
  {} 
  { \smallStepComp{\withhandle{h}{(\cast{\cast{\cast{(\return{\valRes})}{(\coercion_1~!~\coercion_1')}}{\ldots}}{(\coercion_n~!~\coercion_n')})}}
                   {c_r[(\cast{\cast{\cast{\valRes}{\coercion_1}}{\ldots}}{\coercion_n})/ x]}
  } 

\inferrule*[right=]
  {}
  { \smallStepComp{
      \withhandle{h}{(\operation{\valRes}{y : \vty}{c})}
    }{
      c_{\op}[\valRes / x, (\fun{(y : \vty)}{\withhandle{h}{c}}) / k]
    }
  }

\inferrule*[right=]
  {}
  { \smallStepComp{
      \withhandle{h}{(\operation{\valRes}{y : \vty}{c})}
    }{
      \operation{\valRes}{y : \vty}{\withhandle{h}{c}}
    }
  }
\end{mathpar}

\vspace{-5mm}
\caption{\target Operational Semantics (Computations)}
\label{fig:target-opsem-computations}
\end{myfigure}

\paragraph{Reflexivity of Arbitrary Types}

Function $\reflOf{\cdot}$ below shows how to create a reflexivity coercion for
an arbitrary value type, computation type, or dirt:
\begin{center}\begin{shaded}\begin{minipage}{\columnwidth}
\small
\vspace{-1mm}
\[
\begin{array}{l@{~}c@{~}l@{\qquad}l@{~}c@{~}l}
  \reflOf{\tyvar}                         & = & \refl{\tyvar}                           & \reflOf{\vty~!~\dirt} & = & \reflOf{\vty}~!~\reflOf{\dirt} \\
  \reflOf{\tyUnit}                        & = & \trgUnitRefl                            &  &  &  \\
  \reflOf{\vty \to \cty}                  & = & \reflOf{\vty} \to \reflOf{\cty}         &  &  &  \\
  \reflOf{\cty_1 \hto \cty_2}             & = & \reflOf{\cty_1} \hto \reflOf{\cty_2}    &  &  &  \\
  \reflOf{\forall \evar. \vty}            & = & \forall \evar. \reflOf{\vty}            &  &  &  \\
  \reflOf{\forall \tyvar : \ety. \vty}    & = & \forall \tyvar : \ety. \reflOf{\vty}    & \reflOf{\dirtvar}  & = & \dirtRefl{\dirtvar} \\
  \reflOf{\forall \dirtvar. \vty}         & = & \forall \dirtvar. \reflOf{\vty}         & \reflOf{\emptyset} & = & \emptyset_\emptyset \\
  \reflOf{\trgPartialCt \Rightarrow \vty} & = & \trgPartialCt \Rightarrow \reflOf{\vty} & \reflOf{\{ \op \} \cup \dirt} & = & \{ \op \} \cup \reflOf{\dirt} \\
\end{array}
\]
\end{minipage}\end{shaded}\end{center}

\paragraph{Operational Semantics}

The complete small-step, call-by-value operational semantics for \target can be
found in Figures~\ref{fig:target-opsem-values} (values)
and~\ref{fig:target-opsem-computations} (computations).

\section{Type Inference \& Elaboration: Additional Judgements}\label{appendix:inference-additional}

\paragraph{Elaboration of Types, Constraints, and Typing Environments}


Below we give the definitions of elaboration functions $\elabVty{\polyTy}$,
$\elabCty{\cCty}$, $\elabConstraint{\srcFullCt}$, and $\elabTyEnv{\tyEnv}$, for
value types, computation types, constraints, and typing environments.
\begin{center}\begin{shaded}\begin{minipage}{\columnwidth}
\vspace{-3mm}
\small
\[
\begin{array}{@{\hspace{0mm}}c@{\hspace{10mm}}c@{\hspace{0mm}}}
  \begin{array}{@{\hspace{0mm}}lcl@{\hspace{0mm}}}
    \elabVty{\tyvar}                            & = & \tyvar \\
    \elabVty{\aVty \to \cCty}                   & = & \elabVty{\aVty} \to  \elabCty{\cCty} \\
    \elabVty{\cCty \hto \dCty}                  & = & \elabCty{\cCty} \hto \elabCty{\dCty} \\
    \elabVty{\tyUnit}                           & = & \tyUnit \\
    \elabVty{\forall \evar. \polyTy}            & = & \forall \evar. \elabVty{\polyTy} \\
    \elabVty{\forall \tyvar : \ety. \polyTy}    & = & \forall \tyvar : \ety. \elabVty{\polyTy} \\
    \elabVty{\forall \dirtvar. \polyTy}         & = & \forall \dirtvar. \elabVty{\polyTy} \\
    \elabVty{\srcPartialCt \Rightarrow \qualTy} & = & \elabConstraint{\srcPartialCt} \Rightarrow \elabVty{\qualTy} \\
  \end{array}
&
  \begin{array}{@{\hspace{0mm}}lcl@{\hspace{0mm}}}
    \elabCty{\aVty\,!\,\dirt}                   & = & \elabVty{\aVty}\,!\,\dirt \\
    \\
    \elabTyEnv{\epsilon}                        & = & \epsilon \\
    \elabTyEnv{\tyEnv, \evar}                   & = & \elabTyEnv{\tyEnv}, \evar \\
    \elabTyEnv{\tyEnv, \tyvar : \ety}           & = & \elabTyEnv{\tyEnv}, \tyvar : \ety \\
    \elabTyEnv{\tyEnv, \dirtvar}                & = & \elabTyEnv{\tyEnv}, \dirtvar \\
    \elabTyEnv{\tyEnv, x : \polyTy}             & = & \elabTyEnv{\tyEnv}, x : \elabVty{\polyTy} \\
    \elabTyEnv{\tyEnv, \covar : \srcFullCt}     & = & \elabTyEnv{\tyEnv}, \covar : \elabConstraint{\srcFullCt} \\
  \end{array}
\end{array}
\]
\[
\begin{array}{lcl}
  \elabConstraint{\aVty \le \bVty} & = & \elabVty{\aVty} \le \elabVty{\bVty} \\
  \elabConstraint{\cCty \le \dCty} & = & \elabCty{\cCty} \le \elabCty{\dCty} \\
  \elabConstraint{\dirt_1 \le \dirt_2} & = & \dirt_1 \le \dirt_2 \\
\end{array}
\]
\end{minipage}\end{shaded}\end{center}
All four are entirely straightforward and essentially traverse each sort, so
that \source value types $\aVty$ are replaced with \target value types $\vty$.

\paragraph{Skeleton Extraction}

In Section~\ref{sec:inference} we made use of function $\skeletonOf{\aVty}$,
which computes the skeleton of a type. Its formal definition is given below:
\begin{center}\begin{shaded}\begin{minipage}{\columnwidth}
\vspace{-1mm}
\small
\[
\begin{array}{@{\hspace{0mm}}lcl@{\hspace{0mm}}}
  \skeletonOf{\tyvar^\ety}                              & = & \ety    \\
  \skeletonOf{\tyUnit}                                  & = & \tyUnit \\
  \skeletonOf{\aVty \to \bVty\,!\,\dirt}                & = & \skeletonOf{\aVty} \to  \skeletonOf{\bVty} \\
  \skeletonOf{\aVty\,!\,\dirt_1 \hto \bVty\,!\,\dirt_2} & = & \skeletonOf{\aVty} \hto \skeletonOf{\bVty} \\
\end{array}
\]
\vspace{-3mm}
\end{minipage}\end{shaded}\end{center}
A skeleton of a type captures its structure (modulo the dirt information),
which is directly expressed in clauses 2, 3, and 4. Hence, in order to capture
the whole skeleton of a type, the only missing piece of information is the
skeleton of all type variables appearing in the type.

As we mentioned in passing in Section~\ref{subsec:constraint-solving}, each type variable is implicitly annotated
with its skeleton, which allows for the complete determination of the skeleton
of a type (clause 1).

\section{\erased Additional Judgements}\label{appendix:erased-additional}

\paragraph{Typing}

\begin{myfigure}[t]
\[
\begin{array}{r@{~}c@{~}l}
  \text{typing environment}~\ctx & ::= & \epsilon                             \mid
                                         \tyEnv, \evar                       \mid
                                         \tyEnv, x : \ety                     \\
\end{array}
\]
$\ruleform{\tcErsVal{\tyEnv}{v}{\ety}}$ \textbf{Values}
\begin{mathpar}
\inferrule*[right=]
           { (x : \ety) \in \tyEnv }
           { \tcErsVal{\tyEnv}{x}{\ety} }

\inferrule*[right=]
           { }
           { \tcErsVal{\tyEnv}{\tmUnit}{\tyUnit} }

\inferrule*[right=]
           { \tcErsComp{\tyEnv, x : \ety_1}{c}{\ety_2} \\
             \tcErsEty{\tyEnv}{\ety_1}
           } 
           { \tcErsVal{\tyEnv}{(\fun{x:\ety_1}{c})}{\ety_1 \to \ety_2} }

\inferrule*[right=]
           { \tcErsVal{\tyEnv}{v}{\forall \evar. \ety_1} \\
             \tcErsEty{\tyEnv}{\ety_2}
           } %
           { \tcErsVal{\tyEnv}{v~\ety_2}{\ety_1[\ety_2/\evar]} }

\inferrule*[right=]
           { \tcErsVal{\tyEnv, \evar}{v}{\ety} }
           { \tcErsVal{\tyEnv}{\Lambda \evar. v}{\forall \evar.\ety} }

\inferrule*[right=]
           { \tcErsComp{\tyEnv, x : \ety_x}{c_r}{\ety} \\
             \left[
               (\op : \ety_1 \to \ety_2) \in \sig \qquad
               \tcErsComp{\tyEnv, x : \ety_1, k : \ety_2 \to \ety}{c_\op}{\ety}
             \right]_{\op \in \ops}
           } 
           { \tcErsVal{\tyEnv}{\ersShorthand}{\ety_x \hto \ety} }
\end{mathpar}
$\ruleform{\tcErsComp{\tyEnv}{c}{\ety}}$ \textbf{Computations}
\begin{mathpar}
\inferrule*[right=]
           { \tcErsVal{\tyEnv}{v_1}{\ety_1 \to \ety_2} \\
             \tcErsVal{\tyEnv}{v_2}{\ety_1}
           } 
           { \tcErsComp{\tyEnv}{v_1~v_2}{\ety_2} }

\inferrule*[right=]
           { \tcErsVal{\tyEnv}{v}{\ety_1} \\
             \tcErsComp{\tyEnv, x : \ety_1}{c}{\ety_2}
           } 
           { \tcErsComp{\tyEnv}{\letval{x}{v}{c}}{\ety_2} }

\inferrule*[right=]
           { \tcErsVal{\tyEnv}{v}{\ety} }
           { \tcErsComp{\tyEnv}{\return{v}}{\ety} }

\inferrule*[right=]
           { (\op : \ety_1 \to \ety_2) \in \sig \\
             \tcErsVal{\tyEnv}{v}{\ety_1} \\
             \tcErsComp{\tyEnv, y : \ety_2}{c}{\ety}
           } 
           { \tcErsComp{\tyEnv}{\operation{v}{y:\ety_2}{c}}{\ety} }

\inferrule*[right=]
           { \tcErsComp{\tyEnv}{c_1}{\ety_1} \\
             \tcErsComp{\tyEnv, x : \ety_1}{c_2}{\ety_2}
           } 
           { \tcErsComp{\tyEnv}{\doin{x}{c_1}{c_2}}{\ety_2} }

\inferrule*[right=]
           { \tcErsVal{\tyEnv}{v}{\ety_1 \hto \ety_2} \\
             \tcErsComp{\tyEnv}{c}{\ety_1}
           } 
           { \tcErsComp{\tyEnv}{\withhandle{v}{c}}{\ety_2} }
\end{mathpar}

\vspace{-5mm}
\caption{\erased Typing}
\label{fig:erased-typing}
\end{myfigure}

Typing for \erased values and computations is given is
Figure~\ref{fig:erased-typing}. As illustrated by the rules, \erased is
essentially \SystemF extended with term-level (but not type-level) support for
algebraic effects.

\paragraph{Operational Semantics}

\begin{myfigure}[t]
\[
\begin{array}{r@{~}c@{~}l}
  \text{value result}~\valRes & ::= &%
    \tmUnit                                   \mid
    h                                         \mid
    \fun{(x:\ety)}{c}                         \mid
    \Lambda \evar. v                         \\

  \text{computation result}~\compRes & ::= &%
    \return{\valRes}                   \mid
    \operation{\valRes}{y}{c}          \\
\end{array}
\]
$\ruleform{\smallStepVal{v}{v'}}$ \textbf{Values}
\begin{mathpar}
\inferrule
  {\smallStepVal{v}{v'}}
  {\smallStepVal{v~\ety}{v'~\ety}}

\inferrule
  {}
  {\smallStepVal{(\Lambda \evar. v)~\ety}{v[\ety/\evar]}}

\end{mathpar}

$\ruleform{\smallStepComp{c}{c'}}$ \textbf{Computations}
\begin{mathpar}
\inferrule
  { \smallStepVal{v_1}{v_1'} }
  { \smallStepComp{v_1~v_2}{v_1'~v_2} }

\inferrule
  { \smallStepVal{v_2}{v_2'} }
  { \smallStepComp{\valRes_1~v_2}{\valRes_1~v_2'} }

\inferrule
  {}
  { \smallStepComp{
      (\fun{(x:\ety)} c)~\valRes
    }{
      c[\valRes / x]
    }
  }

\inferrule
  { \smallStepVal{v}{v'} }
  { \smallStepComp{\letval{x}{v}{c}}{\letval{x}{v'}{c}} }

\inferrule
  {}
  { \smallStepComp{
      \letval{x}{\valRes}{c}
    }{
      c[\valRes / x]
    }
  }

\inferrule
  { \smallStepVal{v}{v'} }
  { \smallStepComp{\return{v}}{\return{v'}} }

\inferrule
  { \smallStepVal{v}{v'} }
  { \smallStepComp{\operation{v}{y:\ety}{c}}{\operation{v'}{y:\ety}{c}} }

\inferrule
  { \smallStepComp{c_1}{c_1'} }
  { \smallStepComp{\doin{x}{c_1}{c_2}}{\doin{x}{c_1'}{c_2}} }

\inferrule
  {}
  { \smallStepComp{
      \doin{x}{\return{\valRes}}{c_2}
    }{
      c_2[\valRes / x]
    }
  }

\inferrule
  {}
  { \smallStepComp{
      \doin{x}{\operation{\valRes}{y:\ety}{c_1}}{c_2}
    }{
      \operation{\valRes}{y:\ety}{\doin{x}{c_1}{c_2}}
    }
  }

\inferrule
  { \smallStepVal{v}{v'} }
  { \smallStepComp{\withhandle{v}{c}}{\withhandle{v'}{c}} }

\inferrule
  { \smallStepComp{c}{c'} }
  { \smallStepComp{\withhandle{\valRes}{c}}{\withhandle{\valRes}{c'}} }

\inferrule
  {}
  { \smallStepComp{
      \withhandle{h}{(\return{\valRes})}
    }{
      c_r[\valRes / x]
    }
  }

\inferrule
  {}
  { \smallStepComp{
      \withhandle{h}{(\operation{\valRes}{y:\ety}{c})}
    }{
      c_{\op}[\valRes / x, (\fun{(y : \ety)}{\withhandle{h}{c}}) / k]
    }
  }

\inferrule
  {}
  { \smallStepComp{
      \withhandle{h}{(\operation{\valRes}{y : \ety}{c})}
    }{
      \operation{\valRes}{y : \ety}{\withhandle{h}{c}}
    }
  }

\end{mathpar}

\vspace{-5mm}
\caption{\erased Operational Semantics}
\label{fig:erased-opsem}
\end{myfigure}

\begin{myfigure}[t]
\newcommand{\vvctx}{V^\mathrm{v}}
\newcommand{\vhctx}{H^\mathrm{v}}
\newcommand{\vcctx}{C^\mathrm{v}}
\newcommand{\cvctx}{V^\mathrm{c}}
\newcommand{\chctx}{H^\mathrm{c}}
\newcommand{\ccctx}{C^\mathrm{c}}
\textbf{Terms with holes}
\[
\begin{array}{r@{~}c@{~}l}
    \text{value-holed value}~\vvctx & ::= & [\,] \mid x                                         \mid
                           \tmUnit                                   \mid
                           \vhctx                                         \mid
                           \fun{(x:\ety)}{\vcctx}                         \mid
                           \Lambda \evar. \vvctx                         \mid
                           \vvctx~\ety                           \\

    \text{value-holed handler}~\vhctx & ::= & \{ \return{(x : \ety)} \mapsto \vcctx_r, [\call{\op}{x}{k} \mapsto \vcctx_\op]_{\op \in \ops} \} \\

    \text{value-holed computation}~\vcctx & ::=  & \vvctx_1~\vvctx_2                          \mid
                                  \letval{x}{\vvctx}{\vcctx}                 \mid
                                  \return{\vvctx}                       \\
                         & \mid & \operation{\vvctx}{y:\ety}{\vcctx}         \mid
                                  \doin{x}{\vcctx_1}{\vcctx_2}               \\ &\mid&
                                  \withhandle{\vvctx}{\vcctx}                \\
    \\
    \text{computation-holed value}~\cvctx & ::= & x                                         \mid
                           \tmUnit                                   \mid
                           \chctx                                         \mid
                           \fun{(x:\ety)}{\ccctx}                         \mid
                           \Lambda \evar. \cvctx                         \mid
                           \cvctx~\ety                           \\

    \text{computation-holed handler}~\chctx & ::= & \{ \return{(x : \ety)} \mapsto \ccctx_r, [\call{\op}{x}{k} \mapsto \ccctx_\op]_{\op \in \ops} \} \\

    \text{computation-holed computation}~\ccctx & ::=  & [\,] \mid \cvctx_1~\cvctx_2                          \mid
                                  \letval{x}{\cvctx}{\ccctx}                 \mid
                                  \return{\cvctx}                       \\
                         & \mid & \operation{\cvctx}{y:\ety}{\ccctx}         \mid
                                  \doin{x}{\ccctx_1}{\ccctx_2}               \\ &\mid&
                                  \withhandle{\cvctx}{\ccctx}                \\
\end{array}
\]
We define values $\vvctx[v], \cvctx[c]$, and computations $\vcctx[v], \ccctx[c]$ in the obvious way.
\medskip

$\ruleform{\congStepVal{v}{v'}}$ \textbf{Values}
\begin{mathpar}
\inferrule
  {\smallStepVal{v}{v'}}
  {\congStepVal{v}{v'}}

\inferrule
  {}
  {\congStepVal{v}{v}}

\inferrule
  {\congStepVal{v}{v'}}
  {\congStepVal{v'}{v}}

\inferrule
  {\congStepVal{v}{v'} \and \congStepVal{v'}{v''}}
  {\congStepVal{v}{v''}}
\\
\inferrule
  {\congStepVal{v}{v'}}
  {\congStepVal{\vvctx[v]}{\vvctx[v']}}

\inferrule
  {\congStepComp{c}{c'}}
  {\congStepVal{\cvctx[c]}{\cvctx[c']}}
\end{mathpar}

$\ruleform{\congStepComp{c}{c'}}$ \textbf{Computations}
\begin{mathpar}
\inferrule
  {\smallStepComp{c}{c'}}
  {\congStepComp{c}{c'}}

\inferrule
  {}
  {\congStepComp{c}{c}}

\inferrule
  {\congStepComp{c}{c'}}
  {\congStepComp{c'}{c}}

\inferrule
  {\congStepComp{c}{c'} \and \congStepComp{c'}{c''}}
  {\congStepComp{c}{c''}}
\\
\inferrule
  {\congStepVal{v}{v'}}
  {\congStepComp{\vcctx[v]}{\vcctx[v']}}

\inferrule
  {\congStepComp{c}{c'}}
  {\congStepComp{\ccctx[c]}{\ccctx[c']}}
\end{mathpar}

\vspace{-5mm}
\caption{Congruence Closures of the Step Relations}
\label{fig:erased-cong}
\end{myfigure}

Figure~\ref{fig:erased-opsem} presents the small-step, call-by-value
operational semantics of \erased, and Figure~\ref{fig:erased-cong}
gives the congruence closure of the step relations as used in Theorem~\ref{thm:erasure_semantic_preservation}.

\section{\noeff Additional Judgements}\label{appendix:ml-additional}
\paragraph{Type Well-formedness}

Well-formedness for \noeff types is given by judgement
$\tcNoEffTy{\mlEnv}{\mlTyA}$, which is given by the following rules:
\begin{center}\begin{shaded}\begin{minipage}{\columnwidth}
\vspace{-4mm}
\small
\begin{mathpar}
\inferrule*[right=]
           { \tyvar \in \mlEnv }
           { \tcNoEffTy{\mlEnv}{\tyvar} }
\quad
\inferrule*[right=]
           { }
           { \tcNoEffTy{\mlEnv}{\tyUnit} }
\quad
\inferrule*[right=]
           { \tcNoEffTy{\mlEnv}{\mlTyA} \\
             \tcNoEffTy{\mlEnv}{\mlTyB}
           } 
           { \tcNoEffTy{\mlEnv}{\mlTyA \to \mlTyB} }
\quad
\inferrule*[right=]
           { \tcNoEffTy{\mlEnv, \tyvar}{\mlTyA} }
           { \tcNoEffTy{\mlEnv}{\forall \tyvar. \mlTyA} }
\quad
\inferrule*[right=]
           { \tcNoEffTy{\mlEnv}{\mlTyA} \\
             \tcNoEffTy{\mlEnv}{\mlTyB}
           } 
           { \tcNoEffTy{\mlEnv}{\mlTyA \hto \mlTyB} }
\quad
\inferrule*[right=]
           { \tcNoEffCoTy{\mlEnv}{\mlCoTy} \\
             \tcNoEffTy{\mlEnv}{\mlTyA}
           } 
           { \tcNoEffTy{\mlEnv}{\mlCoTy \Rightarrow \mlTyA} }
\end{mathpar}
\end{minipage}\end{shaded}\end{center}
Since \noeff is uni-kinded, the rules simply ensure that types are well-scoped.

\paragraph{Constraint Well-formedness}

Well-formedness for \noeff constraints takes the form
$\tcNoEffCoTy{\mlEnv}{\mlCoTy}$, and is given by the following rule:
\begin{center}\begin{shaded}\begin{minipage}{\columnwidth}
\vspace{-4mm}
\small
\begin{mathpar}
\inferrule*[right=]
           { \tcNoEffTy{\mlEnv}{\mlTyA} \\
             \tcNoEffTy{\mlEnv}{\mlTyB}
           } 
           { \tcNoEffCoTy{\mlEnv}{\mlTyA \le \mlTyB} }
\end{mathpar}
\end{minipage}\end{shaded}\end{center}
Though very similar to the corresponding one for \target, since \noeff features
no skeletons (or kinds), the above rule simply ensures that the types appearing
in the constraint are both well-scoped.

\paragraph{Operational Semantics}

\begin{myfigure}[t]
$\ruleform{\mlSmallStep{\mlTm}{\mlTm'}}$ \textbf{Operational Semantics}
\begin{mathpar}
\inferrule*[right=]
           { \mlSmallStep{\mlTm_1}{\mlTm_1'} }
           { \mlSmallStep{\mlTm_1~\mlTm_2}{\mlTm_1'~\mlTm_2} }

\inferrule*[right=]
           { \mlSmallStep{\mlTm}{\mlTm'} }
           { \mlSmallStep{\mlValue~\mlTm}{\mlValue~\mlTm'} }

\inferrule*[right=]
           {}
           { \mlSmallStep{(\fun{x:\mlTyA}{\mlTm})~\mlValue}{\mlTm[\mlValue/x]} }

\inferrule*[right=]
           { \mlSmallStep{\mlTm}{\mlTm'} }
           { \mlSmallStep{\mlTm~\mlTyA}{\mlTm'~\mlTyA} }

\inferrule*[right=]
           {}
           { \mlSmallStep{(\Lambda \tyvar. \mlTm)~\mlTyA}{\mlTm[\mlTyA/\tyvar]} }

\inferrule*[right=]
           { \mlSmallStep{\mlTm}{\mlTm'} }
           { \mlSmallStep{\mlTm~\mlCoercion}{\mlTm'~\mlCoercion} }

\inferrule*[right=]
           {}
           { \mlSmallStep{(\Lambda (\covar : \mlCoTy). \mlTm)~\mlCoercion}{\mlTm[\mlCoercion/\covar]} }

\inferrule*[right=]
           { \mlSmallStep{\mlTm_1}{\mlTm_1'} }
           { \mlSmallStep{\letval{x}{\mlTm_1}{\mlTm_2}}{\letval{x}{\mlTm_1'}{\mlTm_2}} }

\inferrule*[right=]
           {}
           { \mlSmallStep{\letval{x}{\mlValue}{\mlTm}}{\mlTm[\mlValue/x]} }

\inferrule*[right=]
           { \mlSmallStep{\mlTm}{\mlTm'} }
           { \mlSmallStep{\mlReturn{\mlTm}}{\mlReturn{\mlTm'}} }

\inferrule*[right=]
           { \mlSmallStep{\mlTm_1}{\mlTm_1'} }
           { \mlSmallStep{\operation{\mlTm_1}{y : \mlTyB}{\mlTm_2}}{\operation{\mlTm_1'}{y : \mlTyB}{\mlTm_2}} }

\inferrule*[right=]
           { \mlSmallStep{\mlTm_1}{\mlTm_1'} }
           { \mlSmallStep{\doin{x}{\mlTm_1}{\mlTm_2}}{\doin{x}{\mlTm_1'}{\mlTm_2}} }

\inferrule*[right=]
           {}
           { \mlSmallStep{\doin{x}{\mlReturn{\mlValue}}{\mlTm}}{\mlTm[\mlValue/x]} }

\inferrule*[right=]
           {}
           { \mlSmallStep{\doin{x}{(\operation{\mlValue}{y : \mlTyA}{\mlTm_1})}{\mlTm_2}}
                         {\operation{\mlValue}{y : \mlTyA}{\doin{x}{\mlTm_1}{\mlTm_2}}}
           } 

\inferrule*[right=]
           { \mlSmallStep{\mlTm_h}{\mlTm_h'} }
           { \mlSmallStep{\withhandle{\mlTm_h}{\mlTm_c}}{\withhandle{\mlTm_h'}{\mlTm_c}} }

\inferrule*[right=]
           { \mlSmallStep{\mlTm_c}{\mlTm_c'} }
           { \mlSmallStep{\withhandle{\mlValue_h}{\mlTm_c}}{\withhandle{\mlValue_h}{\mlTm_c'}} }

\inferrule*[right=]
           {}
           { \mlSmallStep{\withhandle{\mlHandler}{(\mlReturn{\mlValue})}}{\mlTm_r[\mlValue/x]} }

\inferrule*[right=]
           { (\call{\op}{x}{k} \mapsto \mlTm_\op) \in h }
           { \mlSmallStep{\withhandle{\mlHandler}{(\operation{\mlValue}{y : \mlTyB}{\mlTm})}}
                         {\mlTm_{\op}[\mlValue / x, (\fun{(y : \mlTyB)}{\withhandle{\mlHandler}{\mlTm}}) / k]}
           } 

\inferrule*[right=]
           { (\call{\op}{x}{k} \mapsto \mlTm_\op) \notin h }
           { \mlSmallStep{\withhandle{\mlHandler}{(\operation{\mlValue}{y : \mlTyB}{\mlTm})}}
                         {\operation{\mlValue}{y : \mlTyB}{\withhandle{\mlHandler}{\mlTm}}}
           } 

\inferrule*[right=]
           { \mlSmallStep{\mlTm}{\mlTm'} }
           { \mlSmallStep{\cast{\mlTm}{\mlCoercion}}{\cast{\mlTm'}{\mlCoercion}} }

\inferrule*[right=]
           {}
           { \mlSmallStep{\cast{\mlValue}{\mlCoUnitRefl}}{\mlValue} }

\inferrule*[right=]
           {}
           { \mlSmallStep{\cast{(\mlReturn{\mlValue})}{(\mkMlCompCo{\mlCoercion})}}{\mlReturn{(\cast{\mlValue}{\mlCoercion})}} }

\inferrule*[right=]
           {}
           { \mlSmallStep{\cast{(\operation{\mlValue}{y : \mlTyB}{\mlTm})}{(\mkMlCompCo{\mlCoercion})}}
                         {\operation{\mlValue}{y : \mlTyB}{(\cast{\mlTm}{(\mkMlCompCo{\mlCoercion})})}}
           } 

\inferrule*[right=]
           {}
           { \mlSmallStep{\cast{\mlValue}{\mlReturn{\mlCoercion}}}{\mlReturn{(\cast{\mlValue}{\mlCoercion})}} }

\inferrule*[right=]
           {}
           { \mlSmallStep{\cast{(\mlReturn{\mlValue})}{(\unsafe{\mlCoercion})}}{\cast{\mlValue}{\mlCoercion}} }

\inferrule*[right=]
           {}
           { \mlSmallStep{(\cast{\mlValue}{(\mlCoercion_1 \to \mlCoercion_2)})~\mlTm}{\cast{(\mlValue~(\cast{\mlTm}{\mlCoercion_1}))}{\mlCoercion_2}} }

\inferrule*[right=]
           {} 
           { \mlSmallStep{\withhandle{(\cast{\mlValue_2}{(\mlCoercion_1 \hto \mlCoercion_2)})}{\mlValue_1}}
                         {\cast{(\withhandle{\mlValue_2}{(\cast{\mlValue_1}{\mlCoercion_1})})}{\mlCoercion_2}}
           } 

\inferrule*[right=]
           {}
           { \mlSmallStep{(\cast{\mlValue_1}{(\handToFun{\mlCoercion_1}{\mlCoercion_2})})~\mlValue_2}
                         {\cast{(\withhandle{\mlValue_1}{(\mlReturn{(\cast{\mlValue_2}{\mlCoercion_1})})})}{\mlCoercion_2}}
           } 

\inferrule*[right=]
           {}
           { {\withhandle{(\cast{\mlValue_2}{(\funToHand{\mlCoercion_1}{\mlCoercion_2})})}{(\operation{\mlValue_1}{y : \mlTyB}{\mlTm})} }
             \\ \leadsto
             {\operation{\mlValue_1}{y : \mlTyB}{\withhandle{(\cast{\mlValue_2}{(\funToHand{\mlCoercion_1}{\mlCoercion_2})})}{\mlTm}}}
           } 

\inferrule*[right=]
           {}
           { \mlSmallStep{\withhandle{(\cast{\mlValue_2}{(\funToHand{\mlCoercion_1}{\mlCoercion_2})})}{(\return{\mlValue_1})} }
                         {\cast{(\mlValue_2~(\cast{\mlValue_1}{\mlCoercion_1}))}{\mlCoercion_2}}
           } 

\inferrule*[right=]
           {}
           { \mlSmallStep{(\cast{\mlValue}{\forall \tyvar. \mlCoercion})~\mlTyA}{\cast{(\mlValue~\mlTyA)}{\mlCoercion[\mlTyA/\tyvar]}} }

\inferrule*[right=]
           {}
           { \mlSmallStep{(\cast{\mlValue}{(\mlCoTy \Rightarrow \mlCoercion_1)})~\mlCoercion_2}{\cast{(\mlValue~\mlCoercion_2)}{\mlCoercion_1}} }
\end{mathpar}

\vspace{-5mm}
\caption{\noeff Operational Semantics}
\label{fig:ml-opsem}
\end{myfigure}

The complete small-step operational semantics for \noeff are presented in
Figure~\ref{fig:ml-opsem}.

\section{\target to \noeff: Additional Judgements}\label{appendix:elab-additional}
\paragraph{Typing Environment Elaboration}

\begin{myfigure}[t]
$\ruleform{\tyEnvToNoEff{\tyEnv}{\mlEnv'}}$ \textbf{Typing Environment Elaboration}
\begin{mathpar}
\inferrule*[right=]
           { }
           { \tyEnvToNoEff{\epsilon}{\epsilon} }

\inferrule*[right=]
           { \tyEnvToNoEff{\tyEnv}{\mlEnv'} }
           { \tyEnvToNoEff{\tyEnv, \evar}{\mlEnv'} }

\inferrule*[right=]
           { \tyEnvToNoEff{\tyEnv}{\mlEnv'} }
           { \tyEnvToNoEff{\tyEnv, \tyvar : \ety}{\mlEnv', \tyvar} }

\inferrule*[right=]
           { \tyEnvToNoEff{\tyEnv}{\mlEnv'} }
           { \tyEnvToNoEff{\tyEnv, \dirtvar}{\mlEnv'} }

\inferrule*[right=]
           { \tyEnvToNoEff{\tyEnv}{\mlEnv'} \\
             \valTyToNoEff{\tyEnv}{\vty}{\ety}{\mlTyA}
           } 
           { \tyEnvToNoEff{\tyEnv, x : \vty}{\mlEnv, x : \mlTyA} }

\inferrule*[right=]
           { \tyEnvToNoEff{\tyEnv}{\mlEnv'} \\
             \valTyToNoEff{\tyEnv}{\vty_1}{\ety}{\mlTyA} \\
             \valTyToNoEff{\tyEnv}{\vty_2}{\ety}{\mlTyB}
           } 
           { \tyEnvToNoEff{\tyEnv, \covar : \vty_1 \le \vty_2}{\mlEnv', \covar : \mlTyA \le \mlTyB} }

\inferrule*[right=]
           { \tyEnvToNoEff{\tyEnv}{\mlEnv'} \\
             \compTyToNoEff{\tyEnv}{\cty_1}{\ety}{\mlTyA} \\
             \compTyToNoEff{\tyEnv}{\cty_2}{\ety}{\mlTyB}
           } 
           { \tyEnvToNoEff{\tyEnv, \covar : \cty_1 \le \cty_2}{\mlEnv', \covar : \mlTyA \le \mlTyB} }

\inferrule*[right=]
           { \tyEnvToNoEff{\tyEnv}{\mlEnv'} \\
             \wfDirt{\tyEnv}{\dirt_1} \\
             \wfDirt{\tyEnv}{\dirt_2}
           } 
           { \tyEnvToNoEff{\tyEnv, \covar : \dirt_1 \le \dirt_2}{\mlEnv'} }
\end{mathpar}

\vspace{-5mm}
\caption{Elaboration of \target Typing Environments to \noeff Typing Environments}
\label{fig:eff-to-ml-environments}
\end{myfigure}

Elaboration of typing environments is given in
Figure~\ref{fig:eff-to-ml-environments}. Essentially the judgement removes all
dirt and skeleton information is removed (including dirt inequalities).

\begin{myfigure}[t]
\small
$\ruleform{\coToNoEff{\tyEnv}{\coercion}{\trgPartialCt}{\mlCoercion'}}$ \textbf{Coercion Elaboration}
\begin{mathpar}
\inferrule*[right=]
           { (\covar : \trgPartialCt) \in \tyEnv }
           { \coToNoEff{\tyEnv}{\covar}{\trgPartialCt}{\covar} }

\inferrule*[right=]
           { }
           { \coToNoEff{\tyEnv}{\trgUnitRefl}{\tyUnit \le \tyUnit}{\mlCoUnitRefl} }

\inferrule*[right=]
           { (\tyvar : \ety) \in \tyEnv }
           { \coToNoEff{\tyEnv}{\refl{\tyvar}}{\tyvar \le \tyvar}{\refl{\tyvar}} }

\inferrule*[right=]
           { \coToNoEff{\tyEnv}{\coercion_1}{\vty_2 \le \vty_1}{\mlCoercion'_1} \\
             \coToNoEff{\tyEnv}{\coercion_2}{\cty_1 \le \cty_2}{\mlCoercion'_2}
           } 
           { \coToNoEff{\tyEnv}{\coercion_1 \to \coercion_2}{(\vty_1 \to \cty_1) \le (\vty_2 \to \cty_2)}{\mlCoercion'_1 \to \mlCoercion'_2} }

\inferrule*[right=]
           { \coToNoEff{\tyEnv}{\coercion_1}{\vty_2~!~\emptyset \le \vty_1~!~\emptyset}{\mlCoercion'_1} \\
             \coToNoEff{\tyEnv}{\coercion_2}{\cty_1 \le \cty_2}{\mlCoercion'_2}
           } 
           { \coToNoEff{\tyEnv}{\coercion_1 \hto \coercion_2}{(\vty_1~!~\emptyset \hto \cty_1) \le (\vty_2~!~\emptyset \hto \cty_2)}{\mlCoercion'_1 \to \mlCoercion'_2} }

\inferrule*[right=]
           { \fullDirt{\dirt_1} \\
             \fullDirt{\dirt_2} \\\\
             \coToNoEff{\tyEnv}{\coercion_1}{(\vty_2~!~\dirt_2 \le \vty_1~!~\dirt_1)}{\mlCoercion'_1} \\
             \coToNoEff{\tyEnv}{\coercion_2}{\vty'_1 \le \vty'_2}{\mlCoercion'_2} \\
             \tcTrgCo{\tyEnv}{\coercion_3}{\dirt'_1 \le \dirt'_2}
           } 
           { \coToNoEff{\tyEnv}
                       {(\coercion_1 \hto (\coercion_2~!~\coercion_3))}
                       {((\vty_1~!~\dirt_1) \hto (\vty'_1~!~\dirt'_1)) \le ((\vty_2~!~\dirt_2) \hto (\vty'_2~!~\dirt'_2))}
                       {\mlCoercion'_1 \hto \mkMlCompCo{\mlCoercion'_2}}
           } 

\inferrule*[right=]
           { \fullDirt{\dirt_1} \\
           	\coToNoEff{\tyEnv}{\coercion_1}{\vty_2 \le \vty_1}{\mlCoercion'_1} \\
             \coToNoEff{\tyEnv}{\coercion_2}{\vty'_1 \le \vty'_2}{\mlCoercion'_2}\\
             \tcTrgCo{\tyEnv}{\coercion_3}{\emptyset \le \dirt_1} \\
             \tcTrgCo{\tyEnv}{\coercion_4}{\emptyset \le \dirt'_2} 
           } 
           { \coToNoEff{\tyEnv}
                       {(\coercion_1~!~\coercion_3 \hto \coercion_2~!~\coercion_4)}
                       {((\vty_1~!~\dirt_1 \hto \vty'_1~!~\emptyset) \le (\vty_2~!~\emptyset \hto \vty'_2~!~\dirt'_2))}
                       {\handToFun{\mlCoercion'_1}{(\unsafe{\mlCoercion'_2})}}
           } 

\inferrule*[right=]
           { \fullDirt{\dirt_1} \\
             \fullDirt{\dirt'_1} \\
             \coToNoEff{\tyEnv}{\coercion_1}{\vty_2 \le \vty_1}{\mlCoercion'_1} \\
             \coToNoEff{\tyEnv}{\coercion_2}{(\vty'_1 \le \vty'_2)}{\mlCoercion'_2} \\
             \tcTrgCo{\tyEnv}{\coercion_3}{\emptyset \le \dirt_1} \\
             \tcTrgCo{\tyEnv}{\coercion_4}{\dirt'_1 \le \dirt'_2}
           } 
           { \coToNoEff{\tyEnv}
                       {(\coercion_1~!~\coercion_3 \hto \coercion_2~!~\coercion_4)}
                       {((\vty_1~!~\dirt_1 \hto \vty'_1~!~\dirt'_1) \le (\vty_2~!~\emptyset \hto \vty'_2~!~\dirt'_2))}
                       {\handToFun{\mlCoercion'_1}{\mlCoercion'_2}}
           } 

\inferrule*[right=]
           { \coToNoEff{\tyEnv, \evar}{\coercion}{\vty_1 \le \vty_2}{\mlCoercion'} }
           { \coToNoEff{\tyEnv}{\forall \evar. \coercion}{\forall \evar. \vty_1 \le \forall \evar. \vty_2}{\mlCoercion'} }

\inferrule*[right=]
           { \tcSkeleton{\tyEnv}{\ety} \\
             \coToNoEff{\tyEnv, \tyvar : \ety}{\coercion}{\vty_1 \le \vty_2}{\mlCoercion'}
           } 
           { \coToNoEff{\tyEnv}{\forall (\tyvar : \ety). \coercion}{\forall (\tyvar : \ety). \vty_1 \le \forall (\tyvar : \ety). \vty_2}{\forall \tyvar. \mlCoercion'} }

\inferrule*[right=]
           { \coToNoEff{\tyEnv, \dirtvar}{\coercion}{\vty_1 \le \vty_2}{\mlCoercion'} }
           { \coToNoEff{\tyEnv}{\forall \dirtvar. \coercion}{\forall \dirtvar. \vty_1 \le \forall \dirtvar. \vty_2}{\mlCoercion'} }

\inferrule*[right=]
           { \coToNoEff{\tyEnv}{\coercion}{\vty_1 \le \vty_2}{\mlCoercion'} \\
             \valTyToNoEff{\tyEnv}{\vty_3}{\ety}{\mlTyA_1} \\
             \valTyToNoEff{\tyEnv}{\vty_4}{\ety}{\mlTyA_2}
           } 
           {\coToNoEff{\tyEnv}{(\vty_3 \le \vty_4) \Rightarrow \coercion}{((\vty_3 \le \vty_4) \Rightarrow \vty_1) \le ((\vty_3 \le \vty_4) \Rightarrow \vty_2)}{(\mlTyA_1 \le \mlTyA_2) \Rightarrow \mlCoercion'}}

\inferrule*[right=]
           { \coToNoEff{\tyEnv}{\coercion}{\vty_1 \le \vty_2}{\mlCoercion'} \\
             \compTyToNoEff{\tyEnv}{\cty_1}{\ety}{\mlTyB_1} \\
             \compTyToNoEff{\tyEnv}{\cty_2}{\ety}{\mlTyB_2}
           } 
           {\coToNoEff{\tyEnv}{(\cty_1 \le \cty_2) \Rightarrow \coercion}{((\cty_1 \le \cty_2) \Rightarrow \vty_1) \le ((\cty_1 \le \cty_2) \Rightarrow \vty_2)}{(\mlTyB_1 \le \mlTyB_2) \Rightarrow \mlCoercion'}}

\inferrule*[right=]
           { \coToNoEff{\tyEnv}{\coercion}{\vty_1 \le \vty_2}{\mlCoercion'} }
           { \coToNoEff{\tyEnv}{(\dirt_1 \le \dirt_2) \Rightarrow \coercion}{(\dirt_1 \le \dirt_2) \Rightarrow \vty_1 \le (\dirt_1 \le \dirt_2) \Rightarrow \vty_2}{\mlCoercion'} }

\inferrule*[right=]
           { \coToNoEff{\tyEnv}{\coercion_1}{\vty_1 \le \vty_2}{\mlCoercion'_1} \\
             \tcTrgCo{\tyEnv}{\coercion_2}{\emptyset \le \emptyset}
           } 
           { \coToNoEff{\tyEnv}{(\coercion_1~!~\coercion_2)}{(\vty_1~!~\emptyset \le \vty_2~!~\emptyset)}{\mlCoercion'_1} }

\inferrule*[right=]
           { \coToNoEff{\tyEnv}{\coercion_1}{\vty_1 \le \vty_2}{\mlCoercion'_1} \\\\
             \tcTrgCo{\tyEnv}{\coercion_2}{\emptyset \le \dirt_2} \\
             \fullDirt{\dirt_2}
           } 
           { \coToNoEff{\tyEnv}{(\coercion_1~!~\coercion_2)}{(\vty_1~!~\emptyset \le \vty_2~!~\dirt_2)}{\mlReturn{\mlCoercion'_1}} }

\inferrule*[right=]
           { \coToNoEff{\tyEnv}{\coercion_1}{\vty_1 \le \vty_2}{\mlCoercion'_1} \\
             \tcTrgCo{\tyEnv}{\coercion_2}{\dirt_1 \le \dirt_2} \\
             \fullDirt{\dirt_1} \\
             \fullDirt{\dirt_2}
           } 
           { \coToNoEff{\tyEnv}{(\coercion_1~!~\coercion_2)}{(\vty_1~!~\dirt_1 \le \vty_2~!~\dirt_2)}{\mkMlCompCo{\mlCoercion'_1}} }
\end{mathpar}

\vspace{-5mm}
\caption{Elaboration of \target Coercions to \noeff Coercions}
\label{fig:eff-to-ml-coercions}
\end{myfigure}

Elaboration of \target coercions to \noeff coercions is given in
Figure~\ref{fig:eff-to-ml-coercions}.

\begin{myfigure}[t!]
$\ruleform{\valToNoEff{\tyEnv}{v}{\vty}{\mlTm}}$ \textbf{Value Elaboration}
\begin{mathpar}
\inferrule*[right=]
           { }
           { \valToNoEff{\tyEnv}{\tmUnit}{\tyUnit}{\tmUnit} }

\inferrule*[right=]
           { \valTyToNoEff{\tyEnv}{\vty}{\ety}{\mlTyA} \\
             \compToNoEff{\tyEnv, x : \vty}{c}{\cty}{\mlTm}
           } 
           { \valToNoEff{\tyEnv}{\fun{(x:\vty)}{c}}{\vty \to \cty}{\fun{(x:\mlTyA)}{\mlTm}} }


\inferrule*[right=]
           { \valTyToNoEff{\tyEnv}{\vty}{\ety}{\mlTyA} \\
             \compToNoEff{\tyEnv,x\!:\!\vty}{c_r}{\cty}{\mlTm}
           } 
           { \valToNoEff{\tyEnv}{\{\return{(x : \vty)} \mapsto c_r\}}{\withdirt{\vty}{\emptyset} \hto \cty}{\fun{(x:\mlTyA)}{\mlTm}} }

\inferrule*[right=]
           { \fullDirt{\ops} \\
             \valTyToNoEff{\tyEnv}{\vty_x}{\ety}{\mlTyA} \\
	     \compToNoEff{\tyEnv, x\!:\!\vty_x}{c_r}{\withdirt{\vty}{\emptyset}}{\mlTm_r} \\
             \left[
               (\op : \vty_1^\op \to \vty_2^\op) \in \sig \quad
               \valTyToNoEff{}{\vty_i^\op}{\ety_i^\op}{\mlTyA_i^\op} \qquad
               \compToNoEff{\tyEnv, x : \vty_1^\op, k : \vty_2^\op \to \vty\,!\,\emptyset}{c_\op}{\vty\,!\,\emptyset}{\mlTm_\op}
             \right]_{\op \in \ops}
           } 
           { \tyEnv \vdashNamedD{v} \trgShorthand : \withdirt{\vty_x}{\ops} \hto \withdirt{\vty}{\emptyset} \\
                  \highlight{\rightsquigarrow 
                        {\handler{\return{(x : \mlTyA)} \mapsto \return{\mlTm_r}
                        ,\big[\call{\op}{x}{k} \mapsto \return{\mlTm_\op[\cast{k}{\refl{\mlTyA_1^\op} \to \unsafe{\refl{\mlTyA_2^\op}}}/k]}\big]_{\op \in \ops}}}
                  }
           }

\inferrule*[right=]
           { \fullDirt{\ops} \\
             \fullDirt{\dirt} \\
             \valTyToNoEff{\tyEnv}{\vty_x}{\ety}{\mlTyA} \\
	     \compToNoEff{\tyEnv, x\!:\!\vty_x}{c_r}{\withdirt{\vty}{\dirt}}{\mlTm_r} \\
             \left[
               (\op : \vty_1^\op \to \vty_2^\op) \in \sig \qquad
               \compToNoEff{\tyEnv, x : \vty_1^\op, k : \vty_2^\op \to \vty\,!\,\dirt}{c_\op}{\vty\,!\,\dirt}{\mlTm_\op}
             \right]_{\op \in \ops}
           } 
           { \tyEnv \vdashNamedD{v} \trgShorthand : \withdirt{\vty_x}{\ops} \hto \withdirt{\vty}{\dirt}
                  \highlight{\rightsquigarrow 
                        {\handler{\return{(x : \mlTyA)} \mapsto {\mlTm_r}
                        ,[\call{\op}{x}{k} \mapsto {\mlTm_\op}]_{\op \in \ops}}}
                  }
           }

%
%

\inferrule*[right=]
           { \valToNoEff{\tyEnv, \evar}{v}{\vty}{\mlTm} }
           { \valToNoEff{\tyEnv}{\Lambda \evar. v}{\forall \evar. \vty}{\mlTm} }

\inferrule*[right=]
           { \valToNoEff{\tyEnv}{v}{\forall \evar. \vty}{\mlTm} }
           { \valToNoEff{\tyEnv}{v~\ety}{\vty[\ety/\evar]}{\mlTm} }

\inferrule*[right=]
           { \valToNoEff{\tyEnv, \tyvar : \ety}{v}{\vty}{\mlTm} }
           { \valToNoEff{\tyEnv}{\Lambda (\tyvar : \ety). v}{\forall (\tyvar : \ety). \vty}{\Lambda \tyvar. \mlTm} }

\inferrule*[right=]
           { \valToNoEff{\tyEnv}{v}{\forall (\tyvar : \ety). \vty}{\mlTm} \\\\
             \valTyToNoEff{\tyEnv}{\vty_1}{\ety}{\mlTyA}
           } 
           { \valToNoEff{\tyEnv}{v~\vty_1}{\vty[\vty_1/\tyvar]}{\mlTm~\mlTyA} }

\inferrule*[right=]
           { \valToNoEff{\tyEnv, \dirtvar}{v}{\vty}{\mlTm} }
           { \valToNoEff{\tyEnv}{\Lambda \dirtvar. v}{\forall \dirtvar. \vty}{\mlTm} }

\inferrule*[right=]
           { \valToNoEff{\tyEnv}{v}{\forall \dirtvar. \vty}{\mlTm} \\\\
             \fromImpureVal{\vty}{\dirt}{\mlCoercion}
           } 
           { \valToNoEff{\tyEnv}{v~\dirt}{\vty[\dirt/\dirtvar]}{\cast{\mlTm}{\mlCoercion}} }

\inferrule*[right=]
           { \valToNoEff{\tyEnv, \covar : \vty_1 \le \vty_2}{v}{\vty}{\mlTm} \\
             \valTyToNoEff{\tyEnv}{\vty_1}{\ety}{\mlTyA} \\
             \valTyToNoEff{\tyEnv}{\vty_2}{\ety}{\mlTyB} \\
           } 
           { \valToNoEff{\tyEnv}{\Lambda (\covar : \vty_1 \le \vty_2). v}{(\vty_1 \le \vty_2 \Rightarrow \vty)}{\Lambda (\covar : \mlTyA \le \mlTyB). \mlTm} }

\inferrule*[right=]
           { \valToNoEff{\tyEnv, \covar : \cty_1 \le \cty_2}{v}{\vty}{\mlTm} \\
             \compTyToNoEff{\tyEnv}{\cty_1}{\ety}{\mlTyA} \\
             \compTyToNoEff{\tyEnv}{\cty_2}{\ety}{\mlTyB} \\
           } 
           { \valToNoEff{\tyEnv}{\Lambda (\covar : \cty_1 \le \cty_2). v}{(\cty_1 \le \cty_2 \Rightarrow \vty)}{\Lambda (\covar : \mlTyA \le \mlTyB). \mlTm} }

\inferrule*[right=]
           { \valToNoEff{\tyEnv, \covar : \dirt_1 \le \dirt_2}{v}{\vty}{\mlTm} }
           { \valToNoEff{\tyEnv}{\Lambda (\covar : \dirt_1 \le \dirt_2). v}{(\dirt_1 \le \dirt_2 \Rightarrow \vty)}{\mlTm} }

\inferrule*[right=]
           { \valToNoEff{\tyEnv}{v}{(\vty_1 \le \vty_2) \Rightarrow \vty}{\mlTm} \\\\
             \coToNoEff{\tyEnv}{\coercion}{\vty_1 \le \vty_2}{\mlCoercion'}
           } 
           { \valToNoEff{\tyEnv}{v~\coercion}{\vty}{\mlTm~\mlCoercion'} }

\inferrule*[right=]
           { \valToNoEff{\tyEnv}{v}{(\cty_1 \le \cty_2) \Rightarrow \vty}{\mlTm} \\\\
             \coToNoEff{\tyEnv}{\coercion}{\cty_1 \le \cty_2}{\mlCoercion'}
           } 
           { \valToNoEff{\tyEnv}{v~\coercion}{\vty}{\mlTm~\mlCoercion'} }

\inferrule*[right=]
           { \valToNoEff{\tyEnv}{v}{(\dirt_1 \le \dirt_2) \Rightarrow \vty}{\mlTm} \\\\
             \tcTrgCo{\tyEnv}{\coercion}{\dirt_1 \le \dirt_2}
           }
           { \valToNoEff{\tyEnv}{v~\coercion}{\vty}{\mlTm} }

\inferrule*[right=]
           { \valToNoEff{\tyEnv}{v}{\vty_1}{\mlTm} \\\\
             \coToNoEff{\tyEnv}{\coercion}{\vty_1 \le \vty_2}{\mlCoercion'}
           } 
           { \valToNoEff{\tyEnv}{\cast{v}{\coercion}}{\vty_2}{\cast{\mlTm}{\mlCoercion'}} }
\end{mathpar}

\vspace{-5mm}
\caption{Elaboration of \target Values to \noeff Terms}
\label{fig:eff-to-ml-values}
\end{myfigure}

Figure~\ref{fig:eff-to-ml-values} shows the elaboration of \target values into
\noeff terms.

\end{document}